%% file: main.tex
\PassOptionsToPackage{dvipsnames}{xcolor} %
\documentclass[manuscript, screen, nonacm]{acmart}

\newcommand{\shortVersion}{false} %

\AtBeginDocument{%
  \providecommand\BibTeX{{%
    \normalfont B\kern-0.5em{\scshape i\kern-0.25em b}\kern-0.8em\TeX}}}

\usepackage{etoolbox}
\makeatletter
\patchcmd{\maketitle}{\@copyrightpermission}{
   \begin{minipage}{0.3\columnwidth}
     \href{https://creativecommons.org/licenses/by/4.0/}{\includegraphics[width=0.90\textwidth]{cc_by4acm.png}}
   \end{minipage}\hfill
   \begin{minipage}{0.7\columnwidth}
     \href{https://creativecommons.org/licenses/by/4.0/}{This work is licensed under a Creative Commons Attribution International 4.0 License.}
   \end{minipage}
 
   \vspace{5pt}
}{}{}

\makeatother

\input{packages}

\usetikzlibrary{tikzmark}    
\usetikzmarklibrary{listings}
\usepackage{subcaption}
\usepackage{multirow}
\usepackage{pgf}

\newcommand{\todoMarcoP}[1]{}
\newcommand{\MP}[1]{}

\newcommand{\todoMarcoG}[1]{}

\newcommand{\todoX}[1]{}

\newcommand{\todoJan}[1]{}

\newcommand{\todoBK}[1]{}

\makeatletter
\def\blfootnote{\xdef\@thefnmark{}\@footnotetext}
\makeatother

\title{Detecting speculative leaks with compositional semantics}

\author{Xaver Fabian}
\affiliation{%
  \institution{CISPA Helmholtz Center for Information Security}
  \city{Saarbr\"ucken}
  \country{Germany}
}
\email{xaver.fabian@cispa.de}

\affiliation{
  \institution{University of Trento}
  \city{Trento}
  \country{Italy}
}

\author{Marco Guarnieri}
\affiliation{%
  \institution{IMDEA Software Institute}
  \city{Madrid}
  \country{Spain}
}
\email{marco.guarnieri@imdea.org}

\author{Boris K\"opf}
\affiliation{%
  \institution{Azure Research, Microsoft}
  \country{UK}
}

\author{Jose F. Morales}
\affiliation{%
  \institution{IMDEA Software Institute}
  \city{Madrid}
  \country{Spain}
}

\author{Marco Patrignani}
\affiliation{%
  \institution{University of Trento}
  \city{Trento}
  \country{Italy}}
\email{marco.patrignani@unitn.it}

\author{Jan Reineke}
\affiliation{%
  \institution{Saarland University}
  \city{Saarbr\"ucken}
  \country{Germany}
}

\author{Andres Sanchez}
\affiliation{%
  \institution{Amazon}
  \city{Madrid}
  \country{Spain}
}

\renewcommand{\shortauthors}{Fabian et al.}

\begin{document}

\begin{abstract}

Speculative execution enhances processor performance by predicting intermediate results and executing instructions based on these predictions. 
However, incorrect predictions can lead to security vulnerabilities, as speculative instructions leave traces in microarchitectural components that attackers can exploit. This is demonstrated by the family of Spectre attacks.
Unfortunately, existing countermeasures to these attacks lack a formal security characterization, making it difficult to verify their effectiveness.

In this paper, we propose a novel framework for detecting information flows introduced by speculative execution and reasoning about software defenses. 
The theoretical foundation of our approach is \emph{speculative non-interference} (SNI), a novel semantic notion of security against speculative execution attacks. SNI relates information leakage observed under a standard non-speculative semantics to leakage arising under semantics that explicitly model speculative execution.
To capture their combined effects, we extend our framework with a mechanism to safely compose multiple speculative semantics, each focussing on a single aspect of speculation.%
This allows us to analyze the complex interactions and resulting leaks that can arise when multiple speculative mechanisms operate together.
On the practical side, we develop Spectector, a symbolic analysis tool that uses our compositional framework and leverages SMT solvers to detect vulnerabilities and verify program security with respect to multiple speculation mechanisms.
We demonstrate the effectiveness of Spectector through evaluations on standard security benchmarks and new vulnerability scenarios.

\end{abstract}



\maketitle

\section{Introduction}\label{sec:intro}
{Speculative execution} avoids pipeline stalls by predicting the results of intermediate computations and by speculatively executing instructions based on such predictions. 
Whenever a prediction turns out to be incorrect, the processor squashes the speculatively executed instructions, thereby rolling back their effects on the architectural state, which consists of registers, flags, and memory.
However, the execution of speculative instructions leaves footprints in a CPU's microarchitectural components (like caches, predictors, and internal buffers) that may persist even after these instructions have been squashed. 
As shown by Spectre~\cite{spectre} and follow-up attacks~\cite{spectreRsb,ret2spec,S_smotherSpectre,S_trans_troj,barberis2022branch, wikner_phantom_2023, retbleed, inception}, these microarchitectural side effects can be exploited to compromise the security of programs and to leak information about speculatively accessed data.\looseness=-1 %

\paragraph{Speculative Leaks}
Modern CPUs employ a wide range of speculation mechanisms (branch predictors, memory disambiguators, etc.) that are used to speculate over different kinds of instructions and intermediate results, such as conditional branches~\cite{spectre}, indirect jumps~\cite{spectre}, store and load operations~\cite{S_specv4},  return instructions~\cite{spectreRsb, sls-whitepaper}, and load addresses and loaded results~\cite{flop, slap, val_pred2}. 
All these speculation mechanisms can be exploited by attackers to leak data.

\begin{figure}
     \begin{minipage}{.40\textwidth}
        \begin{lstlisting}[style=Cstyle, caption=Code vulnerable to branch speculation. , label=lst:v1-vanilla, escapechar=|]
            if (y < size)|\label{line:condition}|
             temp &= B[A[y] * 512]; |\label{line:v1br}|
            \end{lstlisting}
       \vspace{1pt}
        \begin{lstlisting}[basicstyle=\small,style=Cstyle, caption=Code vulnerable to store speculation., label=lst:v4-vanilla1,escapechar=|, captionpos=t]
        p = &secret;            |\label{line:v4sec1}|
        p = &public;            |\label{line:v4pub1}|
        temp = B[*p * 512];     |\label{line:leakv41}|
        \end{lstlisting}
    \end{minipage}
    \hspace{10pt}
    \begin{minipage}{0.45\textwidth}
        \begin{lstlisting}[basicstyle=\small,style=Cstyle,
    caption={Speculative leak arising from speculation over branch and store instructions combined.}, 
    label=lst:v1-v4-combined,escapechar=|, captionpos=t]
x = 0; |\label{line:v14as}|
p = &secret; 
p = &public;    |\label{line:v14pub}|
if (x != 0)     |\label{line:v14branch}| 
    temp &= A[*p];      |\label{line:v14leak}|
\end{lstlisting}
    \end{minipage}
    \caption{Code snippets vulnerable to different kinds of speculation attacks. }
    \label{fig:v1-and-v4-example}
\end{figure}

As an example, the code in \Cref{lst:v1-vanilla} is vulnerable to a Spectre-PHT attack~\cite{spectre}, which exploits speculation over branch instructions.
Whenever the branch prediction mispredicts the outcome of the condition $\inlineCcode{y > size}$ on \Cref{line:condition}, this results in speculatively executing \Cref{line:v1br}, which leaks the out-of-bounds value pointed by $\inlineCcode{A[y]}$ through the data cache.
As another example,  \Cref{lst:v4-vanilla1} depicts a code snippet vulnerable to a Spectre-STL attack~\cite{S_specv4}, which exploits speculation over memory disambiguation checks.
Whenever the memory write on \Cref{line:v4pub1} is predicted to have a different address than the memory read $\inlineCcode{*p}$ on \Cref{line:leakv41},  the content of $\inlineCcode{secret}$ is leaked to an attacker. 
Essentially, the memory write in \Cref{line:v4pub1} is bypassed because of speculation.

The majority of well-known attacks~\cite{spectre, spectreRsb,ret2spec, S_specv4, straight-line-spec} only exploit leaks introduced by  individual speculation mechanisms, e.g., branch predictors in Spectre-PHT (\Cref{lst:v1-vanilla}) and memory disambiguators in Spectre-STL (\Cref{lst:v4-vanilla1}).
However, some speculative leaks only arise due to the interaction of multiple mechanisms.
For example, the code in \Cref{lst:v1-v4-combined} can speculatively leak the value of \inlineCcode{&secret} in \Cref{line:v14leak} whenever (1) the memory write to \inlineCcode{p} in \Cref{line:v14pub} is predicted to have a different address then the memory read \inlineCcode{*p} on \Cref{line:v14leak}, and (2) the branch instruction on \Cref{line:v14branch} is mispredicted as taken.
This leak, therefore, arises from the \emph{combination} of two speculation mechanisms---branch prediction and memory disambiguation prediction---and \emph{it cannot be detected} when considering the two mechanisms in isolation.

\paragraph{Limitations of existing leak-detection approaches}
Since the discovery of Spectre~\cite{spectre}, a number of approaches have been proposed for reasoning about the security of programs against speculative leaks~\cite{ST_binsec, kleeSpectre, ST_specusym, ST_constantTime_Spec, oo7, ST_spectector2, aise, revizor, revizor2, ST_specfuzz, specDoc, revizor_23}.

These approaches employ a wide array of techniques for reasoning about leaks (e.g., symbolic execution~\cite{ST_binsec, kleeSpectre, ST_specusym, ST_constantTime_Spec, oo7, ST_spectector2}, abstract interpretation~\cite{aise}, testing~\cite{revizor, revizor2, ST_specfuzz, specDoc, revizor_23}), but they all share a critical limitation.
They support only \emph{fixed} speculation mechanisms, such as branch prediction~\cite{spectector,ST_spectector2, oo7,kleeSpectre,ST_blade} and memory disambiguation prediction~\cite{ST_constantTime_Spec, ST_binsec, cats}, which are \emph{hard-coded}, e.g., in the underlying formal models.
As a result, extending them to reason about a new speculation mechanism is often complicated (e.g., it might require modifying the underlying formal model and updating the corresponding security proofs), since none of these approaches has been designed in a \emph{compositional} manner.\looseness=-1

As shown by code snippets like \Cref{lst:v1-v4-combined}, sound reasoning about speculative leaks in a program requires accounting for \emph{all} speculation mechanisms, since ignoring some mechanisms might lead to missed leaks.
Unfortunately, (1) we lack a precise understanding of the speculation mechanisms implemented in current CPUs, as demonstrated time and again by attacks discovering previously unknown speculation mechanisms, and (2) new CPU generations often implement refined and improved speculation mechanisms to improve performance~\cite{slap,flop}.
Since we cannot develop an analysis that would account for future microarchitectures, approaches for reasoning about speculative leaks must be \emph{compositional} and easy to \emph{extend} whenever a new speculation mechanism is discovered.
However, we currently lack a precise characterization of security against speculative leaks and an associated program analysis framework that is \emph{extensible and compositional} in the underlying speculation mechanisms.

\paragraph{Our approach:}
In this paper, we develop a novel, principled approach for reasoning about leaks introduced by speculatively executed instructions and for reasoning about software defenses against Spectre-style attacks.
Our approach is backed by a semantic characterization of security against speculative leaks and it comes with an algorithm, based on symbolic execution, for proving the absence of leaks.
Crucially, our approach is \emph{compositional}.
It allows specifying individual operational semantics---each one capturing the effects of a different speculation mechanism---and combining them in a single composed semantics, which captures leaks arising from the interactions of all these mechanisms, thereby leading to simpler formalizations.
Additionally, these semantics can be directly integrated in our verification algorithm, whose soundness proof is compositional and defined in terms of the component semantics, thereby maximizing proof reuse.
Next, we describe our contributions in more detail:
\begin{asparaenum}
\item \textbf{Modeling speculation:}
We propose a generalized template for modeling the effects of speculative execution as an operational semantics.
Our template extends an architectural (non-speculative) semantics (\Cref{sec:bg}) to a \emph{speculative semantics} (\Cref{sec:spec-semantics}) that captures the effects of speculatively executed instructions.
In a nutshell, the speculative semantics follow mispredicted paths for a bounded number of steps before backtracking and restarting the architectural execution, while relying on a \emph{prediction oracle} to model the speculation mechanism.
Note that this template is flexible enough to capture both control-flow and data-flow speculation.
Following this template, we introduce five speculative semantics (denoted $\semb$, $\semj$, $\sems$, $\semr$, and $\semsls$) which capture  speculation over branches, jumps, stores, and return instructions (\Cref{sec:inst-spec-semantics-single}).
This is the most comprehensive collection of individual speculative semantics in a single framework to date. %

\item \textbf{Speculative non-interference:}
We propose \textit{speculative non-interference} (\Cref{sec:sni}), a novel semantic notion of security against speculative execution attacks.
Speculative non-interference is based on comparing a
program with respect to two different semantics:
\begin{inparaenum}[(a)]
\item  a standard, {\em non-speculative semantics}, which we use as a proxy for the intended program behavior, and
\item  a {\em speculative semantics} that captures the effects of speculation induced by one or more speculation mechanisms, which captures the effect of   speculatively executed instructions.
\end{inparaenum}
In a nutshell, speculative non-interference requires that {\em
  speculatively executed instructions do not leak more information
  into the microarchitectural state than what the intended behavior does},
i.e., than what is leaked by the standard, non-speculative semantics.\looseness=-1

To capture ``leakage into the microarchitectural state'', we
consider an observer of the program execution that sees the locations
of memory accesses and jump targets. This observer model is commonly used for
characterizing ``side-channel free'' or ``constant-time''
code~\cite{MolnarPSW05,AlmeidaBBDE16} in the absence of detailed
models of the microarchitecture.
Under this observer model, an adversary may distinguish two initial program states if they yield different traces of memory locations and jump targets.
{\em Speculative non-interference} (SNI)  requires that two initial program states can be distinguished under the speculative semantics only if they can also be distinguished under the standard, non-speculative semantics.

The speculative semantics, and hence SNI, depends on the decisions taken by the prediction oracle. 
We show that one can abstract away from the specific oracle by considering a worst-case oracle that \emph{always mispredicts}. SNI w.r.t. this always-mispredict oracle implies SNI w.r.t. a large class of prediction strategies.
This allows reasoning about speculative leaks while ignoring the details of the specific prediction strategy, which are often undocumented. %

\item \textbf{Composition framework:}
We propose a framework for composing speculative semantics that capture speculation due to different mechanisms (\Cref{sec:frame}).
The framework allows specifying individual semantics for each speculation mechanism and combining them into a composed semantics.
The combination yields a single operational semantics which can be used to reason about leaks involving the different kinds of speculation it comprises (as in \Cref{lst:v1-v4-combined}).
We also formalize the key properties of our framework: if the individual semantics fulfill some (expected) safety conditions (which we prove for all the semantics we combine), then the composed semantics is well-formed and can be used to reason about leakage in a sound manner.
That is, whenever a program is leaky under one of the individual semantics then it is leaky under the composed semantics and, vice versa, whenever a program is SNI under the composed semantics, then it is SNI w.r.t. all individual semantics.
We apply the composition framework to all our individual semantics, thereby obtaining \nrComb{} combined speculative semantics (\Cref{sec:comb-in}).

\item \textbf{Checking speculative non-interference:}
We propose \tool{} (\Cref{sec:impl-spec}), an algorithm to automatically prove that programs satisfy SNI w.r.t. any speculative semantics (or composition) in our framework.
Given a program~$p$, \tool{} uses symbolic execution with respect to the speculative semantics to derive a concise representation of the traces of memory accesses and jump targets during execution along all possible program paths. 
Based on this representation, \tool creates a symbolic formula capturing that, whenever two initial program states produce the
same (instruction and data) memory access patterns in the standard semantics, they also produce the same access patterns in the speculative semantics. 
Validity of this formula for each program path implies SNI.

\item \textbf{Implementation and evaluation:}
We implement a prototype of \tool{}  (\Cref{sec:eval}), with a front-end parsing (a subset of) x86 assembly and a back-end that uses the Z3 SMT solver to perform symbolic execution against all our speculative semantics (and their combinations) and to determine whether programs contain speculative leaks. 
We validate \tool on both existing microbenchmarks (for speculation on branches, indirect jumps, store and return instructions) and on new ones (for speculative leaks arising from combinations of speculation mechanisms) that we introduce.
These microbenchmarks contain both leaky programs as well as programs patched with well-known compiler-level countermeasures against Spectre.
Using \tool{}, we successfully 
(1) detect all leaks pointed out in~\cite{S_koch_mit}, 
(2) detect novel, subtle leaks that are out of the scope of existing approaches that check for known vulnerable code patterns~\cite{oo7}, 
(3) detect leaks arising from speculation over multiple mechanisms that are not detected by existing symbolic approaches~\cite{hauntedBugReport},
and (4) identify cases where compilers unnecessarily inject countermeasures, i.e., opportunities for optimization without sacrificing security.
\end{asparaenum}

\paragraph{Scope of this paper}
This paper provides a unified, extended, and updated version of the results presented in two conference papers by \citet{spectector} and \citet{fabian2022automatic}.
In addition to the original results from~\cite{spectector,fabian2022automatic},  (1) the paper introduces a general template for speculative operational semantics that generalizes existing formalizations, (2) it adds new speculative semantics for straight-line speculation and for speculation over indirect jumps (and their combinations), and (3) it provides additional technical details about the speculative semantics as well as about composition proofs.
The paper is accompanied by an updated version of  \tool{} (with an updated evaluation) that supports all new speculative semantics and combinations presented in the paper.
Overall, we see this paper as a unified reference for the modeling of speculative leaks at program level and for  the \tool{} program analysis tool.

\paragraph{Additional materials}
The \tool program analysis tool is open-source and available at~\cite{tool-og}.
Given the number of different speculative semantics studied in this paper, we leave the full formalization of all semantics, technicalities, and proof details to the associated technical report, which is available at~\cite{techReport}.

\input{src/illustrative-example}

\input{src/background.tex}

\input{src/Semantics/spec_semantics_2}

\input{src/combined_semantics}

\input{src/combined_instances}
\input{src/implementation}
\input{src/Evaluations/evaluation}

\input{src/discussion}

\input{src/related_work}

\input{src/future_work}

\newpage

\bibliographystyle{ACM-Reference-Format}
\balance
\bibliography{Refs/spectre, Refs/spectreTools, Refs/references}

\newpage

\listoftodos[List of suggested changes]{}
\appendix

\input{src/Appendix/code-case-studies}
\input{src/Appendix/spectector_variants}
\input{src/Appendix/trace-projections}

\renewcommand{\sigmaa}{\sigma_{\Symb}}
\todoX{package enumitem needed but clashes and throws errors.
enumitem needed to add label to enumerate environments. Ignore errors for nwo}

\end{document}

%% file: packages.tex
\usepackage[utf8]{inputenc}
\usepackage{balance}
\usepackage{extarrows}
\usepackage{lscape}
\usepackage{amsfonts}
\usepackage{amsmath}
\usepackage{graphicx}
\usepackage{xifthen}
\usepackage{placeins}

\usepackage{halloweenmath}
\usepackage{color}
\usepackage[disable]{todonotes}
\usepackage{amsthm}

\usepackage{pifont}%
\usepackage{mathtools}
\usepackage{mathpartir}
\usepackage{natbib} %
\usepackage{relsize}
\usepackage{listings, lstautogobble}
\usepackage{listingsutf8}
\usepackage{caption}
\usepackage{thmtools, thm-restate} 
\usepackage{booktabs}
\usepackage{paralist}
\usepackage{algorithm}
\usepackage[noend]{algpseudocode}

\usepackage{cleveref} %
\declaretheorem{theorem}

\declaretheorem[]{definition}

\declaretheorem[name=Property]{requirement}

\theoremstyle{definition}

\newtheorem{example}{Example}

\theoremstyle{plain}

\usepackage{tikz}
\usepackage{hf-tikz}
\usetikzlibrary{positioning,shadows,arrows,calc,backgrounds,fit,shapes,snakes,shapes.multipart,decorations.pathreplacing,shapes.misc,patterns}
\usepackage[edges]{forest}
\usepackage{xspace}
\usepackage{theoremref} %

\usepackage{stmaryrd} %

\usepackage[Export]{adjustbox}[2018/04/08]
\usepackage{pdfpages}

\usepackage{xargs}
\usepackage{hyperref}

\usepackage{wrapfig}
\usepackage{extarrows}
\usepackage{soulutf8}
\usepackage{enumitem}
\newcommand{\ulc}[2]{\setulcolor{#1}\ul{#2}}

\colorlet{pred}{Maroon}
\colorlet{pgrn}{SeaGreen}%
\colorlet{pblu}{Cerulean}
\colorlet{porg}{BurntOrange}
\setul{}{1pt}

\usepackage{numprint}
\usepackage[inference]{semantic}
\usepackage{semanticMacros}
\newcounter{typerule}
\crefname{typerule}{rule}{rules}

\newcommand{\typeruleInt}[5]{%
	\def\thetyperule{#1}%
	\refstepcounter{typerule}%
	\label{tr:#4}%
  \ensuremath{\begin{array}{c}#5 \inference{#2}{#3}\end{array}} 
}
\newcommand{\typerule}[4]{%
  \typeruleInt{#1}{#2}{#3}{#4}{\textsf{\scriptsize ({#1})} \\      }
}

\newcommand{\mytoprule}[1]{\vspace{1mm}\noindent\hrulefill\ \raisebox{-0.5ex}{\fbox{\ensuremath{#1}}} \hrulefill\hrulefill\hrulefill\vspace{0.5mm}}

\makeatletter
\xdef\@thefnmark{\@empty}

\makeatother

%% file: src/illustrative-example.tex
\section{Illustrative example}\label{sec:illustration}

To illustrate our approach, we show how  \tool{} applies to the \specb example~\cite{spectre} shown in \Cref{lst:v1-vanilla} using a speculative semantics capturing branch speculation.

\para{\specb}
The program checks whether the index stored in the variable \inlineCcode{y} is less than the size of the array \inlineCcode{A}, stored in the variable \inlineCcode{size}.
If that is the case, the program retrieves \inlineCcode{A[y]}, amplifies it with a multiple (here: \inlineCcode{512}) of the cache line size, and uses the result as an address for accessing the array \inlineCcode{B}.\looseness=-1

If \inlineCcode{size} is not cached, evaluating the branch condition requires traditional processors to wait until \inlineCcode{size} is fetched from main memory.
Modern processors instead speculate on the condition's outcome and continue the computation.
Hence, the memory accesses in line 2 may be executed even if $\inlineCcode{y} \geq \inlineCcode{size}$.\looseness=-1

When \inlineCcode{size} becomes available, the processor checks whether the speculated branch is the correct one.
If it is not, it rolls back the architectural (i.e., ISA) state's changes and executes the correct branch.
However, the speculatively executed memory accesses leave a footprint
in the microarchitectural state, in particular in the cache, which enables an adversary to retrieve 
\inlineCcode{A[y]}, even for $\inlineCcode{y} \geq \inlineCcode{size}$, by probing the array \inlineCcode{B}.

\para{Detecting leaks with \tool{}}
\tool{} automatically detects leaks introduced by speculatively executed instructions, or proves their absence.
Specifically, \tool{} detects a leak whenever executing the program
under the speculative semantics, which captures that the execution can
go down a mispredicted path for a bounded number of steps, leaks more
information into the microarchitectural state than executing the program
under a non-speculative semantics.\looseness=-1

\begin{lstlisting}[style=ASMstyle, 
  float=tp,
  floatplacement=tbp, caption={\specb - Assembly code}, label=figure:spectrev1:asm-code]
 mov	size, %
 mov	y, %
 cmp	%
 jbe	END
 mov	A(%
 shl	$9, %
 mov	B(%
 and	%
\end{lstlisting}

To illustrate how \tool{} operates, we consider the x86 assembly\footnote{We use a simplified AT\&T syntax without operand sizes} translation of \Cref{lst:v1-vanilla}'s program (cf.~\Cref{figure:spectrev1:asm-code}).
\tool{} performs symbolic execution with respect to the speculative semantics to derive a concise representation of the concrete traces of memory accesses and program counter values along each path of the program.
These symbolic traces capture the program's effect on the microarchitectural state. 

During speculative execution, the speculatively executed parts are determined by the predictions of the branch predictor.
As shown in ~\Cref{sec:v1-semantics}, leakage due to speculative execution is maximized under a branch predictor that mispredicts every branch. 
The code in \Cref{figure:spectrev1:asm-code} yields two symbolic traces w.r.t. the speculative semantics that mispredicts every branch:\footnote{For simplicity of presentation, the example traces capture only
  loads but not the program counter. }
\begin{equation}\label{tr:inbounds}
\startObsKywd{} \concat\rollbackObsKywd{}\concat\tauStack \quad \text{when} \quad
  \inlineASMcode{y} < \inlineASMcode{size}
\end{equation}\vspace{-15pt}
\begin{equation} \label{tr:outbounds}
\startObsKywd{} \concat \tauStack  \concat
    \rollbackObsKywd{} \quad \text{when} \quad \inlineASMcode{y} \geq \inlineASMcode{size}
\end{equation}
where
$\tauStack=\loadObs{(\inlineCcode{A} + \inlineASMcode{y} )} \concat
\loadObs{(\inlineASMcode{B} +\inlineASMcode{A[y] * 512})}$.  Here, the
argument of $\loadObsKywd$ is visible to the observer, while
$\startObsKywd{}$ and $\rollbackObsKywd{}$ denote the start and the
end of a misspeculated execution. The traces of the {\em
  non-speculative} semantics are obtained from those of the
speculative semantics by removing all observations in between
$\startObsKywd{}$ and $\rollbackObsKywd{}$. 

Trace~\ref{tr:inbounds} shows that whenever \inlineASMcode{y} is in
bounds (i.e., $\inlineASMcode{y} < \inlineASMcode{size}$) the
observations of the speculative semantics and the non-speculative
semantics coincide (i.e. they are both $\tauStack$). In contrast,
Trace~\ref{tr:outbounds} shows that whenever
$\inlineASMcode{y} \geq \inlineASMcode{size}$, the speculative
execution generates observations $\tauStack$ that depend on \inlineASMcode{A[y]}
whose value is not visible in the non-speculative
execution.
This is flagged as a leak by \tool{}.

\para{Proving Security with \tool{}}
The \clang{} 7.0.0 C++ compiler implements a countermeasure, called speculative load hardening~\cite{spec-hard}, that applies conditional masks to addresses to prevent leaks into the
microarchitectural state. \Cref{figure:spectrev1:asm-code-slh} depicts the protected output of \clang{} on the program from \Cref{lst:v1-vanilla}.

\begin{lstlisting}[style=ASMstyle,
  float=tp,
  floatplacement=tbp, caption={\specb - Assembly code with speculative load hardening.
\clang{} inserted instructions 3, 6, 9, and 11.}, label=figure:spectrev1:asm-code-slh]
 mov    size, %
 mov    y, %
 mov    $0, %
 cmp    %
 jbe    END
 cmovbe $-1, %
 mov    A(%
 shl    $9, %
 or     %
 mov    B(%
 or     %
 and    %
\end{lstlisting}

The symbolic execution of the speculative semantics produces, as before, Trace~\ref{tr:inbounds} and Trace~\ref{tr:outbounds}, but with 
\begin{equation*}
\tauStack=\loadObs{ (\inlineCcode{A}
    + \inlineASMcode{y} )} \concat \loadObs{(\inlineASMcode{B}
    +(\inlineASMcode{A[y] * 512}) \,|\, \mathit{mask})},
\end{equation*}
where
$\mathit{mask} = \ite{\inlineASMcode{y} <
  \inlineASMcode{size}}{\texttt{0x0}}{\texttt{0xFF..FF}}$ corresponds
to the conditional move in line $6$ and $|$ is a bitwise-or operator. 
Here, $\ite{\inlineASMcode{y} <
  \inlineASMcode{size}}{\texttt{0x0}}{\texttt{0xFF..FF}}$ is a symbolic if-then-else expression evaluating to $\texttt{0x0}$ if $\inlineASMcode{y} <
  \inlineASMcode{size}$ and to $\texttt{0xFF..FF}$ otherwise.

The analysis of Trace 1 is as before. For Trace 2, however, \tool{}
determines (via a query to Z3~\cite{z3}) that, for all
$\inlineASMcode{y}\ge \inlineASMcode{size}$ %
there is exactly {\em one} observation that the adversary can make during the speculative execution, namely
$\loadObs{ (\inlineCcode{A} + \inlineASMcode{y} )} \concat \loadObs{(\inlineASMcode{B} + \texttt{0xFF..FF})}$, from which it concludes that no information leaks into the microarchitectural state, i.e., the countermeasure is effective in securing the program.

%% file: src/background.tex
\section{\texorpdfstring{\muasm{}}{uASM} Language}\label{sec:bg}

In this section, we present \muasm{}, a simple assembly-style language serving as the basis for our framework.
We start by describing the attacker model we consider  (\Cref{sec:traces}).
Then, we present the syntax (\Cref{sec:syntax-mu}) and the base non-speculative semantics (\Cref{sec:semantics-mu}) of \muasm{}.

\subsection{Observations and Attacker Model}\label{sec:traces}

We adopt a commonly-used attacker model \cite{spectector, ST_spectector2, ST_binsec, ST_constantTime_Spec, ST_inspectre, ST_jasmin2, S_sec_comp, ST_blade}: a passive attacker observing the execution of a program through events $\tau$.
These events, which we call \emph{observations}, model timing leaks through cache and control flow while abstracting away low-level microarchitectural details. 
Specifically, we consider an attacker that observes the program counter and the locations of memory accesses during computation.
This attacker model is commonly used to formalize timing side-channel free code~\cite{MolnarPSW05,AlmeidaBBDE16}, without requiring microarchitectural models. 
In particular, it captures  through data and instruction caches without requiring an explicit cache model.

\begin{align*}
    \Obs \bnfdef&\ \loadObs{n} \mid \storeObs{n} \mid \pcObs{n} \mid \callObs{f} \mid \retObs{n}  
    &
    \tau \bnfdef&\  \empTr \mid \Obs
    &
        \tauStack \bnfdef&\ \nil \mid \tauStack\cdot \tau
\end{align*}
The $\storeObs{n}$ and $\loadObs{n}$ events denote read and write accesses to memory location $n$, so they capture leaks through the data cache.
In contrast, $\pcObs{n}$, $\callObs{f}$, and $\retObs{n}$ events record the control-flow of the program, thereby capturing leaks through, for instance, the instruction cache. 
An observation $\tau$ is either an event $\Obs$ or the empty observation $\empTr$. 
Traces $\tauStack$ are sequences of observations; we indicate sequences of elements $[e_1; \cdots; e_n]$ as $\bar{e}$, and adding an element $e$ to $\bar{e}$ as $\bar{e} \cdot e$.\looseness=-1

\subsection{Syntax of \texorpdfstring{\muasm{}}{uASM} }\label{sec:syntax-mu}

\begin{figure}
\begin{gather*}
\begin{aligned}
\text{(Programs) } p  \coloneqq 
        &\ 
        n:i \mid p_1;p_2
    &
    \text{(Functions) } \mathcal{F} \coloneqq&\ \varnothing \mid \mathcal{F}; f \mapsto n 
    \\
    \text{(Expressions) }  e \coloneqq
        &\ 
        n \mid x \mid \ominus e \mid e_1 \otimes e_2
    &
    \text{(Registers) } x \in &\ \Reg 
    &
    \text{(Values) }  n, l \in
        &\
        \Val = \Nat \cup \{\bot\}
\end{aligned}
\\
\begin{aligned}
    \text{(Instructions) } i \coloneqq
        &\ 
        \pskip  \mid \passign{x}{e} \mid \pload{x}{e} \mid \pstore{x}{e} 
    \mid 
        \pjmp{e} 
    \mid 
        \pjz{x}{l} 
        \mid 
        \pcondassign{x}{e}{e'} \mid \pbarrier 
        \mid \pcall{f} \mid \pret \mid \pendbr
\end{aligned}
\end{gather*}
\caption{Syntax of the \muasm{} language}\label{fig:uasm:syntax}
\end{figure}

\muasm{} is an assembly-like language whose syntax is presented in \Cref{fig:uasm:syntax}.
In \muasm{}, programs $p$  are sequences of mappings from natural numbers $n$ (i.e., the instruction address or label) to instructions $i$.

Instructions $i$ include skipping $\pskip$, register assignments $\passign{x}{e}$, loads $\pload{x}{e}$, stores $\pstore{x}{e}$, indirect jumps $\pjmp{e} $, conditional branches $\pjz{x}{l} $, conditional assignments $\pcondassign{x}{e}{e'}$, speculation barriers $\pbarrier $, calls $\pcall{f}$, and returns $ \pret$.
Speculation barriers ($\pbarrier$) and conditional assignments ($\pcondassign{x}{e}{e'}$) are both commonly used to implement Spectre countermeasures. 
In particular, speculation barriers stop speculation outright, whereas conditional assignments are often used to replace branching instructions.

Instructions can refer to expressions $e$, which are constructed by combining registers and values with unary $\ominus$ and binary $\otimes$ operators.
Registers come from the set $\Reg$, containing register identifiers and designated registers $\pc$ and $\spR$ modelling the program counter and stack pointer respectively, whereas values come from the set $\Val$, which includes natural numbers and $\bot$. Note that we use $\bot$ to denote the termination of the program.

We say that a program is \textit{well-formed} if (1) it does not contain duplicate labels, (2) it contains an instruction labelled with $0$, i.e., the initial instruction, and (3) it does not contain branch instructions of the form $\cmd{n}{\pjz{x}{n+1}}$.
Note that the last constraint ensures that the all outcomes of branch instructions are always reflected in the value of the program counter.

\begin{example}[
\specb Example]\label{example:v1-vanilla}
	The \specb example from \Cref{lst:v1-vanilla} can be expressed in \muasm{} as follows:
	\begin{align*}
		& 	\cmd{0}{\passign{x}{\mathtt{y} < \mathtt{size}}}\\
		&	\cmd{1}{\pjz{x}{\bot}}\\
		&	\cmd{2}{\pload{z}{\mathtt{A} + \mathtt{y}}}\\
		&	\cmd{3}{\passign{z}{z * 512}}\\
		&	\cmd{4}{\pload{w}{\mathtt{B} + z}}\\
		&	\cmd{5}{\passign{\mathtt{temp}}{\mathtt{temp}\ \&\ w}}
	\end{align*}
	Here, registers $\mathtt{y}$, $\mathtt{size}$, and $\mathtt{temp}$ store the respective variables.
	Similarly, registers $\mathtt{A}$ and $\mathtt{B}$ store the memory addresses of the first elements of the arrays $\mathtt{A}$ and $\mathtt{B}$.
\end{example}

In the following, we use instruction keywords to denote the set of all instructions of a given type.
For instance, $\jzKywd$ is the set of all branch instructions, i.e., $\jzKywd =  \{\pjz{x}{l} \mid x \in \Reg \wedge l \in \Val\}$.

\subsection{Non-speculative Semantics of \texorpdfstring{\muasm{}}{uASM}}\label{sec:semantics-mu}

We now introduce the standard, non-speculative semantics of \muasm{} program, which models their execution on a platform without speculation.
For this, we describe next the small-step operational non-speculative semantics $\nsarrow{}$.

The judgment for this semantics is $\tup{p, \sigma} \nsarrow{\tau} \tup{p, \sigma'}$ and it reads: \emph{``a program state $\tup{p, \sigma}$ steps to a new program state $\tup{p, \sigma'}$ producing observation $\tau$''}.
Program states $\tup{p, \sigma} \in \Omega$ consist of a program $p$ and a configuration $\sigma$.
The program $p$ is used to look up the current instruction, whereas the configuration $\sigma = \tup{m,a}$ records the state of  the memory $m$ and register file $a$. 
Memories map addresses (which are natural numbers) to values, whereas register files map register identifiers to values.

{
\begin{align*}
    \text{Configurations } \sigma \bnfdef
        &\ \tup{m, a} 
            & 
    \text{Prog. States } \Omega \bnfdef
        &\ \tup{p, \sigma}
    \\
    \mi{Register File}~a \bnfdef
        &\ \emptyset \mid r; x \mapsto v
            &
            \mi{Registers}~ x \in
                &\ \Reg
    \\
    \mi{Memory}~m \bnfdef
        &\ \emptyset \mid m; n \mapsto v
            &
            \mi{where }~ n \in
                &\ \mathbb{N}
\end{align*}
}

\begin{figure}[!ht]
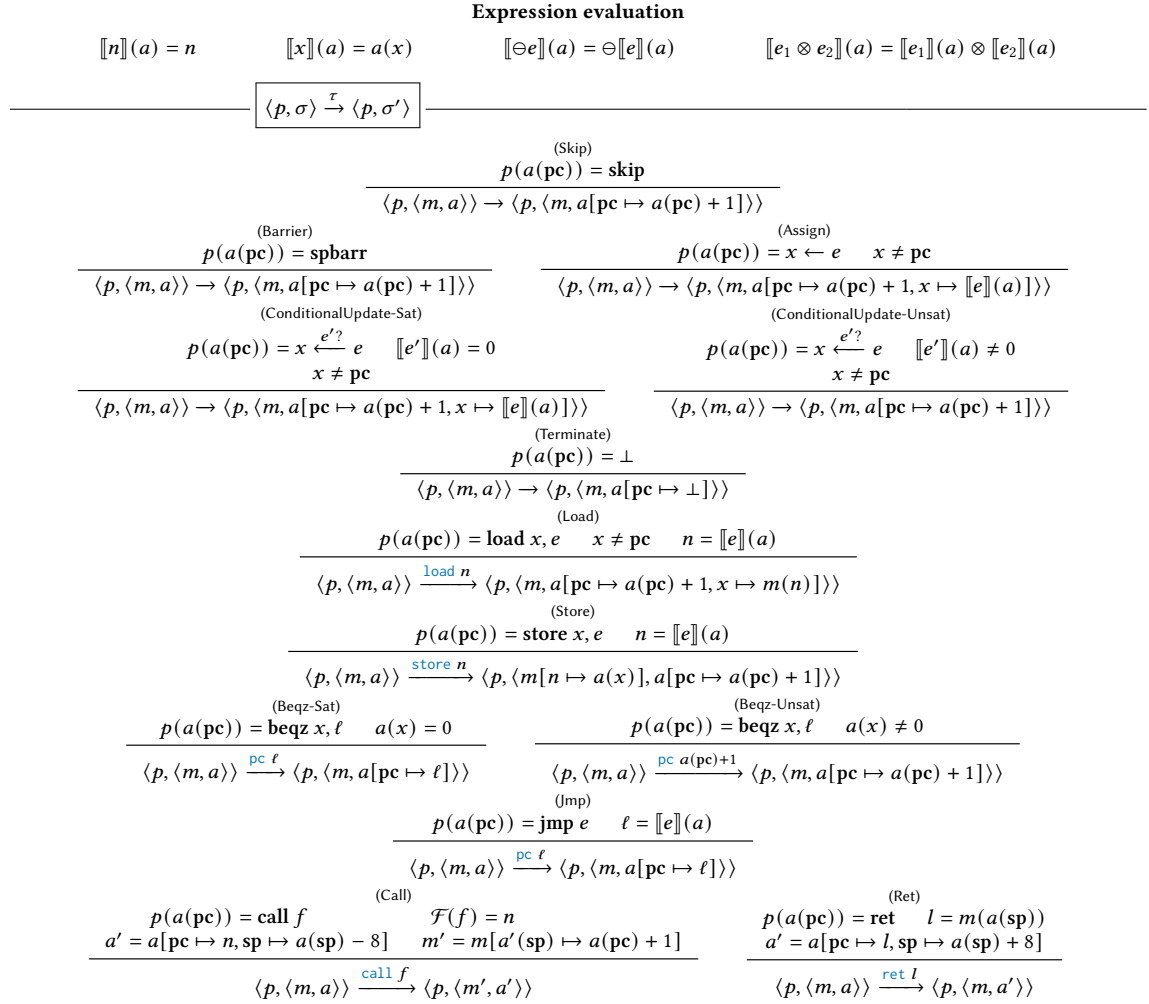
\small

    {\bf Expression evaluation}
\begin{align*}
 \exprEval{n}{a} = n
	    & 
	        & 
         \exprEval{x}{a} = a(x)
	            & 
             &
             \exprEval{\unaryOp{e}}{a} = \unaryOp{\exprEval{e}{a}}
	                & 
	                    & 
                     \exprEval{\binaryOp{e_1}{e_2}}{a} = \binaryOp{\exprEval{e_1}{a}}{\exprEval{e_2}{a}}
\end{align*}

\begin{center}
\mytoprule{\tup{p,\sigma} \nsarrow{\tau} \tup{p, \sigma'}}

\typerule{Skip}
{
\select{p}{a(\pc)} = \pskip
}
{
\tup{p, \tup{m, a}} \eval{p}{} \tup{p, \tup{m, a[\pc \mapsto a(\pc)+1]}}
}{skip-paper}

\typerule{Barrier}
{
	p(a(\pc)) = \pbarrier
}
{
	\tup{p, \tup{m,a}} \eval{p}{} \tup{p, \tup{m,a[\pc \mapsto a(\pc) + 1]}}
}{barr-paper}
\typerule{Assign}
{
\select{p}{a(\pc)} = \passign{x}{e} & x \neq \pc
}
{
\tup{p, \tup{m, a}} \eval{p}{} \tup{p, \tup{m, a[\pc \mapsto a(\pc)+1,x \mapsto \exprEval{e}{a}]}}
}{assign-paper}

\typerule{ConditionalUpdate-Sat}
{
	p(a(\pc)) = \pcondassign{x}{e}{e'} &  \exprEval{e'}{a} = 0\\
	x \neq \pc
}
{
	\tup{p, \tup{m,a}} \eval{p}{} \tup{p, \tup{m,a[\pc \mapsto a(\pc) + 1, x \mapsto \exprEval{e}{a}]}}
}{condup-sat-paper}
\typerule{ConditionalUpdate-Unsat}
{
	p(a(\pc)) = \pcondassign{x}{e}{e'} & \exprEval{e'}{a} \neq 0\\
	x \neq \pc
}
{
	\tup{p, \tup{m,a}} \eval{p}{} \tup{p, \tup{m,a[\pc \mapsto a(\pc) + 1]}}
}{ns-condup-unsat-paper}

\typerule{Terminate}
{
	\select{p}{a(\pc)} = \bot
}
{
\tup{p, \tup{m, a}} \eval{p}{} \tup{p, \tup{m, a[\pc \mapsto \bot]}}
}{terminate-paper}

\typerule{Load}
{
\select{p}{a(\pc)} = \pload{x}{e} & x \neq \pc 
&
n = \exprEval{e}{a}
}
{
\tup{p, \tup{m, a}} \eval{p}{\loadObs{n}} \tup{p, \tup{m, a[\pc \mapsto a(\pc)+1, x \mapsto m(n)]}}
}{ns-load}
\typerule{Store}
{
\select{p}{a(\pc)} = \pstore{x}{e} & n = \exprEval{e}{a}
}
{
\tup{p, \tup{m, a}} \eval{p}{\storeObs{n}} \tup{p, \tup{m[n \mapsto a(x)], a[\pc \mapsto a(\pc)+1]}}
}{ns-store}

\typerule{Beqz-Sat}
{
\select{p}{a(\pc)} = \pjz{x}{\lbl} &  a(x) = 0
}
{
\tup{p, \tup{m, a}} \eval{p}{\pcObs{\lbl}} \tup{p, \tup{m, a[\pc \mapsto \lbl]}}
}{ns-beqz-sat}
\typerule{Beqz-Unsat}
{
\select{p}{a(\pc)} = \pjz{x}{\lbl}  & a(x) \neq 0
}
{
\tup{p, \tup{m, a}} \eval{p}{\pcObs{a(\pc)+1}} \tup{p, \tup{m, a[\pc \mapsto a(\pc) +1]}}
}{ns-beqz-unsat}

\typerule{Jmp}
{
\select{p}{a(\pc)} = \pjmp{e} & \lbl = \exprEval{e}{a}
}
{
\tup{p, \tup{m, a}} \eval{p}{\pcObs{\lbl}} \tup{p, \tup{m, a[\pc \mapsto \lbl]}}
}{jmp-paper}

\typerule{Call}
{
\select{p}{a(\pc)} = \pcall{f} & \mathcal{F}(f) = n & \\
a' = a[\pc \mapsto n, \spR \mapsto a(\spR) - 8] & m' = m[a'(\spR) \mapsto a(\pc) + 1]
}
{
\tup{p, \tup{m, a}} \eval{p}{\callObs{f}} \tup{p, \tup{m', a'}}
}{ns-call}
\typerule{Ret}
{
\select{p}{a(\pc)} = \pret &  l = m(a(\spR)) \\
a' = a[\pc \mapsto l, \spR \mapsto a(\spR) + 8]
}
{
\tup{p, \tup{m, a}} \eval{p}{\retObs{l}} \tup{p, \tup{m, a'}}
}{ns-ret}
\end{center}
\caption{The non-speculative small-step semantics of \muasm{}.}
\label{ns-semantics}
\end{figure}

\Cref{ns-semantics} presents the rules defining the non-speculative semantics $\nsarrow{}$.
The rules rely on the evaluation of expressions (indicated as $\exprEval{e}{a} = n$) where expression $e$ is evaluated to value $n$ under register file $a$.
In the rules, $a[x \mapsto n]$, where $x \in \Reg \cup \Nat$ and $n\in\Val$, denotes the update of a map (memory or registers), whereas  $a(x)$ denotes reading from a map.
Finally, $\sigma(x)$, where $x \in \Reg$ and $\sigma = \tup{m,a}$, denotes $a(x)$. %

Most of the rules of the semantics in \Cref{ns-semantics} are standard; we describe selected rules below.
Branch instructions emit observations recording the outcome of the branch (\Cref{tr:ns-beqz-sat}, \Cref{tr:ns-beqz-sat}), while memory operations emit observations recording the accessed memory (\Cref{tr:ns-store}, \Cref{tr:ns-load}).
A call to function $f$ is a jump to the function's starting line number $n$, as indicated by the function map $\mathcal{F}$. A call stores the return address on the stack at the value of the stack pointer $\spR$ and decreases $\spR$ (\Cref{tr:ns-call}). A return does the inverse: it looks up the return address via the stack pointer $\spR$ and then increases the stack pointer (\Cref{tr:ns-ret}).

\begin{figure}[!ht]
    \centering
    \begin{center} 
        \mytoprule{\Trace{\tup{p,\sigma}}{\tauStack}}
        
        \typerule{NS-Reflection}
        {
        }
        {\tup{p, \sigma}
        \nsbigarrow{\varepsilon}
        \tup{p, \sigma}
        }{ns-reflect-paper}
        \typerule{NS-Single}
        {
        \tup{p, \sigma} \nsbigarrow{\tauStack} \tup{p, \sigma''}    & \tup{p, \sigma''} \nsarrow{\tau} \tup{p, \sigma'}
        }
        {\tup{p, \sigma}
        \nsbigarrow{\tauStack \cdot \tau}
        \tup{p, \sigma'}
        }{ns-single-paper}
    
        \typerule {NS-Trace}
        { \exists \sigma' & \sigma' \in \Final & \tup{p, \sigma} \nsbigarrow{\tauStack} \tup{p, \sigma'}
        }
        { \Trace{\tup{p,\sigma}}{\tauStack}
         }{ns-trace-paper}
        \typerule {NS-Beh}
        {}
        { \behNs{p} =  \{\tauStack \mid \exists \sigma \in \Init.~ \Trace{\tup{p,\sigma}}{\tauStack} \}
         }{ns-beh-paper}
         
    \end{center}
    \caption{The Reflexive-transitive-closure of $\nsarrow{}$ and behaviour of \muasm{} programs.}
    \label{fig:uasm-non-spec:beheavior}
\end{figure}

\Cref{fig:uasm-non-spec:beheavior} presents the rules for deriving the trace of observations associated with an execution according to the non-speculative semantics $\nsarrow{}$.
We denote by $\Trace{\tup{p,\sigma}}{\tauStack}$ the trace $\tauStack$ associated with executing the program $p$  starting from an initial configuration $\sigma$ until reaching a final configuration $\sigma'$ (\Cref{tr:ns-trace-paper}) using the reflexive-transitive-closure $\nsbigarrow{}$ of the non-speculative semantics $\nsarrow{}$ (\Cref{tr:ns-reflect-paper} and \Cref{tr:ns-single-paper}).
To simplify our notation, we introduce the shorthand $\behNs{p, \sigma}$ to denote the specific trace $\tau$ generated by this execution.
Finally, the \emph{non-speculative behaviour} $\behNs{p}$ of a program $p$ is the set of all traces generated by all possible initial states for program $p$ (\Cref{tr:ns-beh-paper}).

%% file: src/Semantics/spec_semantics_2.tex
\section{Modelling Speculative Execution}\label{sec:spec-semantics}

In this section, we propose a general approach for modeling, at program-level, the effects of speculative execution induced by different kinds of speculation mechanisms.
For this, we extend a non-speculative (i.e., architectural) semantics to a \emph{speculative semantics} that is equipped with a \emph{prediction oracle} (which captures how predictions are made) used to explore mispredicted paths.
For this, we first informally explain the core concepts behind our model (\Cref{sec:spec-sem:description}).
Then, we formalize prediction oracles and prediction histories (\Cref{sec:spec-sem:pred-oracle}).
We conclude by presenting the states and the evaluation relation associated with the (speculative) oracle semantics (\Cref{sec:oracle-general}).

We remark that here we present only a general template for defining the oracle semantics.
Such template needs to be instantiated to capture specific speculation mechanisms, and we do so in \Cref{sec:inst-spec-semantics-single} for a large class of mechanisms.
In the following, we indicate a speculative semantics as $\semx$, where the metavariable $x$ is used as a placeholder indicating the type of speculation mechanism.

\subsection{High-level Description}\label{sec:spec-sem:description}
     
The core idea behind our approach is to extend a non-speculative  semantics---like the one from \Cref{sec:bg}---to a \emph{speculative semantics} that captures the effect of speculatively executed instructions.
For this, our speculative semantics is parametric in two components:
(1) a set $\specInstr$ of \lang{} instructions that might trigger speculation, and 
(2) a \emph{prediction oracle} $\orac$ capturing the prediction strategy associated with the modeled speculation mechanism.

In a nutshell, our speculative semantics works as follows.
All instructions not in $\specInstr$ are executed as in the standard non-speculative semantics.
In contrast, whenever an instruction in $\specInstr$ is reached, the oracle $\orac$ is queried to obtain the next predicted state, which records the effects of the prediction.
For instance, for branch prediction, the program counter in the predicted state is updated to decide which of the two branches to execute speculatively.

To enable a subsequent rollback in case of a misprediction, a snapshot of the current program state is taken, before starting a  \emph{speculative transaction}.
In this speculative transaction, the program is executed speculatively starting from the predicted state for a bounded number $w$ of computation steps, called the speculative window.
To abstract from the complexity of estimating the actual speculative window (which depends on the details of a CPU's microarchitecture), in our model the speculation window $w$ is also provided by the prediction oracle.

At the end of a speculative transaction, the correctness of the prediction is evaluated:
\begin{inparaenum}[(1)]
\item If the prediction was {\em correct}, the transaction is committed and the computation continues using the current  configuration.
\item In contrast, if the prediction was {\em incorrect}, the transaction is aborted, the original  configuration is restored, and the computation continues on the correct branch. 
\end{inparaenum}

In the following sections, we formalize the behavior intuitively described above in the general structure of the {\em speculative oracle semantics}, whereas in  \Cref{sec:inst-spec-semantics-single} we will present instantiations of the oracle semantics (and $\specInstr$) for different speculation mechanisms.

\subsection{Prediction Oracles}\label{sec:spec-sem:pred-oracle}

In our semantics, the prediction oracle serves two distinct purposes: (1) predicting the next program state (i.e., modeling how the speculation mechanism works), and (2) determining the length of the speculative transactions (i.e., determining for how long the speculative execution lasts). 
Our notion of prediction oracle is general enough to account for both control-flow speculation, such as branch prediction, and data-flow speculation, such as value prediction.

A \textit{prediction oracle} $\orac$ is a partial function that takes as input a program~$p$, a prediction history~$h$, and the current program state $\sigma$ such that $p(\sigma(\pc))$ is an instruction triggering  speculation, i.e., $p(\sigma(\pc)) \in  \specInstr$.
The prediction oracle returns as output a triple $\tup{\delta, w, \tau}$, where $\delta \in \Conf$ is a partial configuration indicating the predicted values, $w \in \Nat$ represents the speculative transaction's length, and $\tau$ is a (potentially empty) observation.
One can then join the returned partial configuration $\delta$ with the current configuration $\sigma$, indicated as $\sigma \uplus \delta$, to obtain the next speculative configuration. 
Informally, $\sigma \uplus \delta$ simply updates $\sigma$ with the values in $\delta$ at the same places (registers, memory locations, etc).
We say that a prediction oracle $\orac$ has \textit{speculative window at most} $w$ if the length of the transactions generated by its predictions is at most $w$, i.e., for all programs $p$,  histories $h$, and configurations $\sigma$, $\orac(p, h, \sigma) = \tup{\delta,w', \tau}$, for some $\delta, \tau$ such that $w' \leq w$.

Taking into account the prediction history enables us to capture history-based predictors, a general class of predictors that base their decisions on past behaviors.
Formally, a {\em prediction history} is a sequence of triples $\tup{\lbl, \id, \delta}$, where $\lbl \in \Val$ is the label of an instruction related to speculation,  $\id\in\Nat$ is the unique identifier of the transaction that triggered speculation, and $\delta$ are the predicted values.

\begin{example}[Control-flow Prediction]\label{example:speculative-semantics:btfnt-predictor}
The ``backward taken forward not taken'' (BTFNT) branch predictor, implemented in early CPUs~\cite{hennessy2011computer}, predicts the branch as taken if the target instruction address is lower than the program counter.
It can be formalized by the $\mathit{BTFNT}$ oracle below, for a fixed speculative window~$w$, as follows:
\begin{align*}
    \mathit{BTFNT}(p, h, \sigma) = \tup{[\pc \mapsto \lbl'' ],w, \pcObs{\lbl''}} \text{, where } \lbl'' = \mathit{min}(\sigma(\pc)+1, \lbl') \text{ and } p(\sigma(\pc)) = \pjz{x}{\lbl'}
\end{align*}

$\mathit{BTFNT}$ speculatively updates the program counter  to the minimum of the branch target $\lbl'$ and the branch-not-taken target (i.e., $\sigma(\pc)+1$).
Note that the $\mathit{BTFNT}$ oracle is static, that is, it is independent of the prediction history $h$.
Dynamic branch predictors, such as simple 2-bit predictors and more complex correlating or tournament predictors~\cite{hennessy2011computer}, can also be formalized using prediction oracles.
\end{example}

\begin{example}[Data-flow Prediction]\label{example:speculative-semantics:dataflow-predictor}
\citet{revizor2} discovered that certain Intel CPUs speculate on division operations, called Zero Dividend Injection (ZDI), thereby resulting in a restricted form of value prediction.
We model this behavior with the following oracle, which predicts that  the upper bits of a 128-bit by 64-bit division operation are $0$:\looseness=-1
\begin{align*}
\mathit{ZDI}(p, h, \sigma) = \tup{\delta, w, \epsilon}\text{, where } p(\sigma(\pc)) = \passign{k}{ (x:y) \div z} \text{ and } \delta= [k \mapsto (0:y) \div z].
\end{align*}
Above, $(x:y)$ denotes the 128-bit value obtained by concatenating registers $x$ and $y$ whereas $\div$ is the division operator.
\end{example}

\subsection{Oracle Semantics}\label{sec:oracle-general}

Next, we formalize the speculative oracle semantics.
We start by formalizing the states over which the semantics operates and continue by presenting the evaluation relation given a prediction oracle $\orac$.

\paragraph{Speculative States}
The speculative oracle semantics operates on speculative states $\Xx$, each one consisting of  a  stack of speculative instances $\Psix$.
Each instance $\Psix$ contains the program $p$, a counter $\ctr$ that uniquely identifies the speculative transaction associated with the instance, a configuration $\sigma$, the prediction history $h$, and the remaining speculation window $n$ describing the number of instructions that can still be executed speculatively (or $\bot$ when no speculation is happening). 
Depending on the specific speculation that is modelled, additional data might be tracked (indicated with $\cdots$ below as well as in the evaluation rules).
\begin{align*}
    \ti{Spec. States } \Xx  \bnfdef&\ \psiStackx
    &
    \ti{Spec. Instances } \Psix \bnfdef&\ \tup{p, \ctr, \sigma, h, n, \cdots}
\end{align*}

\paragraph{Evaluation rules}
We describe the speculative oracle semantics, given a prediction oracle $\orac_x$, with the relation $\SEspecarrowx{} \subseteq \Xx \times \Obs^* \times \Xx$, which describes how speculative states are modified along the computation while producing observations capturing leaks.
To denote the start and end of speculative transactions, we extend the set of observations $\Obs$ to include three new observations $\startObsx{n}$, $\rollbackObsx{n}$, and $\commitObsx{n}$ marking the start and end (with a commit or a rollback) of speculative transaction with id $n$ respectively.
\begin{align*}
    \Obs \bnfdef&\ \cdots \mid  \startObsx{n} \mid \rollbackObsx{n} \mid \commitObsx{n}
\end{align*}

\begin{figure}[!ht]
\begin{center}
    \mytoprule{\psiStackx \SEspecarrowx{\tau} \psiStackx'}
    
    \typerule {O-Context}
    {\Psix \SEspecarrowx{\tau} \psiStackx' & \mathit{canStep}(\psiStackx{}\cdot \Psix) \\
    \text{if}~ p(\Psix . \sigma(\pc)) =\pbarrier\ \text{then}~ \psiStackx{}'' = \mathit{zeroes}(\psiStackx)\ \text{else}~ \psiStackx{}'' = \decr{\psiStackx}
    }
    {\psiStackx{} \cdot \Psix
    \SEspecarrowx{\tau}
    \psiStackx''{} \cdot \psiStackx' 
     }{o-context}

    \typerule{O-Rollback}
    {\conctauarrow{\tup{p,\sigma''}}{\tup{p,\sigma'''}}  & (\sigma''' \neq \sigma'' \uplus \delta \text{ or } \pFin)}
    { \psiStackx{} \cdot \tup{p, \ctr, \sigma, h, n, \cdots} \cdot  \tup{p, \ctr', \sigma', h', 0, \cdots}^{\tup{\sigma'',\delta}} \cdot  \psiStackx{}'
    \SEspecarrowx{\rollbackObsx{\ctr} \cdot \tau}
     \psiStackx{} \cdot \tup{p, \ctr', \sigma''', h', n-1, \cdots}
    }{o-rollback}

    \typerule{O-Commit}
    {  \conctauarrow{\tup{p,\sigma''}}{\tup{p,\sigma'''}}  & (\sigma''' = \sigma'' \uplus \delta \text{ or } \pFin)  }
    { \psiStackx{} \cdot \tup{p, \ctr, \sigma, h, n, \cdots} \cdot \tup{p, \ctr', \sigma', h', 0, \cdots}^{\tup{\sigma'',\delta}} \cdot \psiStackx{}'
    \SEspecarrowx{\commitObsx{\ctr}}
     \psiStackx{} \cdot \tup{p, \ctr', \sigma', h', n, \cdots} \cdot \psiStackx{}'
    }{o-commit}
\end{center}
\begin{center}
    \mytoprule{\Psix \SEspecarrowx{\tau} \psiStackx'}

     \typerule {O-NoSpec}
    { p(\sigma(\pc)) \notin \specInstr \cup \{\pbarrier \} & \conctauarrow{\tup{p,\sigma}}{\tup{p,\sigma'}} 
    }
    {\tup{p, \ctr, \sigma, h, n + 1, \cdots}
    \SEspecarrowx{\tau}
    \tup{p, \ctr, \sigma', h, n, \cdots}
     }{o-noSpec}
    \typerule {O-barr}
    {\instr{\pbarrier} & \conctauarrow{\sigma}{\sigma'}
    }
    {\tup{p, \ctr, \sigma, h, n, \cdots}
    \SEspecarrowx{\tau}
    \tup{p, \ctr, \sigma', h, 0, \cdots}
     }{o-barr}

    \typerule{O-Spec}
    { p(\sigma(\pc)) \in \specInstr & 
    \orac(p,n,h, \sigma) = \langle \delta, w, \tau \rangle  \\
    h' = h \concat \langle \sigma(\pc), n+1, \delta \rangle & 
     \tauStack = \startObsx{\ctr} \cdot \tau
    }
    {\tup{p, \ctr, \sigma, h, n + 1, \cdots}
    \SEspecarrowx{\tauStack}
    \tup{p, \ctr, \sigma, h, n+1, \cdots} \cdot \tup{p, \ctr + 1, \sigma \uplus \delta, h', w, \cdots}^{\langle \sigma,\delta \rangle}
    }{o-spec}

\end{center}
\caption{Evaluation rules for the speculative oracle semantics $\SEspecarrowx{}$ given an oracle  $\orac_x$}\label{fig:oracle-semantics:general}
\end{figure}
Next, we describe in detail the rules formalizing the oracle semantics, which capture the intuition from \Cref{sec:spec-sem:description} and are depicted in \Cref{fig:oracle-semantics:general}.
Whenever none of the speculative instances in the current state has a speculative window of $0$ or is stuck (denoted by $\mathit{canStep}(\psiStackx{})$), the computation proceeds by doing one step for the instance at the top of the stack (\Cref{tr:o-context}) and by updating the speculative window of the other instances.
In particular, if the instruction to be executed in the topmost instance is a speculative barrier, then all speculative windows are set to 0 ($\psiStackx{}'' = \mathit{zeroes}(\psiStackx)$), otherwise all windows are decremented by $1$ ($\psiStackx{}'' = \decr{\psiStackx}$).

In the topmost speculative instance, instructions that do not trigger speculation (i.e., they are not in $\specInstr$) are executed following the non-speculative semantics while also decrementing the current speculative window by 1 (\Cref{tr:o-noSpec}).
Speculation barriers are handled similarly except for setting the speculative window in the topmost instance to $0$ (\Cref{tr:o-barr}).
Finally, \Cref{tr:o-spec} handles instructions that trigger speculation (i.e., those that belong to $\specInstr$).
In this case, the oracle is queried to obtain the prediction $\delta$ and the next speculative window $w$, and a new speculative instance (starting from the predicted state $\sigma \uplus \delta$ and with an updated prediction history $h'$) is pushed on top of the stack to start the new speculative transaction.
This fresh speculative instance is decorated with the original program state $\sigma$ and with the predicted values $\delta$, which will later on be used to determine whether the transaction should be committed or rolled back. 
The rule also appends the observation $\startObsx{n}$ to the trace to record the beginning of a speculative transaction.

A speculative transaction must be finalized (committed or rolled back) whenever its speculation window reaches $0$ or the execution cannot proceed further. We refer to the latter case as the instance being `stuck', formally indicated by the judgment $\pFin$. This judgment holds whenever the current instruction is undefined, i.e., $p(sigma(\pc)) = \bot$ (see \Cref{tr:terminate-paper}).
To determine whether to commit or rollback, the semantics inspects the snapshot state $\sigma$ and prediction $\delta$ that decorate the instance $\Psix^{\tup{\sigma,\delta}}$ whose window has reached $0$.

\begin{compactitem}
\item If the prediction was {\em correct} (i.e., doing one step from $\sigma$ results indeed in the predicted state $\sigma \uplus \delta$), the transaction is committed and the computation continues using the current configuration (\Cref{tr:o-commit}).
\item If the prediction was {\em incorrect} (i.e., doing one step from $\sigma$ produces a state different from the predicted state $\sigma \uplus \delta$), the transaction is aborted and the computation continues on the correct branch (\Cref{tr:o-rollback}). 
\end{compactitem}
Thus, committing and rolling back speculative transactions happen along the stack of states. 
Rolling back deletes all the instances above the rolled back instance, whereas committing updates the configuration, the counter, the prediction history $h$ and additional data tracked by the semantics of the instance below and the committed instance is deleted.
\Cref{tr:o-commit} and \Cref{tr:o-rollback} also record the  end of speculative transactions on the trace with the $\commitObsx{n}$ and $\rollbackObsx{n}$ observations respectively.

Analogously to the non-speculative case, the behaviour $\SEbehx{p}$ of a program $p$ under the oracle semantics is the set of all traces generated from an initial state until termination.

\paragraph{Non-Speculative Consistency}
Fundamentally, a valid speculative semantics must preserve the program's architectural behaviour. Speculation should only introduce observational side-effects during transient execution, without altering the program's non-speculative behaviour. We formalize this correctness requirement in \Cref{thm:ns-consistent-orac}, stating that the standard non-speculative behaviour can be exactly recovered from the oracle behaviour by applying the non-speculative projection $\nspecProject{}$, which (1) removes from the trace $\tau$ any sub-trace enclosed between  $\startObsx{n}$ and $\rollbackObsx{n}$ for some $n$, and then (2)  drops all remaining $\startObsx{n}$ and $\commitObsx{n}$ observations.
In \Cref{sec:inst-spec-semantics-single}, we prove that all our specualtive semantics instances satisfy \Cref{thm:ns-consistent-orac}.

\begin{requirement}[NS Consistency Oracle]\label{thm:ns-consistent-orac}

    $\behNs{p} = \nspecProject{\SEbehx{p}}$
\end{requirement}

\section{Speculative non-interference}\label{sec:sni}

In this section, we introduce \emph{speculative non-interference} (SNI, \Cref{sec:sni:sni}), a semantic notion characterizing the leaks introduced by speculatively-executed instructions.
Next, we introduce the notion of \emph{always-mispredict speculative semantics} (\Cref{sec:sni:am-semantics}), that is, a speculative semantics that facilitates reasoning about leaks w.r.t. \emph{any} prediction oracle.

\subsection{Speculative Non-Interference}\label{sec:sni:sni}\label{sec:SNI}
Speculative non-interference (SNI) is a semantic notion of security characterizing those information leaks that are introduced by speculative execution.
Intuitively, SNI requires that programs do not leak more information under the speculative semantics than what is leaked under the non-speculative semantics.

SNI is parametric in a policy $\pol$ and in the speculative semantics $x$, which models how the program executes.
The policy $\pol$ describes which parts of the program are public/low information, i.e., those that are known by an adversary.
Formally, a security policy $\policy$ is a finite subset of $\Var \cup \Nat$ specifying the low register identifiers and memory addresses.
Two configurations $\sigma, \sigma'$ are called \textit{low-equivalent} for a policy $\pol$, written $\sigma \backsim_{\pol} \sigma'$, if they agree on all register and memory locations in $\pol$.

\begin{example}[Example Policy for \Cref{example:v1-vanilla}]\label{example:security-condition:policy}
A policy $\policy$ for the program from \Cref{example:v1-vanilla} may state that the content of the registers $\mathtt{y}$, $\mathtt{size}$, $\mathtt{A}$, and $\mathtt{B}$ is non-sensitive, i.e., $\policy = \{\mathtt{y}, \mathtt{size},\mathtt{A}, \mathtt{B}\}$.
\end{example}
Policies need not be manually specified but can in principle be inferred from the context in which a piece of code executes, e.g., whether a variable is reachable from public input or not.

A program $p$ satisfies SNI (\Cref{def:sni}) for a speculative semantics $x$ if 
 \ulc{pred}{any pair of low-equivalent initial configurations $\sigma$ and $\sigma'$} 
that \ulc{pgrn}{generate the same observations under the non-speculative semantics}
 also \ulc{pblu}{generate the same observations under the speculative semantics to}.
\begin{definition}[Speculative Non-Interference]\label{def:sni}
Program $p$ satisfies SNI (denoted $\snix$) for a speculative semantics $x$ if for all $\sigma$, $\sigma'$, 
    {if} \ulc{pred}{$\sigma \backsim_{\pol} \sigma'$} 
    {and} \ulc{pgrn}{$\behNs{p, \sigma}= \behNs{p, \sigma'}$} 
    {then} \expandafter\ulc{pblu}{$\behx{p, \sigma} = \behx{p, \sigma'}$}.
\end{definition}

Speculative non-interference is a variant of non-interference. 
While non-interference compares what is leaked by a program with a policy specifying the allowed leaks, speculative non-interference compares the program leakage under two semantics, the non-speculative and the speculative one. 
The security policy and the non-speculative semantics, together, specify what the program may leak under the speculative semantics.%
\footnote{Conceptually, the non-speculative semantics induces declassification assertions for the speculative semantics~\cite{declas}.} %

\begin{example}[SNI for \Cref{example:v1-vanilla}]
The program $p$ from \Cref{example:v1-vanilla} does not satisfy speculative non-interference for the BTFNT oracle from Example~\ref{example:speculative-semantics:btfnt-predictor} and the policy $\policy$ from Example~\ref{example:security-condition:policy}.
Consider two initial configurations $\sigma:= \tup{m,a}, \sigma':=\tup{m',a'}$ that agree on the values of $\mathtt{y}$, $\mathtt{size}$, $\mathtt{A}$, and $\mathtt{B}$ but disagree on the value of $\mathtt{B}[\mathtt{A}[\mathtt{y}] * 512]$.
Say, for instance, that $m(a(\mathtt{A}) + a(\mathtt{y})) = 0$ and $ m'(a'(\mathtt{A}) + a'(\mathtt{y})) = 1$.
Additionally, assume that $\mathtt{y} \geq \mathtt{size}$.

Executing the program under the non-speculative semantics produces the trace $\pcObs{\bot}$ when starting from $\sigma$ and $\sigma'$.
Moreover, the two initial configurations are indistinguishable with respect to the policy $\policy$.
However, executing $p$ under the speculative semantics produces two distinct traces:
\begin{align*}
    \tau =&\ \startObs{0} \concat \pcObs{3} \concat \loadObs{v_1} \concat \loadObs{(a'(\mathtt{B})+0)} \concat \rollbackObs{0} \concat \pcObs{\bot} \\ 
    \tau' =&\ \startObs{0} \concat \pcObs{3} \concat \loadObs{v_1} \concat \loadObs{(a'(\mathtt{B})+1)} \concat \rollbackObs{0} \concat \pcObs{\bot}
\end{align*} 
where $v_1 = a(\mathtt{A}) + a(\mathtt{y}) = a'(\mathtt{A}) + a'(\mathtt{y})$.
Therefore, $p$ does not satisfy speculative non-interference.
\end{example}

\subsection{Always-Mispredict (AM) Semantics}\label{sec:am-general}\label{sec:sni:am-semantics}

The oracle speculative semantics from \Cref{sec:oracle-general} and, as a result, SNI are parametric in the prediction oracle $\orac$.
Often, however, it is desirable to obtain security guarantees w.r.t \textit{any} prediction oracle, since the details about speculation mechanisms might differ between different CPUs or may even be unknown.
To this end, we introduce a variant of the speculative semantics, which we call the \emph{always-mispredict speculative semantics}, that facilitates reasoning about leaks w.r.t. any prediction oracle.
We formally define the AM semantics as a speculative template rather than a concrete mechanism. This template abstracts the core logic of exploring mispredicted paths and serves as a unified foundation for the specific instantiations defined in \Cref{sec:inst-spec-semantics-single}.

For simplicity, consider the case of branch prediction.
In this case, leakage due to speculative execution is maximized under a predictor that mispredicts every time.
This intuition holds true unless speculative transactions are nested, where a correct prediction of a nested branch sometimes yields more leakage than a misprediction.

\begin{example}\label{ex:mispred}
Consider the following variation of the \spectre{}-\textsc{Pht} example~\cite{spectre} from Figure~\ref{lst:v1-vanilla}, and assume that the function \verb!benign()! runs for longer than the speculative window and does not leak any information.
\end{example}
\begin{wrapfigure}[6]{L}{.45\textwidth}
\begin{lstlisting}[style=Cstyle]
if (y < size)
	if (y-1 < size)
		benign();
	temp &= B[A[y] * 512];
\end{lstlisting}
\end{wrapfigure}
\noindent
Then, under a branch predictor that mispredicts every branch, the speculative transaction corresponding to the outer branch will be rolled back before reaching line 4.
On the other hand, given a correct prediction of the inner branch, line~4 would be reached and a speculative leak would be present.

A simple but inefficient approach to deal with this challenge would be to consider both cases, correct and incorrect predictions, upon every branch.
This, however, would result in an exponential explosion of the number of paths to consider.
Furthermore, it would not be applicable to speculation mechanisms where there might be more than one incorrect prediction, e.g., indirect branch speculation or value speculation.

\paragraph{Intuition}
To address these issues, we  introduce the {\em always-mispredict speculative semantics} that differs from the oracle speculative semantics in three key ways:
\begin{asparaenum}[(1)]
	\item It mispredicts every time, hence its name. 
    For every instruction that might trigger speculation (i.e., the instruction belongs to $\specInstr$), the always-mispredict semantics first speculatively explores all possible  \textit{wrong} paths for a bounded number of steps and then continues with the correct one.
    Thus, the semantics is not parametric in the prediction oracle. 

	\item It initializes the length of every {\em non-nested}  transaction to~$w$, and the length of every {\em nested}  transaction to the remaining length of its enclosing  transaction, decremented by $1$.\looseness=-1
	\item Upon executing instructions, only the remaining length of the innermost transaction is decremented.
\end{asparaenum}
The consequence of these modifications is that nested transactions do not reduce the number of steps that the semantics may explore the correct path for, after the nested transactions have been rolled back.
In Example~\ref{ex:mispred}, after rolling back the nested speculative transaction, the outer transaction continues as if the nested branch had been correctly predicted in the first place, and thus the speculative leak in line~4 is reached.

\newcommand{\pred}[1]{\mathit{Preds}_{#1}}

\begin{figure}
\begin{center}
    \mytoprule{\phiStackx \specarrowx{\tau} \phiStackx'}

    \typerule {AM-NoSpec}
    { p(\sigma(\pc)) \notin \specInstr \cup \{\pbarrier \} & \conctauarrow{\tup{p,\sigma}}{\tup{p,\sigma'}} 
    }
    { \phiStackx{} \cdot \tup{p, \ctr, \sigma, n + 1, \cdots}
    \specarrowx{\tau}
    \phiStackx{} \cdot \tup{p, \ctr, \sigma', n, \cdots}
     }{am-noBranch}
    \typerule {AM-barr}
    {\instr{\pbarrier} & \conctauarrow{\sigma}{\sigma'}
    }
    {\phiStackx{} \cdot \tup{p, \ctr, \sigma, n, \cdots}
    \specarrowx{\tau}
    \phiStackx{} \cdot \tup{p, \ctr, \sigma', 0, \cdots}
     }{am-barr}

    \typerule{AM-Spec}
    {\instr{\specInstr} & \conctauarrow{\tup{p, \sigma}}{\tup{p, \sigma'}} & j = min(\omega, n)  \\
    \tauStack = \tau \cdot \startObsx{\ctr} \cdot \cdots
    }
    {\phiStackx{} \cdot \tup{p, \ctr, \sigma, n + 1}
    \specarrowx{\tauStack}
    \phiStackx{} \cdot \tup{p, \ctr, \sigma', n} \cdot \Phix'
    }{am-spec}
    \typerule{AM-Rollback}
    { \Phix' . n = 0\ \text{or}\ p, \Phix'. \sigma \vdash \text{fin}  }
    { \phiStackx{} \cdot \tup{p, \ctr, \sigma, n} \cdot \Phix'
    \specarrowx{\rollbackObsx{\ctr}}
     \phiStackx{} \cdot \tup{p, \Phix' . \ctr', \sigma, n}
    }{am-rollback}

\end{center}
    \caption{Always-mispredict speculative semantics}\label{fig:am-rules}
\end{figure}

\paragraph{Formalization}
The state $\Sigmax$ of the AM semantics is a stack of speculative instances $\Phix$.
As shown in all rules in \Cref{fig:am-rules}, reductions in the always-mispredict semantics happen only on top of the stack.
Each instance $\Phix$ contains the program $p$, a counter $\ctr$ that identifies the speculative transaction
, a configuration $\sigma$, and the remaining speculation window $n$ describing the number of instructions that can still be executed speculatively (or $\bot$ when no speculation is happening). 
Depending on the specific speculation that is modelled, additional data is tracked (indicated with $\cdots$ in the rules).
Throughout the paper, we fix the maximal speculation window, i.e., the maximum number of speculative instructions, to a global constant $\omega$.\looseness=-1
\begin{align*}
    \ti{Spec. States } \Sigmax  \bnfdef&\ \phiStackx
    &
    \ti{Spec. Instances } \Phix \bnfdef&\ \tup{p, \ctr, \sigma, n, \cdots}
\end{align*}

Modifications (1)--(3) are captured in the four rules given in \Cref{fig:am-rules}. 
The judgement for the AM semantics is: $\Sigmax \specarrowx{\tau} \Sigmax'$.
If the instruction is not related to speculation and it is not a speculation barrier, then the speculative instance on the top of the stack is updated according to the non-speculative semantics $\nsarrow{}$ (\Cref{tr:am-noBranch}).
In contrast, whenever the current instruction is a speculation barrier $\pbarrier$, the remaining speculation window of the topmost instance is set to $0$.

Whenever speculation starts, one or more speculative instances are pushed on top of the stack (\Cref{tr:am-spec}) and when speculation ends, the speculative instance is then popped (\Cref{tr:am-rollback}). 
Finally, we note that how exactly the new speculative instances are created w.r.t. the auxiliary data depends on the specific speculative semantics, as we show in \Cref{sec:inst-spec-semantics-single}.
This is reflected in \Cref{tr:am-spec} not specifying how the new speculative instances $\phiStackx'$ are defined.

The always-mispredict behaviour $\behx{p}$ of a program $p$ is the set of all traces generated from an initial state until termination using the reflexive-transitive closure of $\specarrowx{}$.
Crucially, it is possible to connect the AM semantics and the non-speculative semantics through the non-speculative projection $\nspecProject{}$, which removes from a trace $\tauStack$  all events related to speculation. 
We lift the projection $\nspecProject{}$ to the behaviour of a program $p$ in the natural way.

We require that any instantiation of this template (e.g., for branches or stores) preserves the non-speculative behaviour of the program (Similar to the Oracle Semantics in \Cref{sec:oracle-general}). We formalize this as a correctness requirement that all our instances must satisfy:
\begin{requirement}[NS Consistency AM]\label{thm:ns-consistent-am}
     $\behNs{p} = \nspecProject{\behx{p}}$
\end{requirement}

The goal of the AM semantics is to derive security guarantees that are independent of the choice of a prediction oracle.
\Cref{thm:orac-overapprox} formalizes this intuition by precisely connecting the oracle and the AM semantics of any speculative semantics $\semx$.
In particular, \Cref{thm:orac-overapprox} states that checking SNI w.r.t. the AM semantics is sufficient to obtain security guarantees w.r.t. \textit{all} prediction oracles.
That is, if a program is SNI w.r.t. the always-mispredict semantics, then it is SNI irrespectively of the choice of prediction oracle.
Similarly, if a program violates SNI w.r.t. the always-mispredict semantics, then there is one prediction oracle for which SNI is violated.
As we show in \Cref{sec:inst-spec-semantics-single}, this holds for all instances studied in this paper.

\begin{requirement}[Oracle Overapproximation]\label{thm:orac-overapprox}
    $\snix ~\text{iff}~ \forall \orac\ldotp \SEsnix$
\end{requirement}

\section{Instances of Speculative Semantics}\label{sec:inst-spec-semantics-single}

Here we present specific instances of speculative semantics modelling the effect of speculative execution over branch instructions (\Cref{sec:v1-semantics}), $\storeC$ instructions (\Cref{sec:v4-semantics}), $\retC$ instructions (\Cref{sec:v5-semantics}, \Cref{sec:sls-semantics}) and indirect jump instructions (\Cref{sec:v2-semantics}) using the semantics templates presented in \Cref{sec:spec-semantics}.

Before detailing these specific instances, we must establish the properties that well-formed semantics must satisfy in our framework. We summarise these properties in a single definition:
\begin{definition}[Well-Formed Speculative Semantics]\label{def:sss-paper}
A speculative semantics $\semx$ is safe (denoted \ssssem{\semx}) if:
\begin{itemize}
    \item Oracle Overapproximation:
    $\snix ~\text{iff}~ \forall \orac\ldotp \SEsnix$
    \item NS Consistency:
    $\nspecProject{\behx{p}} = \behNs{p} = \nspecProject{\SEbehx{p}}$
    \item Symbolic Consistency:
    $\behx{p} = \conc(\behxa{p})$
\end{itemize}
\end{definition}

Intuitively, a well-formed speculative semantics is made of three components: an AM semantics, an oracle semantics, and a symbolic AM semantics.

First, the AM semantics must overapproximate the oracle semantics (for any oracle), guaranteeing that it is sufficient to check a program $p$ for \SNI{} w.r.t. the AM semantics.
Next, both the AM and the Oracle semantics must preserve the non-speculative behaviour of a program $p$. 
Applying the non-speculative projection ($\nspecProject{}$) to their traces exactly recovers the standard non-speculative behaviour.
Thus, we can execute $p$ only once to get the (non-)speculative behaviour of that program run.

Finally, to enable automated verification, we require the Symbolic AM semantics. The Symbolic Consistency property mandates that the concrete AM traces exactly match the concretization of the symbolic traces, where $\conc(\behxa{p})$ conceptually denotes the instantiation of the symbolic traces using concrete models $\conc{}$ that satisfy the collected path conditions.

For conciseness' sake, here we only present the concrete AM semantics in full detail and refer to the technical report for the full details of the oracle/symbolic semantics~\cite{techReport}, and defer the detailed discussion of how the symbolic semantics is utilized in practice to the implementation of \tool{} in \Cref{sec:impl-spec}.
For each of the speculative semantics, we prove that they satisfy all validity requirements, that is, that they are well-formed speculative semantics according to \Cref{def:sss-paper}.

\subsection{\texorpdfstring{${\semb}$}{Spec-B}: Speculation on Branch Instructions}\label{sec:v1-semantics}

Modern hardware uses a branch predictor to predict the outcome of branching decisions since this speeds up program execution. However, mispredictions can be steered and exploited by attackers leading to \specb attacks~ \cite{spectre}.

\subsubsection{The AM Semantics}

At every branch instruction, the always-mispredict semantics first speculatively executes the wrong branch for a fixed number of steps and then continues with the correct one. As a result, this semantics is deterministic and agnostic to implementation details of the branch predictor.

The state $\SigmaB$ of the AM semantics is a stack of speculative instances $\PhiB$ where reductions happen only on top of the stack.
Each instance $\PhiB$ contains the program $p$, a counter $\ctr$ that uniquely identifies the speculation instance, a configuration $\sigma$, and the remaining speculation window $n$ describing the number of instructions that can still be executed speculatively (or $\bot$ when no speculation is happening). In this semantics, we have $\specInstr = \jzC{}$ since branches are the source of speculation.

\begin{align*}
    \ti{Spec. States } \SigmaB  \bnfdef&\ \phiStackB
    &
    \ti{Spec. Instances } \PhiB \bnfdef&\ \tup{p, \ctr, \sigma, n}
\end{align*}

The judgement for the AM semantics is: $\SigmaB \specarrowB{\tau} \SigmaB'$.

\begin{center}
    \mytoprule{\phiStackB \specarrowB{\tau} \phiStackB'}
    
    \typerule {$\Bvr$:AM-NoSpec}
    { p(\sigma(\pc)) \notin \jzKywd \cup  \fcolorbox{lightgray}{lightgray}{$\Zb$} \cup \{\pbarrier \}  & \conctauarrow{\tup{p,\sigma}}{\tup{p,\sigma'}} 
    }
    {\tup{p, \ctr, \sigma, n + 1}
    \specarrowB{\tau}
    \tup{p, \ctr, \sigma', n}
     }{v1-noBranch-paper}

     \typerule {$\Bvr$:AM-Barr}
    {\instr{\pbarrier} & \conctauarrow{\sigma}{\sigma'}
    }
    {\tup{p, \ctr, \sigma, n}
    \specarrowB{}
    \tup{p, \ctr, \sigma', 0}
     }{v1-barr-paper}
     \typerule{$\Bvr$:AM-Rollback}
    { n' = 0\ \text{or}\ \text{p is stuck}  }
    { \tup{p, \ctr, \sigma, n} \cdot \tup{p, \ctr', \sigma', n'}
    \specarrowB{\rollbackObsB{\ctr}}
     \tup{p, \ctr', \sigma, n}
    }{v1-rollback-paper}
    
    \typerule{$\Bvr$:AM-Spec}
    {\instr{\pjz{x}{\lbl}} & \conctauarrow{\tup{p, \sigma}}{\tup{p, \sigma'}} & j = min(\omega, n)  \\
     \sigma'' = \sigma[\pc \mapsto l'] &  \tauStack = \tau \cdot \startObsB{\ctr} \cdot \pcObs{l}\\
     l' = {\begin{cases} \sigma´(\pc) + 1 & \text{if $\sigma'(\pc) = l$} \\
                        l & \text{if $\sigma'(\pc) \neq l$}
     \end{cases}
     }
    }
    {\tup{p, \ctr, \sigma, n + 1}
    \specarrowB{\tauStack}
    \tup{p, \ctr, \sigma', n} \cdot \tup{p, \ctr + 1, \sigma'', j}
    }{v1-branch-paper}

\end{center}

As mentioned, \Cref{tr:v1-branch-paper} pushes a new speculative state with the wrong branch, followed by the state with the correct one.
When speculation ends, \Cref{tr:v1-rollback-paper} pops the related state.
All other instructions are handled by delegating back to the non-speculative semantics (\Cref{tr:v1-noBranch-paper}).

\Cref{tr:v1-noBranch-paper} differs slightly from what was presented before in \Cref{sec:am-general}: it applies to instructions that are not branch or barrier instructions \textbf{and} are not in the metaparameter $\Zb$ (in \fcolorbox{lightgray}{lightgray}{gray}).
The latter is a set of instructions and is part of our composition framework (which we explain in \Cref{sec:expl-z}).
Instantiating $\Zb$ allows us to restrict when to apply non-speculative steps in composed semantics.
When we consider ${\semb}$ in isolation, $\Zb$ is the empty set (so, \Cref{tr:v1-noBranch-paper} applies to everything except branch and barrier instructions).
However, we will instantiate $\Zb$ in different manners when building the composed semantics.
In the following, we write $\semb^{\Zb}$ to stress the value of $\Zb$ when needed but we often omit $\Zb$ for simplicity.

The always-mispredict behaviour $\behB{p}$ of a program $p$ is the set of all traces generated from an initial state until termination using the reflexive-transitive closure of $\specarrowB{}$.

Note that \Cref{tr:v1-noBranch-paper}, \Cref{tr:v1-barr-paper} and \Cref{tr:v1-rollback-paper} are direct instantiations of the general rules described in \Cref{sec:am-general}. 
This holds true for all the other speculative semantics that we will present, thus, we omit them.

\subsubsection{Oracle Semantics}\label{sec:v1-oracle}

At every branch instruction, the oracle semantics queries the explicit oracle $\oracB$ for the branching decision.

Here, we summarize the key differences with the AM semantics.
First, speculative instances are extended to track the branching history $h$, which records the outcomes of prior branch instructions.
Second, when executing a $\jzC{}$ instruction, the oracle predicts the branch outcome (based on the branching history $h$) and a new speculative instance is pushed on top of the stack (\Cref{tr:v1-se-skip-paper}). 
Finally, whenever the speculation window of an instance anywhere on the stack reaches $0$, the execution needs to be rolled back or committed. 

\begin{center}
    \mytoprule{\PsiB \SEspecarrowB{\tau} \PsiB'}

     \typerule {$\Bvr$:NoBranch}
    { p(\sigma(\pc)) \notin \cup  \fcolorbox{lightgray}{lightgray}{$\Zb$}  & \conctauarrow{\sigma}{\sigma'} 
    }
    {\tup{p, \ctr, \sigma, h, n + 1}
    \SEspecarrowB{\tau}
    \tup{p, \ctr,\sigma', h, n}
     }{v1-se-noBranch-paper}
    \typerule{$\Bvr$:Spec}
    {\instr{\pjz{x}{\lbl}} & \conctauarrow{\sigma}{\sigma'}  & \orac(p,n,h, \sigma) = (\delta, \omega, \pcObs{\lbl''}) \\
    \delta = [\pc \mapsto l'']  & h' = h \cdot \tup{\sigma(\pc), \ctr, \delta} \\
   \tauStack = \tau \cdot \startObsB{\ctr} \cdot \pcObs{l''} 
    }
    {\tup{p, \ctr, \sigma, h, n + 1}
    \SEspecarrowB{\tauStack}
    \tup{p, \ctr, \sigma', h', n} \cdot \tup{p, \ctr + 1,  \sigma \uplus \delta, h', \omega}^{\langle \sigma,\delta \rangle}
    }{v1-se-skip-paper}
    
\end{center}

As in the oracle-semantics template, rolling back deletes all the instances above the rolled back instance, whereas committing updates the configuration, the counter and the branching history $h$ of the instance below and the committed instance is deleted.
These rules are not shown since they are direct instantiations of the general oracle rules presented in \Cref{sec:oracle-general}.

\subsection{\texorpdfstring{$\sems$}{Spec-S}: Speculation on Store Instructions}\label{sec:v4-semantics}
Modern processors write $\storeC$s to main memory asynchronously to reduce delays caused by the memory subsystem.
Processors employ a \textit{Store Queue} where not-yet-committed $\storeC$ instructions are stored before being permanently written to memory.
When executing a $\loadC{}$ instruction, the processor first inspects the store queue for a matching memory address.
If there is a match, the value is retrieved from the store queue (called \emph{store-to-load forwarding}), and otherwise, the memory request is issued to the memory subsystem. 
To speed up computation, processors employ memory disambiguation predictors to predict if the memory addresses of loads and stores match. 
Since the prediction can be incorrect, processors may speculatively bypass a $\storeC{}$ instruction in the store queue, leading to a $\loadC{}$ instruction retrieving a stale value \cite{S_specv4}.

\begin{example}[\spectre{}-STL]
Consider the code in \Cref{lst:v4-vanilla}, which is the \muasm{} translation of \Cref{lst:v4-vanilla1}:\looseness=-1

\noindent
\begin{minipage}[t]{0.45\textwidth}
\begin{lstlisting}[basicstyle=\small,style=MUASMstyle, caption=\spectre{}-STL in \muasm{}., label=lst:v4-vanilla,escapechar=|, captionpos=t]
store secret, p |\label{line:v4sec}|
store public, p |\label{line:v4pub}|
load eax, p |\label{line:leakv4-1}|
load edx, B + eax |\label{line:leakv4}|
\end{lstlisting}
\end{minipage}%
\hfill
\begin{minipage}[t]{0.5\textwidth}
Assume that the $\storeC{}$ instructions in \Cref{line:v4sec,line:v4pub} are still in the \textit{store queue} and not yet committed to main memory.
A misprediction of the memory disambiguator for the $\loadC{}$ instruction in \Cref{line:leakv4-1} causes it to bypass the $\storeC{}$ instruction in \Cref{line:v4pub} and retrieve the value from the stale $\storeC{}$ instruction in \Cref{line:v4sec}. The speculative access of the memory is then leaked into the microarchitectural state by the array access into \inlineCcode{B} in \Cref{line:leakv4}.
\end{minipage}
\end{example}

Here, we present the speculative AM semantics (\Cref{sec:v4-am}) and the corresponding oracle semantics (\Cref{sec:v4-oracle}).\looseness=-1

\subsubsection{Speculative Semantics}\label{sec:v4-am}
The overall structure of the $\sems$ semantics is similar to that of $\semb$: speculative execution is modelled using a stack of speculative states, instructions that do not start speculative transactions are executed by delegating back to the non-speculative semantics, and speculative transactions are rolled back whenever the speculative window reaches 0.
The key difference between $\sems$ and $\semb$ is the differing source of speculation $\specInstr$. Here $\specInstr = \storeC{}$ for $\sems$ instead of  $\specInstr = \jzC{}$ as in $\semb$.

The states used in $\sems$ are similar to those of $\semb$:
\begin{align*}
    \ti{Spec. States } \SigmaS  \bnfdef&\ \phiStackS
    &
    \ti{Spec. Instance } \PhiS \bnfdef&\ \tup{p, \ctr, \sigma, n}
\end{align*}
We also add a $\skipObs{n}$ observation denoting that the $\storeC{}$ instruction at program counter $n$ was speculatively bypassed.
\begin{align*}
    \ObsS \bnfdef&\ \Obs \mid \skipObs{n}
\end{align*}

The judgement $\SigmaS \specarrowS{\tau} \SigmaS'$ describes how $\SigmaS$ steps to $\SigmaS'$ emitting observation $\tau$. As in $\semb$, reductions only happen on top of the stack.\looseness=-1

\begin{center}
    \mytoprule{\phiStackS \specarrowS{\tau} \phiStackS'}

    \typerule{\Svr:AM-Spec}
        {
            \instr{\pstore{x}{e}} & \conctauarrow{\tup{p, \sigma}}{\tup{p, \sigma'}}  
            &  
            j = min(\omega, n) 
            \\
            \sigma'' = \sigma[\pc \mapsto \sigma(\pc) +1] 
            &
            \tau' = \tau \cdot \skipObs{\sigma(\pc)} \cdot \startObsS{\ctr}
        }
        {
            \tup{p, \ctr, \sigma, n + 1}
        \specarrowS{\tau'}
        \tup{p, \ctr, \sigma', n}
            \cdot \tup{p, \ctr + 1, \sigma'', j}
        }{v4-skip-new}

\end{center}
To model the effect of bypassing a $\storeC{}$ instruction, \Cref{tr:v4-skip-new} bypasses the $\storeC{}$ instruction by increasing the program counter without updating the memory and starts a new speculative transaction by pushing a new speculative instance on top of the state. 
A $\loadC{}$ instruction loading from the same memory location as the bypassed $\storeC{}$ instruction, therefore, retrieves a stale value.

The set $\behS{p}$ contains all traces  generated from an initial state until termination using the reflexive-transitive closure of $\specarrowS{}$.\looseness=-1 %

\subsubsection{Oracle Semantics}\label{sec:v4-oracle}
Instead of bypassing $\storeC{}$, the oracle semantics employs an oracle $\orac$ that decides if the $\storeC{}$ instruction should be speculatively bypassed or not. 
As before, the behaviour $\SEbehS{p}$ of a program $p$ is the set of all traces starting from an initial state until termination using the reflexive-transitive closure of the oracle semantics.\looseness=-1

\subsection{\texorpdfstring{$\semr$}{Spec-R}: Speculation on Return Instructions}\label{sec:v5-semantics}

The return-stack buffer (RSB) is a small stack the CPU uses to save return addresses upon $\pcall{}$ instructions. 
These saved return addresses are speculatively used when the function returns because accessing the RSB is faster than looking up the return address on the stack (stored in main memory).
This works well because return addresses rarely change during function execution. 
However, mispredictions can be exploited by an attacker \cite{ret2spec, spectreRsb}. 

\begin{example}[Return Speculation Vulnerability]
Consider the example in \Cref{lst:v5-example} and recall that register $\spR$ is used to find return addresses saved on the stack.
\begin{figure}[h]
    \begin{minipage}{.5\textwidth}
        \begin{lstlisting}[style=MUASMstyle, caption={A program exploiting RSB speculation.}, label={lst:v5-example}, escapeinside=!!]
        Manip_Stack:
            sp !$\xleftarrow{}$! sp + 8  !\label{v5:line1}!
            ret
        Speculate:
            call Manip_Stack !\label{v5:line4}!
            load eax, secret !\label{v5:line5}!
            load edx, eax   !\label{v5:line6}!
            ret
        Main:
            call Speculate
            skip !\label{v5:line11}!
        \end{lstlisting}
    \end{minipage}
    \qquad
    \begin{minipage}{.4\textwidth}
        \begin{tikzpicture}[arrow/.style = {dotted,-stealth, thick}]
        \node[rectangle, draw] (start) at (0,0) {Main};
        \node[rectangle, draw, right=of start] (middle) {Speculate};
        \node[rectangle, draw, right=of middle] (end)  {Manip\_Stack};
        
        \draw[arrow] (start) -- (middle);
        \draw[arrow] (middle) -- (end);
        \draw[arrow, red] (end.south) |- ++(0,-0.5) -|  (middle.south);
        \draw[arrow] (end.south east) |- ++(0,-1) -| (start.south); %
        \end{tikzpicture}
    \end{minipage}
    \caption{Control Flow of the vulnerable program in \Cref{lst:v5-example}. A red arrow indicates a speculative control flow that happens because of misprediction with the RSB.}
    \label{fig:rsb-control-flow}
\end{figure}
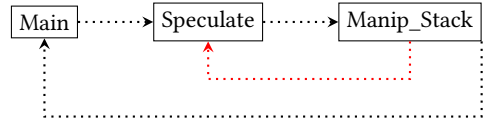

Each function call pushes a return address on the stack and decrements the $\spR$ register. 
After reaching the function \textit{Manip\textunderscore Stack}, the $\spR$ register is incremented (\cref{v5:line1}). 
Thus, $\spR$ points to the previous return address on the stack (i.e., \cref{v5:line11}), and the non-speculative execution continues in \textit{Main} and terminates.
However,  the return address of the call in \cref{v5:line4} is \cref{v5:line5} and it is on top of the RSB.
Thus, the CPU speculatively executes lines~\ref{v5:line5}--\ref{v5:line6} and leaks the secret.
\end{example}

This section describes the AM semantics (\Cref{sec:v5-am}) and the oracle semantics (\Cref{sec:v5-oracle}), %
Then, it discusses formalising different implementations of the RSB in the CPU (\Cref{sec:v5-rsb}).

\subsubsection{Speculative Semantics}\label{sec:v5-am}

Unlike before, the state of $\semr$ contains a model of the RSB which is used to retrieve return addresses instead of relying on the stack.

Thus, speculative instances of $\semr$ are extended with an additional entry $\Rsb$ for tracking the RSB, whose size is limited by a global constant $\Rsb_{size}$ denoting the maximal RSB size. 
A speculative instance $\PhiR$ now consists of the program $p$, the counter $\ctr$, the configuration $\sigma$, the speculation window $\omega$, and the RSB $\Rsb$. 
As before, a state $\SigmaR$ is a stack of speculative instances $\phiStackR$.

\begin{align*}
    \ti{Spec. States } \SigmaR  \bnfdef&\ \phiStackR
    &
    \ti{Spec. Instance } \PhiR \bnfdef&\ \tup{p, \ctr, \sigma, \Rsb, n}
\end{align*}

As before, in  $\SigmaR \specarrowR{\tau} \SigmaR$ reductions happen on the top of the stack and in this semantics we have $\specInstr = \callC{} \cup \retC{}$ since both $\callC{}$ and $\retC$ instructions interact with the RSB and $\retC$ instructions are the source of speculation.
\begin{center}

\mytoprule{\phiStackR \specarrowR{\tau} \phiStackR'}

\typerule{$\Rvr$:AM-Spec}
    {
        \instr{\pret} & 
        \sigma = \tup{m,a} &
        \conctauarrow{\tup{p,\sigma}}{\tup{p,\sigma'}}  
        \\
        \Rsb = \Rsb' \cdot l & 
        j = min(\omega, n) 
        &
        l \neq m(a(\spR))
        \\
        \sigma'' = \sigma[\pc \mapsto l, \spR \mapsto a(\spR) + 8] & 
        \tauStack = \tau \cdot \startObsR{\ctr} \cdot \retObs{l}
    }
    {\tup{p, \ctr, \sigma, \Rsb, n + 1}
    \specarrowR{\tauStack}
    \tup{p, \ctr, \sigma', \Rsb', n} \cdot \tup{p, \ctr + 1, \sigma'', \Rsb', j}
    }{v5-ret-paper}
    \typerule{$\Rvr$:AM-Call}
    {
        \instr{\pcall{f}} &  
        \conctauarrow{\tup{p,\sigma}}{\tup{p,\sigma'}}
        \\
        \Rsb' = \Rsb \cdot \tup{a(\pc) + 1} & 
        \vert \Rsb \vert < \Rsb_{size}
    }
    {
    \tup{p, \ctr, \sigma, \Rsb, n+ 1} 
    \specarrowR{\tau}
    \tup{p, \ctr, \sigma', \Rsb', n} 
    }{v5-call-paper}

\end{center}
During $\pcall{}$ instructions (\Cref{tr:v5-call-paper}), the return address is pushed on top of the RSB (if there is space available) and during $\pret$ instructions, the return address stored on the RSB is used if the entry on top of the RSB is different from the one stored on the stack (\Cref{tr:v5-ret-paper}). 
Then, the rule creates a new speculative instance that uses the return address from the RSB $\Rsb$. 
Note that speculation only happens when the return address from the RSB differs from the one on the stack (stored in $m(a(\spR))$). 

Here, we overview how our semantics behaves with empty and full RSB.
Whenever the RSB is empty, executing a $\pret{}$ instruction does not cause speculation and we return to the address pointed by $\spR$.
In contrast, whenever the RSB is full, executing a $\pcall{}$ instruction does not add entries to the RSB, i.e., we model an \emph{acyclic} RSB.%
\footnote{We follow the way AMD processors handle this kind of speculation~\cite{ret2spec}.}

The behaviour $\behR{p}$ is the set of all traces generated from an initial state until termination using $\specarrowR{}$.

\subsubsection{Oracle Semantics}\label{sec:v5-oracle}

Unlike before, the oracle cannot decide the outcome of the $\pret$ instruction, because the CPU always uses the return address stored in the RSB (if there is one) and it does not speculate otherwise~\cite{ST_constantTime_Spec}. 
The only thing the oracle decides here is the size of the speculation window $\omega$.

\subsubsection{Different Behaviours of Empty and Full RSBs}\label{sec:v5-rsb}
Modern CPUs use different RSB implementations that differ in the way they handle underflows and overflows, i.e., when the RSB is empty or full \cite{ret2spec}. 
For example, cyclic RSB implementations overwrite old entries when the RSB is full.
Alternatively, CPUs can fallback to other predictors (like the indirect branch predictor) to predict return addresses whenever the RSB is empty.

In our model, the RSB is not cyclic and there is no speculation when the RSB is empty (\Cref{tr:v5-retE-paper}).
\begin{center}
    \typerule{$\Rvr$:AM-Ret-Empty}
    {
    \instr{\pret} & \conctauarrow{\tup{p,\sigma}}{\tup{p,\sigma'}} \\
    }
    {
    \tup{p, \ctr, \sigma, \mathbb{\emptyset}, n + 1}
    \specarrowR{\tau} 
    \tup{p, \ctr, \sigma', \mathbb{\emptyset}, n}
    }{v5-retE-paper}
\end{center}

We remark that extending $\semr$ to support different RSBs implementations can be done with minimal effort.

\input{src/Semantics/sls-semantics}

\input{src/Semantics/v2-semantics}

%% file: src/Semantics/sls-semantics.tex
\subsection{\texorpdfstring{${\semsls}$}{Spec-SLS}: Straightline Speculation}\label{sec:sls-semantics}

Modern CPUs use the Branch Target Buffer (BTB) to assist with branch prediction. The BTB is indexed by possible jump instructions and predicts the next program counter. However, it also stores information about the type of branch (i.e. no branch, direct branch, indirect branch, or return) encountered at that location. Thus, it can happen that the BTB correctly predicts the location of a branch but mispredicts the type of the branch. For example, a $\pret$ instruction can be mispredicted to have the type no branch which in turn means that the CPU speculatively executes code right past the $\pret$ instruction. This kind of speculation is called straight-line speculation (SLS) \cite{sls-whitepaper, sls-whitepaper2}.
Note that a non-branch instruction can also be predicted as a branch which results in ghost jumps. 
Here, however, we focus only on straight-line speculation.

\begin{example}[Straightline Speculation Vulnerability]
Consider the example in \Cref{lst:sls-vanilla}:

\noindent %
\begin{minipage}[t]{0.40\textwidth} %
\vspace{0pt} %
\begin{lstlisting}[basicstyle=\small,style=MUASMstyle, caption=Code vulnerable to straightline speculation., label=lst:sls-vanilla,escapechar=|, captionpos=t]
ret |\label{line:slsret}|
load eax, p |\label{line:leaksls-1}|
load edx, B + eax |\label{line:sls-2}|
\end{lstlisting}
\end{minipage}%
\hfill %
\begin{minipage}[t]{0.55\textwidth} %
    \vspace{0pt} %
    Assume that $\inlineMUASMcode{p}$ is an attacker-controlled value during the execution.
After the execution of the $\retC$ instruction, the BTB of the processor predicts a no branch for the $\pret$ instruction. Thus, the processor speculatively bypasses the $\pret$ instruction and executes the following $\loadC{}$ instruction in \Cref{line:leaksls-1}.
The speculative access of the memory is then leaked into the microarchitectural state by the array access into $\inlineCcode{B}$ in \Cref{line:sls-2}.
\end{minipage}

\end{example}

This section describes the AM semantics (\Cref{sec:sls-am}) and the oracle semantics (\Cref{sec:sls-oracle}).

\subsubsection{Speculative Semantics}\label{sec:sls-am}
\begin{center}
    \centering
    \small
    \mytoprule{\phiStackSLS \specarrowSLS{\tau} \phiStackSLS'}

    \typerule{$\SLSvr$:AM-Spec}
    {\instr{\pret} &  \sigma \nsarrow{\tau} \sigma' & j = min(\omega, n) \\
     \sigma'' = \sigma[\pc \mapsto \sigma(\pc) + 1] & \tau' = \tau \cdot \skipObs{\sigma(\pc)} \cdot \startObsSLS{\ctr} 
    }
    {\tup{p, \ctr, \sigma, n + 1} \specarrowSLS{\tau'} \tup{p, \ctr, \sigma', n} \cdot \tup{p, \ctr + 1, \sigma'', j}
    }{sls-spec-paper}
 
\end{center}

Judgement $\SigmaSLS \specarrowSLS{\tau} \SigmaSLS'$ describes how $\SigmaSLS$ steps to $\SigmaSLS'$ emitting observation $\tau$. As in $\semb$, reductions only happen on top of the stack. Here we have $\specInstr =\retC{}$ because $\retC$ instructions are the source of speculation.

To model the effect of bypassing a $\retC{}$ instruction, \Cref{tr:sls-spec-paper} bypasses the $\retC{}$ instruction by increasing the program counter instead of using the return address to update the program counter and starts a new speculative transaction by pushing a new speculative instance on top of the state. 
This is in contrast to speculation on $\retC$ in $\semr$, which uses the RSB to update the program counter accordingly.

The set $\behSLS{p}$ contains all traces generated from an initial state until termination using the reflexive-transitive closure of $\specarrowSLS{}$.\looseness=-1 %

\subsubsection{Oracle Semantics}\label{sec:sls-oracle}

Instead of bypassing every $\retC{}$ instruction, the oracle semantics employs an oracle $\orac$ that decides if the $\retC{}$ instruction should be speculatively bypassed or not. 
As before, the behaviour $\SEbehSLS{p}$ of a program $p$ is the set of all traces starting from an initial state until termination using the reflexive-transitive closure of the oracle semantics.

%% file: src/Semantics/v2-semantics.tex
\subsection{\texorpdfstring{${\semj}$}{Spec-J}: Speculation on (Indirect) Jump Instructions}\label{sec:v2-semantics}

Modern processors use an indirect branch predictor to predict the outcome of indirect branches. Indirect branches are branches where the outcome is not known until the execution of the program. 
For example, $\pjmp{\inlineMUASMcode{r1}}$, which jumps to the location specified by register $\inlineMUASMcode{r1}$. Instead of waiting for the value of $r1$ to be available, the processor predicts the jump target based on the branch target buffer. However, attackers can poison the content of  the branch target buffer, which allows  the attacker to speculatively divert the control flow of the program \cite{spectre}.

\begin{figure*}[!h]
    \centering
    \begin{minipage}[t]{.45\textwidth}
        \begin{lstlisting}[basicstyle=\small,style=MUASMstyle, caption=Code vulnerable to indirect jump speculation., label=lst:v2-vanilla,escapechar=|, captionpos=t, keepspaces, showtabs=true]
        Main:
            beqz x, L1 |\label{v2-setup1}|
            r1 <- J1 
            jmp End
        L1:
            r1 <- J2  
        End:        |\label{v2-setup2}|
            jmp r1
        J1:
            x <- 0
            jmp Fin   |\label{v2-jump}|
        J2:
            x <- 1
            jmp Fin
            load eax, p |\label{v2-leak1}|
            load edx, B + eax |\label{v2-leak2}|
        Fin: |\label{v2-end}|
        \end{lstlisting}
        
        \end{minipage}\hfill
        \begin{minipage}[t]{.45\textwidth}
        \begin{lstlisting}[basicstyle=\small,style=MUASMstyle, caption=Example addition of endbr instructions., label=lst:v2-vanilla-endbr,escapechar=|, captionpos=t]
        Main:
            beqz x, L1 
            r1 <- J1 
            jmp End
        L1:
            r1 <- J2  
        End: 
            jmp r1
            |\color{PineGreen}\textbf{endbr}| |\label{v2-endbr1}|
        J1:
            x <- 0
            jmp Fin 
            |\color{PineGreen}\textbf{endbr}| |\label{v2-endbr2}|
        J2:
            x <- 1
            jmp Fin
            load eax, p |\label{v2-leak1-endbr}|
            load edx, B + p 
        Fin: 
        \end{lstlisting}
        
        \end{minipage}

    \caption{Example program vulnerable to jump speculation (\Cref{lst:v2-vanilla}) and a version restricting the jump speculation using $\color{PineGreen}\textbf{endbr}$ instructions (\Cref{lst:v2-vanilla-endbr}).}
    \label{fig:enter-label}
\end{figure*}

\begin{example}[A Program Exploiting Jump Speculation]
Consider the example in \Cref{lst:v2-vanilla} implementing a small jump table. In \Cref{v2-setup1} until \Cref{v2-setup2} the program assigns the jump target to $\inlineMUASMcode{r1}$ depending on the value of $\inlineMUASMcode{x}$. Next, the indirect jump in \Cref{v2-jump} executes and non-speculative execution continues at either \inlineMUASMcode{J1} or \inlineMUASMcode{J2}. In both cases, execution terminates by jumping to the end in \Cref{v2-end},
By exploiting jump speculation, an attacker could guide the indirect jump in \Cref{v2-jump} to \Cref{v2-leak1} thus leaking the contents of the private variable $\inlineMUASMcode{p}$ in \Cref{v2-leak2}.
\end{example}

This section describes the AM semantics (\Cref{sec:v2-am}), the oracle semantics (\Cref{sec:v2-oracle}). %

\subsubsection{Speculative Semantics}\label{sec:v2-am}
The attacker can inject any address as the new jump target, making analysis especially tricky. 
Consider a hypothetical speculation rule capturing the speculative behaviour of indirect jumps (note, this is not the version we use, which is presented below):
\begin{center}
    \centering
    \small

    \typerule{$\Jvr$:AM-Spec (Non-Final-Version)}
    {\instr{\pjmp{e}} & \sigma \nsarrow{\tau} \sigma' \\
    \exists x \subset e \ldotp x \in \Reg & j = min(\omega, n)  \\
     S = \{l \mid l \in p \setminus \{\sigma' . \pc \} \}  & \tup{p, \ctr, \sigma, j} \vdash^{S} \phiStackJ
    }
    {\tup{p, \ctr, \sigma, n + 1} \specarrowJ{\tau} \tup{p, \ctr, \sigma, n} \cdot \phiStackJ
    }{v2-spec-wrong}
    
    \typerule{Helper-Base} 
    {}
    {\tup{p, \ctr, \sigma, n} \vdash^{\varnothing} \varnothing
    }{v2-helper-base-paper}
    \typerule{Helper-Ind} 
    {
      l \in S &  \tup{p, \ctr + 1, \sigma, n} \vdash^{S \setminus l} \phiStackJ
    }
    {\tup{p, \ctr, \sigma, n} \vdash^{S} \tup{p, \ctr + 1, \sigma[\pc \mapsto l], n} \cdot \phiStackJ
    }{v2-helper-ind-paper}
\end{center}

\Cref{tr:v2-spec-wrong} allows the execution to speculatively jump to any location and we would need to create a speculative instance for \textbf{each} of these locations.
Here, \Cref{tr:v2-helper-ind-paper} and \Cref{tr:v2-helper-base-paper} are helpers creating the speculative transactions and we use $x \subset e$ to denote the subexpressions of $e$, thereby ensuring that only indirect jumps are used for speculation.

However, creating a speculative instance for each location in the program makes the analysis of the program infeasible because of the amount of states that need to be explored. %
Furthermore, against such a model, almost any ``interesting'' program would be considered insecure since speculation is, in practice, unrestricted. 

One of the techniques employed by modern CPUs to restrict the scope of indirect-jump speculation is (hardware) control-flow-integrity (CFI) \cite{cfi}.
CFI leverages tags to mark valid jump targets in the code and ensures that the code can only jump to these tagged targets.
There are different ways tagging and enforcement mechanisms can be implemented (see \citet{cfi_sok} for a comprehensive survey). We will focus on the hardware implementations of CFI of Intel (Intel-CET \cite{intel_cet}) and ARM (branch target identification \cite{arm-cet}) because they are available on new hardware and are supposed to apply also to transient instructions. %

Both implementations add a new instruction $\pendbr$ (in the case of ARM this instruction is called $\mathbf{bti}$) that marks the legal targets of indirect jumps in a program and enables coarse-grained forward edge control flow integrity. Enforcement is done by a state machine in hardware that ensures only valid jumps are allowed.
Furthermore, all available standard compilers (i.e. \gcc{}, \clang{}) already emit these $\pendbr$ instructions and these instructions are handled as no-ops if the current hardware does not support CFI, making it backwards-compatible.
Thus, we similarly add a $\pendbr$ instruction to \muasm{}, which marks legal targets for indirect jumps in the program and otherwise behaves like a $\pskip$ instruction (\Cref{tr:endbr-paper2}).
This restricts the scope of indirect-jump speculation and allows for a tractable analysis.
\begin{center}
$\begin{aligned}
     \text{(Instructions) } i \coloneqq \cdots \mid \pendbr &
 \typerule{Endbr}
{
\select{p}{a(\pc)} = \pendbr
}
{
\tup{p, \tup{m,a}} \nsarrow{} \tup{p, \tup{m, a[\pc \mapsto a(\pc)+1]}}
}{endbr-paper2}
\end{aligned}$
\end{center}

First, we keep track of all possible allowed jump targets in the program.
We collect the instruction number for each $\pendbr$ instruction into a labelset $\labelset$ defined as follows: $\labelset_p = \{i  \mid \forall i \in \Nat \ldotp  p(i) = \pendbr \}$. We will omit $p$ in $\labelset_p$ if it is clear from context.
One possible way to modify the program in \Cref{lst:v2-vanilla} with $\pendbr$ instruction is the program in \Cref{lst:v2-vanilla-endbr}.
Then, the labelset is defined as: $\labelset = \{\Cref{v2-endbr1}, \Cref{v2-endbr2} \}$. 
Next, we modify the rule for non-speculative jumps by requiring that the jump target is part of the allowed jump targets $\labelset$ (\Cref{tr:jmp-paper-v2}):
\begin{center}
\typerule{Jmp-Mod}
{
\select{p}{a(\pc)} = \pjmp{e} & \lbl = \exprEval{e}{a} \\
 \exists x \subset e \ldotp x \in \Reg & l \in \labelset
}
{
\tup{p, \tup{m, a}} \eval{p}{\pcObs{\lbl}} \tup{p, \tup{m, a[\pc \mapsto \lbl]}}
}{jmp-paper-v2}
\end{center}
Note that for all other semantics, we can define the set of possible jump targets to be all instruction labels in the program to recover the original rule for non-speculative jumps.

Next, we define the new speculative semantics:
\begin{center}
    \centering
    \small

    \mytoprule{\PhiJ \specarrowJ{\tau} \phiStackJ'}

     \typerule{$\Jvr$:AM-Spec}
    {
    \instr{\pjmp{x}} & \sigma \nsarrow{\tau} \sigma' & j = min(\omega, n) \\
    \exists x \subset e \ldotp x \in \Reg &   \tau' = \tau \cdot \pcObs{\sigma'(\pc)} \cdot \startObsJ{\ctr} \\
    \tup{p, \ctr, \sigma, j} \vdash^{\labelset \setminus \{\sigma' . \pc \}} \phiStackJ
    }
    {\tup{p, \ctr, \sigma, n + 1} \specarrowJ{\tau'} \tup{p, \ctr, \sigma', n} \cdot \phiStackJ
    }{v2-spec-paper}

\end{center}

We replace the unmitigated \Cref{tr:v2-spec-wrong} with the constrained \Cref{tr:v2-spec-paper}. 
The main difference lies in the definition of the allowed jump target set. The set $S$ in the unmitigated rule allows jumps to arbitrary addresses in the program. In the new rule this is restricted to $\labelset$, allowing jumps only to targets designated by $\pendbr$ instructions.
When an indirect branch is encountered, \Cref{tr:v2-spec-paper} creates a speculative instance for each target in $\labelset$, excluding the correct non-speculative target. Thus, the rule creates exactly $\vert\labelset\vert - 1$ speculative instances.
Consider the example in \Cref{lst:v2-vanilla-endbr} where $\labelset = \{\Cref{v2-endbr1}, \Cref{v2-endbr2} \}$.
Because \Cref{tr:v2-spec-paper} can only speculatively jump to targets in $\labelset$, execution cannot speculatively reach \Cref{v2-leak1-endbr}, successfully preventing the leakage of the private variable $\inlineMUASMcode{p}$. 

Crucially, because this rule creates a set of instances rather than a single misprediction (diverging from the standard single-target template in \Cref{sec:am-general}), it introduces an asymmetry between start and rollback observations: A single $\startObsJ{\ctr}$ observation is now associated with multiple rollback observations (exactly $\vert \labelset \vert -1$ many).\footnote{We could have decided to add matching $\startObsJ{\ctr}$ observations to \Cref{tr:v2-spec-paper} to balance the $\startObsJ{i}$ and the $\rollbackObsJ{i}$.}
However, because the speculative instances are pushed onto the execution stack sequentially, their execution is strictly nested. The resulting trace exhibits the following structure, where inner transactions are rolled back before the outer transaction concludes:
\begin{align*}
    \startObsJ{\ctr} \cdots \rollbackObsJ{\ctr + \vert \labelset \vert -2} \cdots \rollbackObsJ{\ctr}
\end{align*}

This nesting guarantees the correctness of the non-speculative projection function $\nspecProject{}$. Since $\nspecProject{}$ erases the trace segment between a start event and its matching rollback (here, the final $\rollbackObsJ{\ctr}$), all intermediate speculative events—including the nested rollbacks—are correctly removed from the trace.

\subsubsection{Oracle Semantics}\label{sec:v2-oracle}

Instead of predicting all of the possible indirect $\jC$ targets, the oracle chooses one of the possible jump targets in $\labelset$ and continues only with this one choice.
As before, the behaviour $\SEbehJ{p}$ of a program $p$ is the set of all traces starting from an initial state until termination using the reflexive-transitive closure of the oracle semantics.

\subsection{Safety of Semantics}\label{sec:prop-semantics}
We conclude this section by proving the core properties satisfied by all semantics.
\Cref{thm:sss-semantics} characterizes that all the speculative semantics presented in the paper are well-formed speculative semantics, i.e., (1) their always-mispredict version over-approximates the corresponding oracle version, (2) they are consistent under non-speculative trace projection, and (3) their symbolic version is consistent with the corresponding non-symbolic version.

\begin{theorem}[Well-Formed Speculative Semantics]\label{thm:sss-semantics}
The following statements hold:
\begin{itemize}
    \item ($\semb$ is \sss)
        $\ssssem{\semb}$
    \item ($\sems$ is \sss)
        $\ssssem{\sems}$
    \item ($\semr$ is \sss)
        $\ssssem{\semr}$
    \item ($\semsls$ is \sss)
        $\ssssem{\semsls}$
    \item ($\semj$ is \sss)
        $\ssssem{\semj}$
\end{itemize}

\end{theorem}

%% file: src/combined_semantics.tex
\section{A Framework for Composing Speculative Semantics}\label{sec:frame}

Although the instances of speculative semantics presented in \Cref{sec:inst-spec-semantics-single} each capture one specific aspect of speculation (e.g., $\semb$ captures branch speculation) 
, they do not capture the vulnerability in \Cref{lst:v1-v4-combined} (restated here for clarity) as the traces of \Cref{ex:comp-sni-iso} show.

\begin{example}[\SNI{} for \Cref{lst:v1-v4-combined-restate}]\label{ex:comp-sni-iso}
The traces generated are:
\begin{align*}
\begin{split}
\tauStack_{\Bv}^1 = \tauStack_{\Bv}^2 := {}& \storeObs{p} \concat \storeObs{p} \concat 
\startObsB{0} \concat \loadObs{p} 
\concat \loadObs{A + public} \concat \rollbackObsB{0} \concat \pcObs{9}
\end{split}\\
\begin{split}
\tauStack_{\Sv}^1 = \tauStack_{\Sv}^2 := {}& ... \concat \storeObs{p} \concat \startObsS{1} \concat \skipObs{1} \concat \pcObs{\perp}  \concat 
\rollbackObsS{1} \concat \pcObs{\perp}
\end{split}
\end{align*}
The program in \Cref{lst:v1-v4-combined-restate} seems secure since there is no secret value leaked in the speculative transaction; thus the program satisfies \SNI{} for $\semb$ and $\sems$ in isolation.
However, this program speculatively leaks when considering speculation over $\jzC{}$ and $\storeC{}$ instructions, but we would need a combined semantics to detect this vulnerability.
\end{example}

\begin{wrapfigure}[8]{L}{0.35\textwidth}
\vspace{-20pt}
\begin{lstlisting}[basicstyle=\small,style=Cstyle, caption= $\sembs$ example.,
    label=lst:v1-v4-combined-restate,escapechar=|, captionpos=t]
x = 0; 
p = &secret; 
p = &public;    
if (x != 0)     
    temp &= A[*p];      
\end{lstlisting}
\end{wrapfigure}

The vulnerability only appears when the branch predictor (\Cref{sec:v1-semantics}) and the memory disambiguator (\Cref{sec:v4-semantics}) are used \emph{together}.
Intuitively, we know that CPUs use all the speculation mechanisms described here (and many others as well) at the same time. 
Thus, we should not only focus on these individual speculation mechanisms in \textit{isolation} but we need to look at their \textit{combinations} as well.
That is, we need a way to compose the different semantics into new semantics that can reason about these ``combined'' leaks.

This section presents a novel, general framework for composing two speculative semantics $x$ and $y$, each one capturing the effects of a single speculation mechanism, to allow for speculation from both mechanisms $x$ and $y$. 
The semantics $x$ and $y$ are also called the \textit{source} semantics of the composition. 
Next, we first introduce the new composed semantics, which consists of an always-mispredict semantics, an oracle semantics, and a symbolic semantics (\Cref{sec:comb-am}). 
Then, we present the notion of \emph{well-formed} composition which we use  to study the properties of composed semantics  (\Cref{sec:comb-correctness}). %

\paragraph{New Notation}
The states $\Sigmaxy$, instances $\Phixy$, and the trace model $\Obsxy$ are defined as the union of the source parts.
Furthermore, we define a projection function $\specProjectxy{}$ and two projections $\specProjectxyx{}$ and $\specProjectxyy{}$ that return the first and second projection of the pair from $\specProjectxy{}$. These functions are lifted to states by applying them pointwise:
\begin{align*}
    &\Obsxy \eqdef\ \Obsx \cup \Obsy  
    &
    &\Phixy \eqdef\ \Phix \cup \Phiy
    &
     \Sigmaxy \eqdef\ \Sigmax \cup \Sigmay\\
    &\specProjectxy{} \colon \Phixy \mapsto (\Phix, \Phiy)
    &
    &\specProjectxyx{} \colon\ \Phixy \mapsto \Phix 
    &
    \specProjectxyy{} \colon\ \Phixy \mapsto \Phiy 
\end{align*}

For example, the $\PhiSR$ state resulting from the union of $\PhiS$ and $\PhiR$ states (from \Cref{sec:v4-am} and \Cref{sec:v5-am} respectively) is $\tup{p, \ctr, \sigma, \Rsb, n}$, as it contains all common elements (the program $p$, the counter $\ctr$, the state $\sigma$, and the speculation count $n$) plus the return stack buffer $\Rsb$ from $\PhiR$ only.
Taking the $\specProjectSSR{\cdot}$ of a $\PhiSR$ state returns the $\PhiS$ subpart, i.e., all but the return stack buffer.

We overload $\specProjectxyx{}$ and $\specProjectxyy{}$ to also work on traces $\tauStack$.
The projection $\specProjectxyx{\tau}$ deletes all speculative transactions (marked by $\startObsy{\id}$ and $\rollbackObsy{\id}$) not generated by the source semantics $x$. The definition of $\specProjectxyy{}$ is similar by replacing $x$ with~$y$:
\begin{align*}
    \specProjectxyx{\empTr} =~ \empTr 
    \qquad\qquad
    \specProjectxyx{(\tau \cdot \tauStack)} =&~ \tau \cdot \specProjectxyx{(\tauStack)}
    \\
    \specProjectxyx{(\startObsy{\id} \cdot \cdots \rollbackObsy{\id} \cdot \tauStack)} =&~ \specProjectxyx{\tauStack} 
\end{align*}
We indicate source semantics for $x$ and $y$ as $\semx$ and $\semy$ respectively and use $\semxy$ to indicate the composed semantics.

\subsection{Combined Speculative Semantics}\label{sec:comb-am}\label{sec:expl-z}

The combined semantics delegates back to the source semantics of $x$ and $y$ to model the effects of both speculation mechanisms (modeled by $x$ and $y$). This is captured in the two core rules below:
\begin{center}
    \typerule {AM-x-Step}
    {\specProjectxyx{\Phixy} \specarrowxZ{\tau} \specProjectxyx{\phiStackxy'}
    }
    {\Phixy
    \specarrowxyZ{\tau}
    \phiStackxy'
     }{comb-x-step-Z}
     \typerule {AM-y-Step}
    {\specProjectxyy{\Phixy} \specarrowyZ{\tau} \specProjectxyy{\phiStackxy'}
    }
    {\Phixy
    \specarrowxyZ{\tau}
    \phiStackxy'
     }{comb-y-step-Z}
\end{center}

The combined semantics does a step by either delegating back to the $x$ source semantics (\Cref{tr:comb-x-step-Z}) or to the $y$ one (\Cref{tr:comb-y-step-Z}).\footnote{To simplify notation, we omit that the $\Phix \setminus \Phiy$ parts of state $\Phixy$ in x-step (similar $\Phiy \setminus \Phix$ in y-step) do not change between $\Phixy$ and $\phiStackxy'$.}
The rules rely on metaparameter $\Zxy$, which is a pair of two metaparameters $\Zxy \eqdef (\Zx, \Zy)$ --- one for $x$ and one for $y$. 
We overload the projections $\specProjectxyx{}$ and $\specProjectxyy{}$ to extract the corresponding metaparameter from $\Zxy$, e.g., $\specProjectxyx{\Zxy} = \Zx$.

The role of $\Z$ is central to making the composed semantics work as expected. %
It restricts how the combined semantics delegates execution to the components to ensure that the correct rule is applied.\looseness=-1

With $\Z = (\emptyset,\emptyset)$, consider the execution of the $\jzC{}$ instruction in \Cref{line:v14branch} in \Cref{lst:v1-v4-combined-restate}. 
The combined semantics $\sembs$ can use \Cref{tr:comb-x-step-Z} to delegate back to $\semb$ for $\jzC{}$ instructions, creating a new speculative transaction (\Cref{tr:v1-branch-paper}).
However, $\sembs$ can also use \Cref{tr:comb-y-step-Z}, because $\jzC{}$ instructions are also handled by $\sems$. 
Unfortunately, this does not start speculation, which happens only on $\storeC{}$ instructions (\Cref{tr:v4-skip-new}). 

Intuitively, $\sembs$ should delegate back to $\semb$, so \Cref{tr:comb-y-step-Z} should not be applicable.
This can be obtained by instantiating $\Zbs = (\storeC{}, \jzC{})$, so that its projections are $\Zb = {\storeC{}}$ and $\Zs = {\jzC{}}$. 
Now, $\sembs$ can only apply \Cref{tr:comb-x-step-Z} on the $\jzC{}$ of \Cref{line:v14branch}, because $\Zs$ ensures that $\sems$ cannot execute $\jzC{}$ instructions, as depicted in the full rule for $\semsZ{{\jzC{}}}$ below (where we indicate the instructions derived from $\Zs = \jzC{}$ in blue):

\begin{center}
    \typerule {$\Svr$:AM-NoSpec}
    {p(\sigma(\pc)) \notin \storeKywd \cup \textcolor{RoyalBlue}{\mathbf{beqz}}
    & \conctauarrow{\tup{p,\sigma}}{\tup{p,\sigma'}} 
    }
    {\tup{p, \ctr, \sigma, n + 1}
    \specarrowS{\tau}^{\textcolor{RoyalBlue}{\mathbf{beqz}}}
    \tup{p, \ctr, \sigma', n}
     }{v4-noBranch-paper}
\end{center}

Having clarified the intuition behind the semantics, we can define the behaviour $\behxy{}$ as the set of all traces generated from initial states until termination using  $\specarrowxy{}$.

\subsubsection{Oracle Combination}\label{sec:comb-oracle}
Instead of using one oracle, the combination uses a pair of oracles, one from each source semantics. When delegating back to either source, the correct oracle of the source is handed over to the source semantics.

\subsubsection{Symbolic Combination}\label{sec:comb-symb}
Instead of using the AM semantics for delegation, the combined symbolic semantics $\semxya$ uses the symbolic source semantics for delegation. Furthermore, the new notation (union, projections) is lifted to the symbolic combination to create the symbolic states $\Sigmaxya$. The behaviour $\behxya{p}$ of program $p$ is the set of all traces generated using the symbolic semantics.

\subsection{Properties of Composition}\label{sec:comb-correctness}

We now illustrate the benefits of our composition framework.
For this, we first introduce a notion of well-formed composition (\Cref{sec:comb:correctness:wellformedness}), which intuitively tells when a combined semantics  ``makes sense''.
Then, we show that for well-formed compositions, if the source semantics are \sss, so is the combined semantics (\Cref{sec:comb:correctness:preservation}).
Since we proved this property for \emph{any} well-formed composition in our framework, all (well-formed) compositions we present in \Cref{sec:comb-in} are \sss \emph{for free}.
This proof reuse and extensibility is our framework's key advantage over having ad-hoc semantics combining multiple speculation mechanisms, which requires one to manually prove the \sss results we instead obtain for free.\looseness=-1

\subsubsection{Well-formed Compositions}\label{sec:comb:correctness:wellformedness}
The well-formedness conditions for the composition in \Cref{def:wellformed-paper}  ensures that the delegation between the source semantics is done properly.
Note that these conditions  are the \emph{minimal} set of assumptions that let us derive \sss of the combined semantics for free:

\begin{definition}[Well-Formed Composition]\label{def:wellformed-paper}
A composition $\semxy$ of two speculative semantics $\semx$ and $\semy$ is \emph{well-formed}, written $\wfc{\semxy}$, if:
\begin{asparaenum}
    \item \emph{(Confluence)} Whenever $\Sigmaxy \specarrowxy{\tau} \Sigmaxy'$ and $\Sigmaxy \specarrowxy{\tau} \Sigmaxy''$, then  $\Sigmaxy' = \Sigmaxy''$.
    \item \emph{(Projection preservation)} For all  $p$, $\behx{p} = \specProjectxyx{\behxy{(p)}}$ and $\behy{p} = \specProjectxyy{\behxy{(p)}}$.
    \item \emph{(Relation preservation)} If $\Sigmaxy\srelxy\Xxy$ and $\Sigmaxy\specarrowxy{\tauStack}^* \Sigmaxy'$ then $\Xxy \SEspecarrowxy{\tauStack'}\!\!^*\ \Xxy'$ and $\Sigmaxy'\srelxy\Xxy'$.
    \item \emph{(Symbolic preservation)} If $\Sigmaxya \specarrowxya{\taua} \SigmaxyaP$ and $\conc{\Sigmaxya} = \Sigmaxy$, then there is $\Sigmaxy'$ s.t. $\Sigmaxy \specarrowxy{\conc{\taua}} \Sigmaxy'$ and $\conc{\SigmaxyaP} = \Sigmaxy'$. \looseness=-1
\end{asparaenum}
\end{definition}

Next, we explain the well-formedness conditions:
\begin{asparaitem}
\item Confluence (point 1) ensures that the non-determinism of the combined semantics (that non-deterministically delegates back to its sources) is not harmful. 
Consider the assignment in \Cref{line:v14as} in \Cref{lst:v1-v4-combined}. 
$\sembs$ can delegate to either $\semb$ or $\sems$ to reduce the assignment.
If the combined semantics is \emph{confluent}, then it does not matter which source rule executes the assignment in  \Cref{line:v14as} in \Cref{lst:v1-v4-combined}, the semantics reaches the same state either way.

\item Projection preservation (point 2) ensures that the combined semantics is not hiding or forgetting traces of its sources.
Any observable emitted by a source semantics must be propagated to the combined one, this is also the reason why $\Obsxy$ is defined as the union of the source $\Obs$.

\item To explain relation preservation (point 3), we need to mention a technical detail: the state relation (denoted $\srelxy$ and defined in our technical report) between the AM states ($\Sigmaxy$) and the Oracle ones ($\Xxy$).
Intuitively, two states are related if they are the same or if one is waiting on a speculation of the other to end. %
Then, point (3) ensures that whenever we start from related states ($\Sigmaxy\srelxy\Xxy$) and we do one or more steps of the AM composed semantics ($\Sigmaxy\specarrowxy{\tauStack}^* \Sigmaxy'$), then we can \emph{always} find a related state ($\Sigmaxy'\srelxy\Xxy'$)  that is reachable by performing one or more steps of the composed oracle semantics ($\Xxy \SEspecarrowxy{\tauStack'}\!\!^*\ \Xxy'$).
This fact is used when proving that \SNI{} of a program under the composed AM semantics implies \SNI{} under the composed oracle semantics (point 1 of \Cref{def:sss-paper}).
Thus, it is not important for the AM and the Oracle semantics to produce the same traces, just that the two AM traces and the two Oracle traces are pairwise equivalent -- which follows from the state relation.

\item Finally, symbolic preservation (point 4) ensures that any step of the always-mispredict  composed semantics corresponds to the concretization of a step of the symbolic composed semantics (and vice versa\footnote{For space reasons, \Cref{def:wellformed-paper} only reports one direction (with a simplified notation).}).
Note that proving symbolic preservation is almost trivial whenever both source semantics enjoy the same property (like our semantics $\semb$, $\semj$, $\sems$, $\semr$ and $\semsls$).
\end{asparaitem}

\subsubsection{\sss Preservation}\label{sec:comb:correctness:preservation}
The key result of this section is that well-formed compositions whose sources are well-formed speculative semantics (\sss) are also \sss (\Cref{thm:comp-sss-paper}).
Note that our proof of \Cref{thm:comp-sss-paper}\onlyShortVersion{, available in the companion technical report~\cite{techReport}, }\xspace holds for \emph{any} well-formed composition in our framework and, therefore, it applies \emph{for free} to all the compositions in \Cref{sec:comb-in}.

\begin{theorem}[$\semxy$ is \sss]\label{thm:comp-sss-paper}
If $\ssssem{\semx}$ and $\ssssem{\semy}$ and $\wfc{\semxy}$, then $\ssssem{\semxy}$.
\end{theorem}

As a corollary of \Cref{thm:comp-sss-paper}, we obtain that the security of well-formed compositions is related to the security of their components (\Cref{thm:comb:sni-preservation}).
In particular, whenever a program is insecure w.r.t. one of the components, then it is insecure w.r.t. the composed semantics.
Dually, if a program is secure w.r.t. the composed semantics, then it is secure w.r.t. the single components.
Note, however, that there are programs that are secure for the single components but insecure w.r.t. the composed semantics like \Cref{lst:v1-v4-combined}.

\begin{theorem}[Combined \SNI{} Preservation]\label{thm:comb:sni-preservation}
Whenever $\wfc{\semxy}$ holds: 
\begin{compactitem}
    \item If  $\nsnix$ or $\nsniy$, then $\nsnixy$.
    \item If  $\snixy$, then $\snix$ and $\sniy$.
\end{compactitem}
\end{theorem}

These results have an immediate practical impact on \tool:
(1) the analysis of \tool{} relies on the (symbolic) speculative semantics, (2) the source semantics $\semb$, $\semj$, $\sems$, $\semr$ and $\semsls$ are \sss, (3) well-formed compositions are also \sss, and (4) the composition of $\semb$, $\semj$, $\sems$, $\semr$ and $\semsls$ are well-formed.
So, \tool{} equipped with any combination of the $\semb$, $\semj$, $\sems$, $\semr$ and $\semsls$ produces sound results, i.e., whenever the tool proves that a program is leak-free then the program satisfies \SNI{}.
In the next section, we describe all the compositions and prove they are well-formed (this implies that they are \sss thanks to \Cref{thm:comp-sss-paper}). %

%% file: src/combined_instances.tex
\section{Instantiating our Framework}\label{sec:comb-in}

\newcommand{\lineCol}{gray}
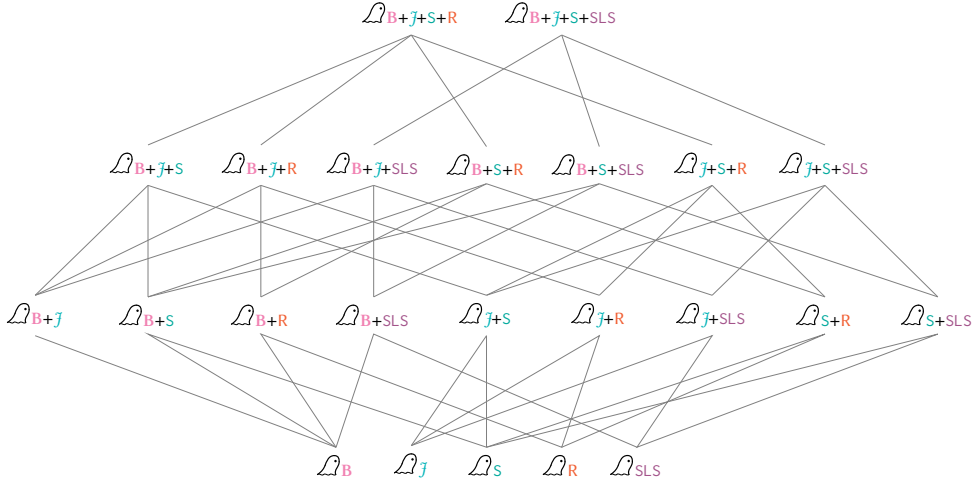
\begin{figure}
    \centering
    \begin{tikzpicture}[line width=0.1 mm, arrows={[red]}]
          \node (BJSR) at (-1, 4) {$\sembjsr$};
          \node (BJSSLS) at (1, 4) {$\sembjssls$};
          \node (BJS) at (-4.5, 2) {$\sembjs$};
          \node (BJR) at (-3, 2) {$\sembjr$};
          \node (BJSLS) at (-1.5, 2) {$\sembjsls$};
          \node (BSR) at (0, 2) {$\sembsr$};
          \node (BSSLS) at (1.5, 2) {$\sembssls$};
          \node (JSR) at (3, 2) {$\semjsr$};
          \node (JSSLS) at (4.5, 2) {$\semjssls$};
          \node (BJ) at (-6,0) {$\sembj$};
          \node (BS) at (-4.5,0) {$\sembs$};
          \node (BR) at (-3,0) {$\sembr$};
          \node (BSLS) at (-1.5,0) {$\sembsls$};
          \node (JS) at (0,0) {$\semjs$};
          \node (JR) at (1.5,0) {$\semjr$};
          \node (JSLS) at (3,0) {$\semjsls$};
          \node (SR) at (4.5,0) {$\semsr$};
          \node (SSLS) at (6,0) {$\semssls$};
          \node (B) at (-2,-2) {$\semb$};
          \node (J) at (-1,-2) {$\semj$};
          \node (S) at (0,-2) {$\sems$};
          \node (R) at (1,-2) {$\semr$};
          \node  (SLS) at (2,-2) {$\semsls$};
          \draw[\lineCol]  (B.north) -- (BJ.south);
          \draw[\lineCol]  (J.north) -- (JS.south);
          \draw[\lineCol]  (S.north) -- (JS.south);
          \draw[\lineCol]  (S.north) -- (SR.south) -- (R.north);
          \draw[\lineCol]  (B.north) -- (BS.south) -- (S.north) -- (SSLS.south) -- (SLS.north) -- (JSLS.south) -- (J.north);
          \draw[\lineCol]  (B.north) -- (BR.south) -- (R.north) -- (JR.south) -- (J.north);
          \draw[\lineCol]  (B.north) -- (BSLS.south) -- (SLS.north);
          \draw[\lineCol] (BJ.north) -- (BJS.south);
          \draw[\lineCol] (BJ.north) -- (BJR.south);
          \draw[\lineCol]  (BJ.north) -- (BJSLS.south);
          \draw[\lineCol]  (BS.north) -- (BSR.south);
          \draw[\lineCol]  (BS.north) -- (BSSLS.south);
          \draw[\lineCol]  (BS.north) -- (BJS.south);
          \draw[\lineCol]  (BR.north) -- (BSR.south);
          \draw[\lineCol]  (BR.north) -- (BJR.south);
          \draw[\lineCol]  (BSLS.north) -- (BSSLS.south);
          \draw[\lineCol]  (BSLS.north) -- (BJSLS.south);
          \draw[\lineCol]  (JS.north) -- (BJS.south);
          \draw[\lineCol]  (JS.north) -- (JSR.south);
          \draw[\lineCol]  (JS.north) -- (JSSLS.south);
          \draw[\lineCol]  (JR.north) -- (BJR.south);
          \draw[\lineCol]  (JR.north) -- (JSR.south);
          \draw[\lineCol]  (JSLS.north) -- (BJSLS.south);
          \draw[\lineCol]  (JSLS.north) -- (JSSLS.south);
          \draw[\lineCol]  (SR.north) -- (BSR.south);
          \draw[\lineCol]  (SR.north) -- (JSR.south);
          \draw[\lineCol]  (SSLS.north) -- (BSSLS.south);
          \draw[\lineCol]  (SSLS.north) -- (JSSLS.south);
          \draw[\lineCol]  (BSR.north) -- (BJSR.south);
          \draw[\lineCol]  (BSSLS.north) -- (BJSSLS.south);
          \draw[\lineCol]  (BJS.north) -- (BJSR.south);
          \draw[\lineCol]  (BJR.north) -- (BJSR.south);
          \draw[\lineCol]  (BJSLS.north) -- (BJSSLS.south);
          \draw[\lineCol]  (JSR.north) -- (BJSR.south);
          \draw[\lineCol]  (JSSLS.north) -- (BJSSLS.south);
    \end{tikzpicture}
    \caption{Partial order of the different combinations that we instantiate using our framework}
    \label{fig:instantiate}
\end{figure}

We instantiate our combination framework with all possible combinations of $\semb$, $\semj$, $\sems$, $\semr$ and $\semsls$, yielding \nrComb{} combinations depicted in \Cref{fig:instantiate}. Combinations of $\semsls$ and $\semr$ are impossible because they speculate on the same instructions ($\pret$). Thus, we cannot set the metaparameter $Z$ in a sensible way (more details in \Cref{sec:limitations}).
Since we cannot describe all of the combinations in detail, we focus on two representative combinations: $\semsr$ (\Cref{sec:comp-45}) and $\sembjsr$ (\Cref{sec:comp-1245}); the other combinations can be instantiated in a similar way.

For each of these, we overview the combined AM semantics using examples
and we prove that the combined semantics is well-formed, i.e., it satisfies \Cref{def:wellformed-paper}.

\onlyShortVersion{Full details and well-formedness proofs of all other combinations are available in the companion technical report~\cite{techReport}}.

\subsection{\texorpdfstring{$\semsr$}{Spec-S+R} Composition }\label{sec:comp-45}

To combine semantics using our framework, we need to define the states, observations, and metaparameter $\Zsr$ for the composed semantics $\semsr$.
The combined state $\SigmaSR$ is the union of the states $\SigmaS$ and $\SigmaR$; thus it contains the RSB $\Rsb$ as well. 
\begin{gather*}
\begin{aligned}
\ti{Spec. States } \SigmaSR  \bnfdef&\ \phiStackSR
&
\ti{Spec. Instance } \PhiSR \bnfdef&\ \tup{p, \ctr, \sigma, \Rsb, n}
\end{aligned}
\end{gather*}
The union $\ObsSR{}$ of the trace models $\ObsS{}$ and $\ObsR{}$ is defined as:
\begin{gather*}
\begin{aligned}
\ObsSR{} \bnfdef&\ \startObsS{n} \mid \startObsR{n} \mid \rollbackObsS{n} \mid \rollbackObsR{n} \mid \skipObs{n} \mid ... 
\end{aligned}
\end{gather*}

To define the metaparameter  $\Zsr$, we need to identify the instructions that are related with speculative execution for each component semantics.
For $\sems$, the only instruction associated with speculative execution is $\storeC{}$, since the semantics can only speculatively bypass stores.
For $\semr$, even though the semantics speculates only over $\retC{}$ instructions,  $\callC{}$ instructions also affect speculative execution since $\semr$ pushes return addresses onto the RSB $\Rsb$ when executing $\callC$s. 
Therefore, we set the metaparameter $\Zsr$ to $( \callC{} \cup \retC{}, \storeC{})$.
This ensures that in $\semsr$, $\storeC{}$ instructions are only executed by delegating back to $\semsZ{\callC{} \cup \retC{}}$ whereas $\callC{}$ and $\retC{}$ instructions are only executed by delegating back to $\semrZ{\storeC{}}$.\looseness=-1

\Cref{thm:v45-goodcomp-paper} states the combination of $\sems$ and $\semr$ described above is well-formed.
Given that $\sems$ and $\semr$ are \sss (\Cref{thm:sss-semantics}), we can derive ``for free'' that $\semsr$ is \sss  (\Cref{thm:comp-sss-paper}).

\begin{theorem}[$\semsr$ is well-formed]\label{thm:v45-goodcomp-paper}
$\!\wfc{\semsr}$
\end{theorem}

\begin{lstlisting}[style=MUASMstyle, caption={$\semsr$ example.}, label={lst:v4-v5-comb}, escapechar=|,
  float=tp,
  floatplacement=tbp,]
Manip_Stack:
    sp |$\xleftarrow{}$| sp + 8 |\label{line:v45-add}|
    ret
Speculate:
    call Manip_Stack |\label{line:v45-start-manip}|
    store secret, p |\label{line:v45x}|
    store pub, p |\label{line:v45s}|
    load eax, p |\label{line:v45l}|
    load edi, eax |\label{line:v45l2}|
    ret
Main:
    call Speculate |\label{line:v45start}|
    skip |\label{line:v45end}|
\end{lstlisting}

\Cref{lst:v4-v5-comb} presents a program that contains a leak that can be detected only by $\semsr$ but not by its components $\sems$ and $\semr$.
Execution starts on \Cref{line:v45start} by calling the function $\speculate$ and it continues at \Cref{line:v45-start-manip}. 
Next, the function $\manipStack{}$ is called and the stack pointer $\spR$ is incremented (\Cref{line:v45-add}).
This modifies the return address of the function $\manipStack{}$ to now point to \Cref{line:v45end} (the return address of the $\callC{}$ to $\speculate$).
Under $\semr$, mispredicting the return address of $\manipStack{}$ using the RSB leads to continuing the execution at \Cref{line:v45x}. 
However, the $\storeC{}$ instructions in \Cref{line:v45s} overwrites the secret value stored in \Cref{line:v45x}.
Then, the $\loadC{}$ instructions in \Cref{line:v45l} and \Cref{line:v45l2} emit only public values. 
As a result, no secret is leaked and speculation ends.
Similarly, under $\sems$, speculation over store bypasses has no effect in  \Cref{lst:v4-v5-comb} because the $\storeC{}$ instruction in \Cref{line:v45x} is never reached and function $\manipStack{}$ returns to \Cref{line:v45end}.
Therefore, the leak is missed under $\sems$ and $\semr$, i.e., $\text{\Cref{lst:v4-v5-comb}} \vdash_{\Sv} \text{\SNI}$ and $\text{\Cref{lst:v4-v5-comb}} \vdash_{ \Rv} \text{\SNI}$.\looseness=-1

However, under the combined semantics $\semsr$, the $\storeC{}$ instruction on \Cref{line:v45s} is now speculatively bypassed and when returning from function $\manipStack{}$ the execution speculatively continues from \Cref{line:v45l}. 
Now, the $\loadC{}$ instructions are executed and the secret is leaked, as shown in the traces below.
Since $\mi{secret}$ is a high value, there are low-equivalent configurations $\sigma^1, \sigma^2$ that differ in the value of $\mi{secret}$.
Thus, there are two traces that differs in the observation $\loadObs{secret}$ (highlighted in gray).
Hence, the program is not secure under the combined semantics, i.e., $\text{\Cref{lst:v4-v5-comb}} \nvdash_{\Sv + \Rv} \text{\SNI}$.
\begin{align*}
\tau_{\Sv + \Rv}^2 \neq \tau_{\Sv + \Rv}^1 \isdef &\  
    \callObs{\mathit{Speculate}} \cdots \startObsR{0} \cdots \startObsS{1} \cdots \rollbackObsS{1} 
    \cdots
    \startObsS{2} \concat \skipObs{7} \concat \loadObs{p} \concat  \fcolorbox{lightgray}{lightgray}{$\loadObs{secret}$} 
    \cdots
\end{align*}
 
\newcommand\tikzmarkT[1]{%
  \tikz[remember picture,overlay]\node (#1) {};%
} 

The relation between the source semantics and their composition is visualised in \Cref{fig:comb-sem}, which shows the insecure programs (with respect to \SNI{}) detected under the  different semantics.
The combined semantics encompasses all vulnerable programs of $\sems$ and $\semr$ \textit{and} additional programs like \Cref{lst:v4-v5-comb}.
These additional programs are the reason why the semantics $\semsr$ is ``stronger than the sum of its parts'' $\sems$ and $\semr$.
\begin{figure}[!ht]
    \centering
\begin{tikzpicture}[remember picture]

\tikzset{
 dot/.style = {circle, fill=black, minimum size=3pt,
               inner sep=0pt, outer sep=0pt},
 }

\draw [black, dashed, line width=0.5mm] plot [smooth cycle] coordinates {(-2,0) (-1,1) (1,1.5) (3,1.5) (6,0) (2,-1)};

\draw [\scol] plot [smooth cycle] coordinates {(-1,0) (1,0.5) (2,0.5) (2.5,0) (2,-0.5) };

\draw [\rcol] plot [smooth cycle] coordinates {(2,0.8) (4,0.5) (5,0.3) (5,0) (4,-0.5) };

\node[draw] at (3, 1.9)  {$\semsr$};

\node[draw, \scol] at (-0.6, 0.7)    {$\sems$};
\node[draw, \rcol] at (3.3, 1.0)    {$\semr$};

\node[dot,\scol] at (1,0) (V4E) {}; %
\node[dot, \rcol] at (4,0) (V5E) {}; %
\node[dot,black] at (1.5,1.1) (V45E) {}; %

\node[draw] at (-0.6, 1.7)   (V4S) {\Cref{lst:v4-vanilla}};
\node[draw] at (5.5, 1.4)    (V5S){\Cref{lst:v5-example}};
\node[draw] at (1.5, 2.1)    (V45S){\Cref{lst:v4-v5-comb}};

 \draw[->, overlay,\scol, thick, dotted,>=latex,] (V4S) -- (V4E);
 \draw[->, overlay, \rcol, thick, dotted,>=latex,] (V5S) -- (V5E);
 
 \draw[->, overlay,black, thick, dotted,>=latex,] (V45S) -- (V45E);

\end{tikzpicture}
\caption{Relating $\sems$, $\semr$ and $\semsr$ w.r.t. \SNI{}.}
\label{fig:comb-sem}
\end{figure}

\subsection{\texorpdfstring{$\sembjsr$}{Spec-B+J+S+R} Composition }\label{sec:comp-1245}

We conclude this section by combining four semantics $\semb$, $\semj$, $\sems$ and $\semr$ (as mentioned, a combination with five semantics is not possible because $\semsls$ and $\semr$ speculate on the same instruction). 
Our framework (\Cref{sec:frame}) allows us to only combine a pair of source semantics into a combined one.
For simplicity, we present $\sembjsr$ as a direct combination of the four source semantics (technically, we obtain $\sembjsr$ by combining $\sembj$ with $\semsr$).

The metaparameter $\Zbjsr$ (which we represent as a quadruple of values) is :
\begin{align*}
    \Zb \isdef& (\callC{} \cup \retC{} \cup \storeC{} \cup \jC) \\ 
    \Zj \isdef& (\callC{} \cup \retC{} \cup \storeC{} \cup \jzC) \\ 
    \Zs \isdef& (\callC{} \cup \retC{} \cup \jzC{} \cup \jC) \\ 
    \Zr \isdef& (\jzC \cup \storeC{} \cup \jC) \\ 
    \Zbjsr \isdef& (\Zb, \Zj, \Zs, \Zr)
\end{align*}
As a result, the combined semantics $\sembjsr$ can only delegate to the corresponding speculative semantics for the appropriate speculation sources.

As stated in \Cref{thm:v1245-goodc}, $\sembjsr$ is well-formed and we derive that $\sembjsr$ is \sss through \Cref{thm:comp-sss-paper}.\looseness=-1

\begin{theorem}[$\sembsr$ is well-formed]\label{thm:v1245-goodc}
$\wfc{\sembjsr}$
\end{theorem}

\begin{figure}
\centering
\begin{minipage}[t]{0.48\textwidth}
\begin{lstlisting}[style=MUASMstyle, name=combined, escapechar=|]
Manip_Stack:
    sp <- sp + 8
    ret
Speculate:
    call Manip_Stack
    r1 <- L2 |\label{line:v1245r}| 
    jmp r1 |\label{line:v1245jump}|
L1: 
    endbr |\label{line:v1245jumpT}|
    x <- 0 
    beqz x, L2 |\label{line:v1245branch}|
    load eax, p
    load edi, eax
\end{lstlisting}
\end{minipage}%
\hfill
\begin{minipage}[t]{0.48\textwidth}
\begin{lstlisting}[style=MUASMstyle, name=combined, escapechar=|]
L2:
    ret
Main:
    store secret, p 
    store pub, p |\label{line:v1245s}|
    call Speculate
\end{lstlisting}
\end{minipage}
    \caption{$\sembjsr$ example.}
    \label{lst:v1-v2-v4-v5-comb}
\end{figure}

\Cref{lst:v1-v2-v4-v5-comb} depicts a leaky program that can be detected only under $\sembjsr$, since the program satisfies \SNI{} under $\semb$, $\semj$, $\sems$ and $\semr$. 
Under $\sembjsr$, the $\storeC{}$ instruction in \Cref{line:v1245s} is bypassed.
Therefore, when returning from the $\manipStack$ function, the program mispredicts the return address and speculatively returns to \Cref{line:v1245r}. 
Here, the indirect jump in \Cref{line:v1245jump} mispredicts and execution continues speculatively in \Cref{line:v1245jumpT}.
Finally, the $\jzC{}$ instruction in \Cref{line:v1245branch} is mispredicted and the $\loadC{}$ instructions are executed, which leaks the secret value.

The resulting traces, differing in the value of \textit{secret} exposed by the $\loadObs{secret}$ observation (highlighted in gray), are given below:
\begin{align*}
\tau_{\Bv + \Jv + \Sv + \Rv}^2 \neq \tau_{\Bv + \Jv + \Sv + \Rv}^1 \isdef&\
    \cdots \startObsS{1} \concat \skipObs{18} \concat \callObs{\mathit{Speculate}} \cdot \callObs{\mathit{Manip\textunderscore Stack}} \concat  \startObsR{2} \concat \retObs{6} \concat \startObsJ{3} \\
     &\ \concat \pcObs{9} \concat
    \startObsB{4} \concat \pcObs{12} \concat \loadObs{p} \concat \fcolorbox{lightgray}{lightgray}{$\loadObs{secret}$} \concat \rollbackObsB{4} \cdot \rollbackObsJ{3} \concat \rollbackObsR{2} \concat \rollbackObsS{1} \cdots
\end{align*}
Thus, the program is not secure, i.e., $\text{\Cref{lst:v1-v2-v4-v5-comb}} \nvdash_{\Bv + \Jv + \Sv + \Rv} \text{\SNI}$.

%% file: src/implementation.tex
\section{Detecting Speculative Information Flows : The \tool algorithm}\label{sec:impl-spec}

This section presents \tool{}, a program analysis for detecting speculative leaks or proving their absence.
The core idea behind \tool is to formally compare a program's behavior under speculative execution against its standard, non-speculative execution. 
A leak is identified if the speculative execution reveals information that is not revealed by the non-speculative execution.
To make this comparison comprehensive, analyzing every possible concrete execution is infeasible. Therefore, Spectector leverages symbolic execution to represent all possible program behaviors concisely as a set of symbolic traces. Our approach requires two key formalisms:

\begin{enumerate}
    \item A symbolic non-speculative semantics to establish the program's intended, baseline behavior (\Cref{sec:impl-symbolic}).

    \item A symbolic architectural model (AM) semantics to model the processor's speculative behavior (\Cref{sec:sym-general-am}).
\end{enumerate}
By analyzing the symbolic traces generated from both semantics with an SMT solver, \tool can identify discrepancies that correspond to memory or control-flow leaks. 
If no such discrepancies are found, the program is proven secure against the modeled speculative behaviors (\Cref{sec:spectector-alg}). 
Finally, we discuss the implementation of \tool within the \ciao{} logic programming system~\cite{ciao} (\Cref{sec:tool}).

\subsection{Symbolic Non-Speculative Semantics for \texorpdfstring{\muasm{}}{uASM}}\label{sec:impl-symbolic}
To establish a ground truth for our analysis, we first define the program's intended behavior using a symbolic non-speculative semantics, denoted by the relation $\tup{p, \sigmaa} \nsarrowa{\taua} \tup{p, {\sigmaa}'}$.

This semantics lifts the concrete non-speculative semantics $\tup{p, \sigma} \nsarrow{\tau} \tup{p, \sigma'}$ to operate over symbolic values instead of concrete ones. We now detail the key extensions.

Symbolic program states consist of a program $p$ and a symbolic configuration $\sigmaa \bnfdef \tup{sm, sa}$.

Concrete memories $m$ are replaced with symbolic memories $sm$, modeled as symbolic memories using the standard theory of arrays \cite{calc-comp}. 
We model memory updates as triples of the form $\symWrite{sm}{se}{se'}$, which update the symbolic memory $sm$ by assigning the symbolic value $\sexpr'$ to the symbolic location $\sexpr$. Memory reads $\symRead{sm}{\sexpr}$ retrieve the value at symbolic memory location $\sexpr$ in $sm$.
$\symRead{sm}{\sexpr}$ and $\symWrite{sm}{\sexpr}{se'}$ are used in \Cref{tr:load-sym} and \Cref{tr:store-sym} respectively, which is the only difference to the concrete non-speculative semantics.
Concrete register assignments $a$ are replaced with symbolic register assignments $sa$, which are functions mapping registers to symbolic expressions.\looseness=-1

Symbolic expressions represent computations over symbolic values.
A \textit{symbolic expression} $\sexpr$ is a concrete value $n \in \Val$, a symbolic value $s \in \SymVal$, an if-then-else expression $\ite{\sexpr}{\sexpr'}{\sexpr''}$, or the application of a unary   $\unaryOp{}$  or a binary operator $\binaryOp{}{}$.
\begin{align*}
 \sexpr := n \mid s \mid \ite{\sexpr}{\sexpr'}{\sexpr''} \mid \unaryOp{\sexpr} \mid \binaryOp{\sexpr}{\sexpr'}
\end{align*}

Most operational rules of $\nsarrowa{}$ are straightforward extensions of the concrete semantics, and we omit them. 
The primary differences w.r.t. the concrete \muasm{} semantics arise when handling memory operations and control-flow statements with symbolic values, which are formalized in the operational semantics rules below.

\begin{center}\small
\mytoprule{\sigmaa \nsarrowa{\tau} {\sigmaa}'}

\typerule{Load-Symb}
{
\select{p}{sa(\pc)} = \pload{x}{e} & x \neq \pc \\
\sexpr = \exprEval{e}{sa} & \sexpr' = \symRead{sm}{\sexpr}
}
{
\tup{p, sm, sa} \nsarrowa{\loadObs{se}} \tup{p, sm, sa[\pc \mapsto sa(\pc)+1, x \mapsto \sexpr']}
}{load-sym}
\typerule{Store-Symb}
{
\select{p}{sa(\pc)} = \pstore{x}{e} & \sexpr = \exprEval{e}{sa} \\ 
sm' = \symWrite{sm}{se}{sa(x)}
}
{
\tup{p, sm, sa} \nsarrowa{\storeObs{\sexpr}} \tup{p, sm', sa[\pc \mapsto a(\pc)+1]}
}{store-sym}

\typerule{Beqz-Conc-Sat}
{
\select{p}{sa(\pc)} = \pjz{x}{\lbl} &  sa(x) = 0
}
{
\tup{p, sm, sa} \nsarrowa{\pcObs{\lbl} \cdot \symPcObs{\top}} \tup{p, sm, sa[\pc \mapsto \lbl]}
}{beqz-concval-sat-sym}
\typerule{Beqz-Unsat}
{
\select{p}{sa(\pc)} = \pjz{x}{\lbl}  & a(x) \neq 0
}
{
\tup{p, sm, sa} \nsarrowa{\pcObs{a(\pc)+1}} \tup{p, sm, sa[\pc \mapsto a(\pc) +1]}
}{beqz-unsat-sym}

\typerule{Beqz-Symb-Sat}
{
\select{p}{sa(\pc)} = \pjz{x}{\lbl} \\
sa(x) \not\in \Val 
}
{
\tup{p, sm, sa} \nsarrowa{\pcObs{\lbl} \cdot  \symPcObs{sa(x) = 0}} \tup{p, sm, sa[\pc \mapsto \lbl]}
}{beqz-symval-sat-sym}
\typerule{Beqz-Symb-Unsat}
{
\select{p}{sa(\pc)} = \pjz{x}{\lbl} \\
sa(x) \not\in \Val 
}
{
\tup{p, sm, sa} \nsarrowa{\pcObs{sa(\pc)+1} \cdot \symPcObs{sa(x) \neq 0}} \tup{p, sm, sa[\pc \mapsto sa(\pc) +1]}
}
{beqz-symval-unsat-sym}

\end{center}

When symbolically executing a program, we may produce observations whose values are symbolic. To account for this, we introduce symbolic observations of the form $\loadObs{se}$ and $\storeObs{se}$ for symbolic load and store instructions.

Furthermore, because of the symbolic execution, we need to keep track if certain paths in our program are feasible. We encode this information using $\symPcObs{\sexpr}$ observations, which record the choices made at branches and jumps.
We can see these observations in action in \Cref{tr:beqz-symval-unsat-sym} and \Cref{tr:beqz-symval-sat-sym}. Because $x$ cannot be evaluated to a value, the symbolic semantics decides if the branch is taken or not and records this decision in the observation $\symPcObs{sa(x) = 0}$
Without these observations, we could later concretize x to a value unequal to $0$ even though the branch was not taken, making the path infeasible in a concrete program execution.
To explore all paths of the program, our symbolic tool will negate these symbolic branching observations to explore the other branches as well.
The path condition $\pathCond{\tauStacka}\!\!=\!\!\bigwedge_{\symPcObs{\sexpr} \in \tau} \sexpr$ of trace~$\tau$ is the conjunction of all symbolic branching conditions in $\tau$.\looseness=-1

The value of a symbolic expression $\sexpr$ depends on a \textit{valuation function} $\conc:\SymVal \to \Val$ mapping symbolic values to concrete ones. The evaluation $\conc(se)$ is also standard.
We write $\conc \vDash \sexpr$ to denote that symbolic expression $\sexpr$ is \textit{satisfiable} for a valuation $\conc$, i.e.,  $\conc(\sexpr) \neq 0$.
Every valuation that satisfies a symbolic run's path condition maps the symbolic run to a concrete one.
Finally, we write $\vDash \sexpr$ to denote that there exists a valuation $\conc$ such that $\conc \vDash \sexpr$.

The \emph{symbolic non-speculative behaviour} $\behNsa{p}$ of a program $p$ is the set of all traces generated by all possible initial states for program $p$.

\Cref{thm:ns-symbolic} connects the symbolic and concrete non-speculative semantics via the concretization function $\conc$:
\begin{theorem}[NS: Symbolic Consistency]\label{thm:ns-symbolic}
     $\behNs{p} = \conc(\behNsa{p})$.
\end{theorem}

This allows us to use the symbolic non-speculative semantics in our tool \tool with confidence that it behaves equivalently to the concrete non-speculative semantics.

\subsection{Symbolic Always-Mispredict Semantics}\label{sec:sym-general-am}

To enable automated verification, \tool implements a symbolic always-mispredict semantics. Structurally, this semantics perfectly mirrors the concrete always-mispredict semantics introduced earlier, with the sole distinction that it operates over symbolic expressions and relies on the symbolic non-speculative step relation $\nsarrowa{}$ rather than its concrete counterpart.
\onlyShortVersion{Thus, we omit details related to this semantics and refer to the technical report \cite{techReport}}.

We denote by $\conc({(\tup{p, \sigmaa})}\statesem^{\spw}_{x}{\tauStacka})$ the set $\{{(\tup{p, \sigmaa})}\statesem^{\spw}_{x}{(\conc(\tauStacka))} \mid \mu \vDash \pathCond{\tauStacka} \}$ and lift it to $\behxa{p}$.
Here ${(\tup{p, \sigmaa})}\statesem^{\spw}_{x}{\tauStacka}$ denotes the run of the program $p$ with initial configuration $\sigmaa$ generating $\tauStacka$ for the symbolic always-mispredict semantics $x$.

\paragraph{Symbolic Consisteny}

Fundamentally, a valid symbolic analysis must faithfully capture all possible concrete executions of a program.
The symbolic always-mispredict semantic should explore the exact same set of execution paths and generate the same observations as its concrete counterpart, merely operating over symbolic values and constraints. We formalize this correctness requirement in \Cref{thm:am-consistent-symb}, stating that the concrete always-mispredict behaviour can be exactly recovered from the symbolic behaviour by applying the concretization function.

\begin{requirement}[Symbolic Consistency AM]\label{thm:am-consistent-symb}
    $\behx{p} = \conc(\behxa{p})$
\end{requirement}

Because we have proven that all our speculative semantics and their combinations satisfy the requirements of our framework (i.e., they are well-formed speculative semantics), they inherently satisfy \Cref{thm:am-consistent-symb}.
This allows us to use all of the symbolic semantics in our tool \tool with confidence that they behave equivalently to the concrete semantics.

\begin{example}[Symbolic Trace for \Cref{example:v1-vanilla}]\label{example:symbolic-semantics}
		Executing the program from \Cref{example:v1-vanilla} under the symbolic speculative semantics $\semb$ with speculative window~$2$ yields the following two symbolic traces:
    \begin{align*}
        \tau_1 := \symPcObs{{\mathtt{y}} < {\mathtt{size}}} \concat \startObsB{0} \concat \pcObs{2} \concat \pcObs{10} \concat \rollbackObsB{0} \concat \pcObs{3} \concat \loadObs{{\mathtt{A}} + {\mathtt{y}}}  \concat \loadObs{{\mathtt{B}} + \symRead{sm}{ ({\mathtt{A}} + {\mathtt{y}}) }*512} \\
        \tau_2 := \symPcObs{{\mathtt{y}} \geq {\mathtt{size}}} \concat  \startObsB{0} \concat \pcObs{3} \concat \loadObs{{\mathtt{A}} + {\mathtt{y}}}  \concat \loadObs{{\mathtt{B}} + \symRead{sm}{ ({\mathtt{A}} + {\mathtt{y}}) }*512} \concat \rollbackObsB{0} \concat \pcObs{2} \concat \pcObs{10}
    \end{align*}

\end{example}

\subsection{The Spectector Algorithm}\label{sec:spectector-alg}

\begin{algorithm}
    \caption{\tool{} program analysis}
    \label{algorithm:tool}
    \begin{algorithmic}[1]
	\Require A program $p$, a security policy $\policy$, a speculative window  $w \in \Nat$.
	\Ensure \textsc{Secure} if $p$ satisfies speculative non-interference with respect to the policy $\policy$ and speculative semantics $\semx$; \textsc{Insecure} otherwise
		\Statex{}
	\Procedure{\tool}{$p,\policy,w$}
		\For{each symbolic run $\tauStack \in \behxa{p}$}
			\If{$\memcheck(\tauStack,P)  \vee \pccheck(\tauStack,P)$}
				\State{\Return{\textsc{Insecure}}}
			\EndIf
		\EndFor
		\State{\Return{\textsc{Secure}}}
        \EndProcedure
    \Statex{}
    \Procedure{$\memcheck$}{$\tauStack, \policy$}
    		\State{$\psi \gets \pathCond{\tauStack}_{1 \wedge 2} \wedge \policyEqv{\policy} \wedge $}
    		\Statex{$\qquad \qquad \cstrs{\nspecProject{\tauStack}} \wedge \neg \cstrs{\specProject{\tauStack}}$}
			\State{\Return{$\textsc{Satisfiable}(\psi)$}}

    \EndProcedure
    \Statex{}
    \Procedure{$\pccheck$}{$\tauStack, \policy$}
    		\For{each prefix $\nu \concat \symPcObs{\mathit{se}}$ of $\specProject{\tauStack}$}
				\State{$\psi \gets \pathCond{\nspecProject{\tauStack}\concat \nu}_{1 \wedge 2} \wedge \policyEqv{\policy} \wedge$}
				\Statex{$\qquad \qquad \quad \cstrs{\nspecProject{\tauStack}} \wedge \neg \sameSymPc{se}$}
				\If{$\textsc{Satisfiable}(\psi)$}
					\State{\Return{$\top$}}
				\EndIf
			\EndFor
    		\State{\Return{$\bot$}}
    \EndProcedure
    \end{algorithmic}
\end{algorithm}
The \tool{} program analysis approach is presented in Algorithm~\ref{algorithm:tool}.
It relies on two procedures: \memcheck{} and \pccheck{}, to detect leaks resulting from memory  and  control-flow instructions, respectively.
We start by discussing the \tool{} algorithm and next explain the \memcheck{} and \pccheck{} procedures.

\tool{} takes as input a program $p$, a policy $\policy$ specifying the non-sensitive information, and a speculative window $w$.
The algorithm iterates over all symbolic runs produced by the symbolic always-mispredict speculative semantics (lines 2-4).
For each trace $\tauStack \in \behxa{p}$, the algorithm checks whether $\tauStack$ speculatively leaks information through memory accesses or control-flow instructions.
If this is the case, then \tool{} has found a witness of a speculative leak and it reports $p$ as \textsc{Insecure}.
If none of the traces contains speculative leaks, the algorithm terminates returning \textsc{Secure} (line 5).

\para{Detecting leaks caused by memory accesses}
The procedure \memcheck{} takes as input a  trace~$\tauStack$ and a policy $\policy$, and it determines whether $\tauStack$ leaks information through symbolic $\loadObs{}$ and $\storeObs{}$ observations. 
The check is expressed as a satisfiability check of a  constraint $\psi$. The construction of~$\psi$ is inspired by self-composition~\cite{R_self_comp}, which reduces reasoning about {\em pairs} of program runs to reasoning about single runs by replacing each symbolic variable $x$ with two copies $x_1$ and~$x_2$. We lift the subscript notation to symbolic expressions.

The constraint $\psi$ is the conjunction of four formulas:
\begin{asparaitem}
\item $\pathCond{\tau}_{1 \wedge 2}$ stands for $\pathCond{\tauStack}_{1} \wedge \pathCond{\tauStack}_{2}$, which ensures that both runs follow the path associated with $\tauStack$.\looseness=-1
\item $\policyEqv{\policy}$ introduces constraints $x_1 = x_2$ for each register $x\in \policy$ and $\symRead{sm_1}{n} = \symRead{sm_2}{n}$ for each memory location $n \in P$, which ensure that both runs agree on all non-sensitive inputs.
\item $\cstrs{\nspecProject{\tauStack}}$ introduces a constraint $\sexpr_1 = \sexpr_2$ for each $\loadObs{\sexpr}$ or $\storeObs{\sexpr}$ in $\nspecProject{\tau}$, which ensures that the non-speculative observations associated with memory accesses are the same in both runs. 
\item $\neg \cstrs{\specProject{\tauStack}}$  ensures that at least one speculative observation associated with memory accesses differs among the two runs. 
\end{asparaitem}

If $\psi$ is satisfiable, there are two $\policy$-indistinguishable configurations that produce the same non-speculative traces (since $\pathCond{\tauStack}_{1 \wedge 2} \wedge \policyEqv{\policy} \wedge \cstrs{\nspecProject{\tauStack}}$ is satisfied) and whose speculative traces differ in a memory access observation (since $\neg \cstrs{\specProject{\tauStack}}$ is satisfied), i.e. a violation of SNI.

\para{Detecting leaks caused by control-flow instructions}
To detect leaks caused by control-flow instructions, \pccheck{} checks whether there are two traces in $\tau$'s concretization that agree on the outcomes of all non-speculative branch and jump instructions, while differing in the outcome of at least one speculatively executed branch or jump instruction.

In addition to $\pathCond{\tauStack}$, $\cstrs{\tauStack}$, and $\policyEqv{\policy}$, the procedure relies on the  function $\sameSymPc{\sexpr}$ that introduces the constraint $\sexpr_1 \leftrightarrow \sexpr_2$ ensuring that  $\sexpr$ is satisfied in one concretization iff it is satisfied in the other.

\pccheck{} checks, for each prefix $\nu \concat \symPcObs{\sexpr}$ in $\tauStack$'s speculative projection $\specProject{\tauStack}$,  the satisfiability of the conjunction of $\pathCond{\nspecProject{\tauStack} \concat \nu}_{1 \wedge 2}$, $\policyEqv{\policy}$, $\cstrs{\nspecProject{\tauStack}}$, and $\neg \sameSymPc{\sexpr}$.
Whenever the formula is satisfiable, there are two $\policy$-indistinguishable configurations that produce the same non-speculative traces, but whose speculative traces differ on program counter observations, i.e. a violation of SNI.

\begin{example}[Example Application of \tool]
	Consider  the  trace from \Cref{example:symbolic-semantics}:
 \begin{align*}
     \tauStack := \symPcObs{{\mathtt{y}} \geq {\mathtt{size}}} \concat  \startObsB{0} \concat \pcObs{3} \concat \loadObs{{\mathtt{A}} + {\mathtt{y}}}  \concat \loadObs{{\mathtt{B}} + \symRead{sm}{  ({\mathtt{A}} + {\mathtt{y}})  }*512 } \concat \rollbackObsB{0} \concat \pcObs{2} \concat \pcObs{10}
 \end{align*} 
 \memcheck{} detects a leak caused by the observation $\loadObs{{\mathtt{B}} + \symRead{sm}{  ({\mathtt{A}} + {\mathtt{y}})  }*512 }$.
	Specifically, it detects that there are distinct symbolic valuations that agree on the non-speculative observations but disagree on the value of $\loadObs{{\mathtt{B}} + \symRead{sm}{  ({\mathtt{A}} + {\mathtt{y}})  }*512 }$. That is, the observation depends on sensitive information that is not disclosed by $\tauStack$'s non-speculative projection.	
\end{example}

\para{Soundness and Completeness}

\Cref{theorem:spectector-soundness-and-completeness-paper} states that \tool{} deems secure only speculatively non-interferent programs, and all detected leaks are actual violations of SNI.\looseness=-1

\begin{theorem}[Correctness \tool]\label{theorem:spectector-soundness-and-completeness-paper}
If $\tool{}(p,\policy,w)$ terminates, then  $\tool{}(p,\policy,w) = \textsc{Secure}$ iff  the program $p$ satisfies speculative non-interference w.r.t. the  policy~$\policy$ and all prediction oracles~$\orac$ with speculative window at most~$w$.
\end{theorem}

The theorem follows from Oracle Overapproximation and Symbolic Consistency of the speculative semantics. Importantly, both conditions are part of our well-formed speculative semantics definition (\Cref{def:sss-paper}). Since $\semb$, $\semj$, $\sems$, $\semr$ and $\semsls$ are well-formed speculative semantics $\ssssem{\semx}$ and our combinations are well-formed (\Cref{def:wellformed-paper}), we have that all the combinations are \sss as well. Thus, we get \Cref{theorem:spectector-soundness-and-completeness-paper} for free for all the combinations.

\subsection{\tool Implementation}\label{sec:tool}
We implement our approach in our tool \tool{}, which is available at \cite{tool-og}.
The tool, which is implemented on top of the \ciao{} logic
programming system~\cite{ciao}, consists of three components: a front end that translates x86 assembly programs into \lang{}, a core engine implementing Algorithm~\ref{algorithm:tool}, and a back end  handling SMT queries.

\para{x86 Front End}
The front end translates AT\&T/GAS and Intel-style assembly files into \lang{}. 
It currently supports over 120 instructions: data movement instructions
($\kywd{mov}$, etc.), logical, arithmetic, and comparison instructions
($\kywd{xor}$, $\kywd{add}$, $\kywd{cmp}$, etc.),
branching and jumping instructions
($\kywd{jae}$, $\kywd{jmp}$, etc.), conditional moves
($\kywd{cmovae}$, etc.), stack manipulation ($\kywd{push}$,
$\kywd{pop}$, etc.), and function calls\footnote{We model the so-called ``near calls'', where the callee is in the same code segment as the caller.}
($\kywd{call}$, $\kywd{ret}$).

It currently does not support privileged x86 instructions, e.g., for handling model specific registers and virtual memory.
Further, it does not support sub-registers (like $\mathtt{eax}$,
$\mathtt{ah}$, and $\mathtt{al}$) and unaligned memory accesses, i.e.,
we assume that only 64-bit words are read/written at each address
without overlaps.
Finally, the translation currently maps symbolic address names
to \lang{} instruction addresses, limiting arithmetic on code addresses.

\para{Core Engine}
The core engine implements Algorithm~\ref{algorithm:tool}.
It relies on a concolic approach to implement symbolic execution that performs
a depth-first exploration of the symbolic runs. 
Starting from a concrete initial configuration, the engine executes the program under the symbolic always-mispredict speculative semantics while keeping track of the symbolic configuration and path condition.
It discovers new runs by iteratively negating the last (not previously negated) conjunct in the path condition until it finds a new initial configuration, which is then used to re-execute the program concolically.
In our current implementation, indirect jumps are \emph{not} included in the path conditions, and thus new symbolic runs and corresponding inputs are only discovered based on negated branch conditions.\footnote{We plan to remove this limitation in a future release of our tool.}
This process is interleaved with the \memcheck{} and \pccheck{} checks and iterates until a leak is found or all paths have been explored.

\para{SMT Back End}
The Z3 SMT solver~\cite{z3} acts as a back end for checking satisfiability and finding models of symbolic expressions using the \textsc{Bitvector} and \textsc{Array} theories, which are used to model registers and memory. 
The implementation currently does not rely on incremental solving, since it was less efficient than one-shot solving for the selected theories.\looseness=-1

\para{Implementation of the Speculative Semantics and the Combinations in \tool}
We implemented all our semantics (the symbolic versions of $\semb$, $\sems$, $\semj$, $\semr$ and $\semsls$ plus all \nrComb{} compositions from  \Cref{sec:comb-in}) in \tool. 
The implementation of compositions closely follows the structure of our framework.
As in \Cref{sec:comb-in}, selecting one of the composed semantics in \tool sets the metaparameter Z, which is used to delegate back to the correct individual semantics.\looseness=-1

%% file: src/Evaluations/evaluation.tex
\section{Evaluation}\label{sec:eval}

This section reports on two case studies in which we apply \tool to analyze the security of programs. 
The goals of the first case study are (1) to determine whether speculative non-interference realistically captures speculative leaks and (2) to assess \tool’s precision
Therefore, we analyze various examples targeting the different Spectre versions we want to capture with our speculative semantics (\Cref{sec:case-study-semantics}).

The goal of our second case study is to investigate if the combined speculative semantics allow us to capture stronger attacks that are not captured by individual semantics.
Thus, we create code snippets that are only exploitable using a combination of different Spectre attacks and check if \tool using the combined semantics can detect leaks  (\Cref{sec:case-study-comb}).\looseness=-1

All code snippets as well as all scripts to reproduce our results are available in our public repository at \url{https://spectector.github.io/}.

\input{src/Evaluations/case-study-semantics}

\input{src/Evaluations/case-study-combinations}

%% file: src/Evaluations/case-study-semantics.tex
\subsection{Case Study: Detecting Leaks w.r.t. Individual Speculative Semantics}\label{sec:case-study-semantics}

\newcommand{\nrExamplesSLS}{2}
\newcommand{\nrExamplesVJ}{2}
\newcommand{\nrExamplesSemantics}{41}
\renewcommand{\P}{\fcirc}
\newcommand{\Q}{?}
\newcommand{\clm}{0.2cm}
\newcommand{\unp}{\textsc{Unp}}
\newcommand{\fen}{\textsc{Fen}}
\newcommand{\slh}{\textsc{Slh}}
\newcommand{\opt}{\texttt{-O2}}
\newcommand{\unopt}{\texttt{-O0}}
Using \tool{}, we analyze a corpus of \nrExamplesSemantics{} microbenchmarks containing speculative leaks generated by different speculation mechanisms in isolation (\Cref{sec:inst-spec-semantics-single}).
With these experiments, we aim to show that speculative non-interference and our individual speculative semantics can correctly identify speculative leaks associated with speculation over branches, indirect jumps, store-bypasses and return instructions.

\newcommand{\nrExamplesCompiled}{291}
\subsubsection{Benchmarks}
We start from \nrExamplesSemantics{} snippets of code containing leaks resulting from speculation over $\jzC{}$, $\storeC{}$, $\loadC{}$, $\jC{}$, and $\retC{}$ instructions (and their combinations).
We compile these snippets using various compilers, compiler options, and mitigations, thereby obtaining a corpus of \nrExamplesCompiled{} x64 assembly programs, which we analyse with \tool{}.
Next, we describe next in detail the composition of our benchmark corpus:
\begin{asparaitem}
    \item  \textbf{Spectre-PHT:} 15 snippets are variants of the Spectre-PHT vulnerability by Kocher~\cite{S_koch_mit}.
    For each of the 15 snippets, we analyse the assembly programs obtained using different compilers and compiler options.
    For compilers, we rely on three state-of-the-art compilers: Microsoft \vcc{} versions v19.15.26732.1 and v19.20.27317.96, Intel \icc{} v19.0.0.117, and \clang{} v7.0.0 and compile each snippet using two different \textit{optimization levels} (\unopt{} and \opt{}) and three \textit{mitigation levels}: 
    \begin{inparaenum}[(a)]
        \item \unp{}: we compile  without any \spectre{} mitigations. 
        \item \fen{}: we compile with automated injection of speculation
        barriers.\footnote{Fences are supported by \clang{} with
        the flag \texttt{-x86-speculative\-load-hardening-lfence}, by \icc{}
        with \texttt{-mconditional\-branch=all-fix}, and by \vcc{} with
        \texttt{/Qspectre}.\looseness=-1}
        \item \slh{}: we compile using speculative load hardening.\footnote{Speculative load hardening is supported by \clang{} with the flag \texttt{-x86-speculative-load-hardening}.}
        \end{inparaenum}
    Compiling each of the 15 examples from \cite{S_koch_mit} with each of the 3 compilers, each of the 2 optimization levels, and each of the 2-3 mitigation levels, yields a corpus of 240 x64 assembly programs.
    For each program, we specify a security policy that flags as ``low'' all registers and memory locations that can either be
    controlled by the adversary or can be assumed to be public. This includes variables \inlineASMcode{y} and \inlineASMcode{size}, and the base addresses of the arrays $\inlineASMcode{A}$ and $\inlineASMcode{B}$ as well as the stack pointer.

    \item  \textbf{Spectre-BTB:} \nrExamplesVJ{} snippets are variants of the \specj vulnerability.
    They exploit speculation over indirect jump instructions and were created by ourselves.
    For each snippet, we also analyze manually patched versions obtained by inserting \textsc{lfence}s after every \textsc{endbr} instruction and automatically patched versions using the retpoline \cite{retpoline} countermeasure.\footnote{Retpoline is supported by \gcc{} with the flag \texttt{-mindirect-branch=thunk}.}
    This results in a corpus of 6 x64 assembly programs.
    
    To make the analysis tractable, we use the \texttt{endbr} instruction for CFI (see \Cref{sec:v2-semantics}). Note that these \texttt{endbr} instructions were automatically added by the compiler.

    \item \textbf{Spectre-STL:} 13 snippets are variants of the Spectre-STL vulnerability. 
    They exploit speculation over memory disambiguation, and they have been used as benchmarks in prior work~\cite{ST_binsec,cats}.
    For each snippet, we also analyze a patched version where a manually inserted \textsc{lfence} instruction stops speculation over store-bypasses and prevents the leak.
    This results in a corpus of 26 x64 assembly programs.

    \item  \textbf{Spectre-RSB:} 5 snippets are variants of the Spectre-RSB vulnerability. 
    They exploit speculation over return instructions, and they are obtained from the \texttt{safeside}~\cite{safeside}  and \texttt{transientfail}~\cite{transientfail} projects\footnote{Out of the three Spectre-RSB examples from \texttt{safeside}~\cite{safeside}, we analyze the only one that works against an acyclic RSB like the one supported by $\semr$. Programs \textit{ca\_ip}, \textit{ca\_oop}, and \textit{sa\_ip} from \texttt{transientfail}~\cite{transientfail} rely on concurrent execution.
Since \tool{} does not support concurrency, we hardcode the worst-case interleaving in terms of speculative leakage in our benchmark.}.
    For each snippet, we also analyze manually patched versions obtained by (1) inserting \textsc{lfence}s after call instructions (i.e., at the instruction address where $\pret$ speculatively returns), and (2) using the modified \texttt{retpoline} defense proposed in \cite[Section 6.1]{ret2spec}. 
    This results in a corpus of 15 x64 assembly programs.\looseness=-1

   \item \textbf{Spectre-SLS:}  \nrExamplesSLS{} snippets are variants of the Spectre-SLS vulnerability.
    They exploit speculation over return instructions and were created by ourselves.

    For each snippets, we also analyze a manually patched version obtained by inserting \textsc{lfence}s after every return instruction, resulting in a corpus of 4 x64 assembly programs.%
    \footnote{
        We note that compilers like \clang{} have the option \texttt{mharden-sls-all} to automatically protect the code. 
        However, this countermeasure inserts an \texttt{int3} instruction after every return on x86. Since interrupts and exceptions are not handled by \tool, they cannot be translated to \muasm{}. An \texttt{lfence} behaves similarly in this case.
    }
\end{asparaitem}

\subsubsection{Experimental Setup}
The benchmarks for \textbf{Spectre-PHT} are compiled as described in the section above.
The benchmarks for \textbf{Spectre-STL}, \textbf{Spectre-RSB}, and \textbf{Spectre-SLS} are implemented in C and compiled with \gcc{} 11.1.0 and we manually inserted \texttt{lfence}/\texttt{modified retpoline} countermeasures in the patched versions.
The benchmarks for \textbf{Spectre-BTI} are also implemented in C and \texttt{lfence}s were inserted manually while retpoline was added automatically by the compiler into the patched programs.

The benchmarks for \textbf{Spectre-PHT} were run on a Linux machine (kernel 4.9.0-8-
amd64) with Debian 9.0, a Xeon Gold 6154 CPU, and 64 GB of RAM.
All other experiments were run on a laptop with a Dual Core Intel Core i5-7200U CPU and 8GB of RAM.

\begin{figure*}[h]
\centering
\begin{tabular}
{c  c c c c c c  c c c c  c c c c c c  }
\toprule
\multirow{3}{*}{Ex.}	&	\multicolumn{6}{c}{\textsc{Vcc}}  &  \multicolumn{4}{c }{\icc} & \multicolumn{6}{c }{\clang{}}	\\ 
 \cmidrule(lr){2-7}
 \cmidrule(lr){8-11}
 \cmidrule(lr){12-17}
	& \multicolumn{2}{c}{\unp}	& \multicolumn{2}{c }{\fen{} 19.15}		& \multicolumn{2}{c }{\fen{} 19.20}	&  \multicolumn{2}{c }{\unp}	& \multicolumn{2}{c }{\fen} &  \multicolumn{2}{c }{\unp}	& \multicolumn{2}{c }{\fen} & \multicolumn{2}{c }{\slh} \\
 \cmidrule(lr){2-3}   \cmidrule(lr){4-5}  \cmidrule(lr){6-7}
 \cmidrule(lr){8-9}  \cmidrule(lr){10-11}
 \cmidrule(lr){12-13}  \cmidrule(lr){14-15}
 \cmidrule(lr){16-17} 
	&	\unopt 	&	 	\opt & \unopt 	&	 	\opt	 	&	\unopt		& 		\opt		&	\unopt		& 	\opt			& 	\unopt 	&	 	\opt	 	&	\unopt 	&	 	\opt	  & 	\unopt 	&	 	\opt	 	& 	\unopt 	&	 	\opt	\\ 
 \midrule
01	&	$\N$	&	$\N$	&	$\P$	&	$\P$	& 	$\P$	&	$\P$ &	$\N$	&	$\N$	&	$\P$	&	$\P$	&	$\N$	&	$\N$	&	$\P$	&	$\P$	&	$\P$	&	$\P$ \\ 
02	&	$\N$	&	$\N$	&	$\P$	&	$\P$	& 	$\P$	&	$\P$	 &	$\N$	&	$\N$	&	$\P$	&	$\P$	&	$\N$	&	$\N$	&	$\P$	&	$\P$	&	$\P$	&	$\P$ \\ 
03	&	$\N$ 	&	$\N$	&	$\P$	&	$\N$	& 	$\P$	&	$\P$	 &	$\N$	&	$\N$	&	$\P$	&	$\P$	&	$\N$	&	$\N$	&	$\P$	&	$\P$	&	$\P$	&	$\P$ \\ 
04	&	$\N$	&	$\N$	&	$\N$	&	$\N$	& 	$\P$	&	$\P$ &	$\N$	&	$\N$	&	$\P$	&	$\P$	&	$\N$	&	$\N$	&	$\P$	&	$\P$	&	$\P$	&	$\P$ \\ 
05	&	$\N$	&	$\N$	&	$\P$	&	$\N$	& 	$\P$	&	$\N$	 &	$\N$	&	$\N$	&	$\P$	&	$\P$	&	$\N$	&	$\N$	&	$\P$	&	$\P$	&	$\P$	&	$\P$ \\ 
06	&	$\N$	&	$\N$	&	$\N$	&	$\N$	& 	$\N$	&	$\N$ &	$\N$	&	$\N$	&	$\P$	&	$\P$	&	$\N$	&	$\N$	&	$\P$	&	$\P$	&	$\P$	&	$\P$ \\ 
07	&	$\N$	&	$\N$	&	$\N$	&	$\N$	& 	$\N$	&	$\N$ &	$\N$	&	$\N$	&	$\P$	&	$\P$	&	$\N$	&	$\N$	&	$\P$	&	$\P$	&	$\P$	&	$\P$ \\ 
08	&	$\N$	&	$\P$	&	$\N$	&	$\P$	& 	$\N$	&	$\P$ &	$\N$	&	$\P$	&	$\P$	&	$\P$	&	$\N$	&	$\P$	&	$\P$	&	$\P$	&	$\P$	&	$\P$ \\ 
09	&	$\N$	&	$\N$	&	$\N$	&	$\N$	& 	$\N$	&	$\N$ &	$\N$	&	$\N$	&	$\P$	&	$\P$	&	$\N$	&	$\N$	&	$\P$	&	$\P$	&	$\P$	&	$\P$ \\ 
10	&	$\N$	&	$\N$	&	$\N$	&	$\N$	&	$\N$	&	$\N$	 &	$\N$	&	$\N$	&	$\P$	&	$\P$	&	$\N$	&	$\N$	&	$\P$	&	$\P$	&	$\P$	&	$\N$ \\ 
11	&	$\N$	&	$\N$	&	$\N$	&	$\N$	& 	$\N$	&	$\N$ &	$\N$	&	$\N$	&	$\P$	&	$\P$	&	$\N$	&	$\N$	&	$\P$	&	$\P$	&	$\P$	&	$\P$ \\ 
12	&	$\N$	&	$\N$	&	$\N$	&	$\N$	& 	$\P$	&	$\P$	 &	$\N$	&	$\N$	&	$\P$	&	$\P$	&	$\N$	&	$\N$	&	$\P$	&	$\P$	&	$\P$	&	$\P$ \\ 
13	&	$\N$	&	$\N$	&	$\N$	&	$\N$	& 	$\N$	&	$\N$	 &	$\N$	&	$\N$	&	$\P$	&	$\P$	&	$\N$	&	$\N$	&	$\P$	&	$\P$	&	$\P$	&	$\P$ \\ 
14	&	$\N$	&	$\N$	&	$\N$	&	$\N$	& 	$\P$	&	$\P$	 &	$\N$	&	$\N$	&	$\P$	&	$\P$	&	$\N$	&	$\N$	&	$\P$	&	$\P$	&	$\P$	&	$\P$ \\ 
15	&	$\N$	&	$\N$	&	$\N$	&	$\N$	& 	$\N$	&	$\N$	 &	$\N$	&	$\N$	&	$\P$	&	$\P$	&	$\N$	&	$\N$	&	$\P$	&	$\P$	&	$\N$	&	$\P$ \\ 
\bottomrule
\end{tabular}
\caption{Analysis of Kocher's examples~\cite{S_koch_mit},
  compiled with different compilers and options.
For each of the 15 examples, we analyzed the unpatched version (denoted by \unp), the version patched with speculation barriers (denoted by \fen), and the version patched using speculative load hardening (denoted by \slh).
Programs have been compiled without optimizations (\unopt{}) or with compiler optimizations (\opt) using the compilers \vcc{} (two versions), \icc{}, and \clang{}. 
$\N$ denotes that \tool{} detects a speculative leak, whereas 
$\P$ indicates that \tool{} proves the program secure.
}\label{figure:case-studies:results}
\end{figure*}

\subsubsection{Results for Spectre-PHT}
\Cref{figure:case-studies:results} depicts  the results of applying \tool{} to the 240 \textbf{Spectre-PHT} examples. We highlight the following findings:

\begin{asparaitem}
\item \tool{} detects the speculative leaks in almost all unprotected
  programs, for all compilers (see the \unp{} columns). The exception
  is Example \#8, which uses a conditional expression instead of the if statement of
  \Cref{lst:v1-vanilla}:
\begin{lstlisting}[style=Cstyle]
temp &= B[A[y<size?(y+1):0]*512];
\end{lstlisting}
At optimization level \unopt{}, this is translated to a (vulnerable) branch instruction by all compilers, and at level \opt{} to a (safe) conditional move, thus closing the leak. See Appendix~\ref{secs:example8} for the 
corresponding \clang{} assembly.
\item The \clang{} and Intel \icc{} compilers defensively insert fences after
  each branch instruction, and \tool{} can prove security for
  all cases (see the \fen{} columns for \clang{} and \icc{}).  In
  Example \#8 with options \opt{} and \fen{}, \icc{} inserts an \textbf{lfence} instruction, even
  though the baseline relies on a conditional move, see line 10 below. This \textbf{lfence} is unnecessary according to our semantics, but may close leaks on processors that speculate over conditional moves. 
  \begin{lstlisting}[style=ASMstyle]
    mov 	y, %
    lea 	1(%
    mov 	size, %
    xor 	%
    cmp 	%
    cmovb 	%
    mov 	temp, %
    mov 	A(%
    shl 	$9, %
    lfence
    and 	B(%
    mov 	%
\end{lstlisting}

\item For the \vcc{} compiler, \tool{} automatically detects all
  leaks pointed out in~\cite{S_koch_mit} (see the \fen{} 19.15 \opt{} column for \textsc{Vcc}). Our analysis differs
  from Kocher's only on Example \#8, where the compiler v19.15.26732.1 introduces a safe conditional move, as explained above. Moreover, without compiler optimizations (which is not considered in~\cite{S_koch_mit}), \tool{} establishes the security of Examples \#3 and \#5 (see the \fen{} 19.15 \unopt{} column).
The latest \textsc{Vcc} compiler additionally mitigates the leaks in Examples \#4, \#12, and \#14 (see the \fen{} 19.20 column).\looseness=-1

\item \tool{} can prove the security of speculative load hardening in
  Clang (see the \slh{} column for \clang{}), except for Example \#10 with \opt{}
  and Example \#15 with \unopt{}.
\end{asparaitem}

\subsubsection*{Example 10 with Speculative Load Hardening}
Example \#10 %
differs from \Cref{lst:v1-vanilla} in that it leaks sensitive
information into the microarchitectural state by conditionally reading
the content of \inlineCcode{B[0]}, depending on the value of
\inlineCcode{A[y]}.
\begin{lstlisting}[style=CStyle,]
 if (y < size) 
   if (A[y] == k)
     temp &= B[0];
\end{lstlisting}

\tool{} proves the security of the program produced with \clang{} \unopt{} and speculative load hardening.

However, at optimization level \opt{}, \clang{} outputs the following
code that \tool{} reports as insecure. 
\begin{lstlisting}[style = ASMstyle] 
 mov     size, %
 mov     y, %
 mov     $0, %
 cmp     %
 jbe     END
 cmovbe  $-1, %
 or      %
 mov     k, %
 cmp     %
 jne     END
 cmovne  $-1, %
 mov     B, %
 and     %
 jmp     END
\end{lstlisting}
The reason for this is that \clang{} masks only the register
\inlineASMcode{\%rbx} that contains the index of the memory access
\inlineCcode{A[y]}, cf.~lines 6--7. However, it does {\em not} mask
the value that is read from \inlineCcode{A[y]}.  As a result, the
comparison at line 9 speculatively leaks (via the jump target) whether
the content of \inlineCcode{A[}\texttt{0xFF...FF}\inlineCcode{]} is \inlineCcode{k}.
\tool{} detects this subtle leak and flags a violation of speculative
non-interference.

While this example nicely illustrates the scope of \tool{}, it is likely not a problem in practice: 
First, the leak may be mitigated by how data dependencies are handled in modern out-of-order CPUs. Specifically, the conditional move in line 6 relies on the comparison in Line 4. If executing the conditional leak effectively terminates speculation, the reported leak is spurious.  
Second, the leak can be mitigated at the OS-level by ensuring that \texttt{0xFF...FF} is not mapped in the page tables, or that the value of \inlineCcode{A[}\texttt{0xFF...FF}\inlineCcode{]} does not contain any secret. \footnotemark Such contextual information can be expressed with policies (see \Cref{sec:SNI}) to improve the precision of the analysis.
\footnotetext{Personal Communication with C. Marinas}

\begin{figure*}

\begin{subtable}[t]{.45\linewidth}
\vspace{0pt}
\centering
\begin{tabular}{llccc}
\toprule
\multirow{2}{*}{Test case} &  & \multicolumn{2}{c}{$\sems$} \\  \cline{3-4}
    &  &      None & Fence \\
\midrule
case01 &(\minusR)  & $\N$ & $\P$ \\
case02 &(\minusR) & $\N$ & $\P$ \\
case03 & (\plusG) & $\P$ & $\P$ \\
case04 &  (\minusR)& $\N$ & $\P$ \\
case05 &(\minusR) & $\N$ & $\P$ \\
case06 &(\minusR) & $\N$ & $\P$\\
case07 &(\minusR) & $\N$ & $\P$ \\
case08 &(\minusR) & $\N$ & $\P$ \\
case09&     (\plusG) & $\P$ & $\P$ \\
case10      &(\minusR) & $\N$ & $\P$ \\
case11      &(\minusR) & $\N$ & $\P$ \\
case12      &(\plusG) & $\P$ & $\P$ \\
case13 &   (\minusR) & $\N$ & $\P$ \\
\midrule

\end{tabular}
\vspace{-5pt}
\subcaption{Results for the Spectre-STL programs under the $\sems$ semantics against unpatched programs (column ``None'') and programs patched with \texttt{lfence} (column ``Fence'').}
\label{t:table-v4}
\end{subtable}
\hspace{2em}
\begin{subtable}[t]{.45\linewidth}
\centering
\vspace{0pt}
\begin{tabular}{llcccc}
\toprule
\multirow{2}{*}{Test case}  &  &  \multicolumn{3}{c}{$\semr$} \\ \cline{3-5}
&  & None & Fence & Mod. Retpoline \\
\midrule
$ret2spec\_c\_d$ &(\minusR)  & $\N$ & $\P$ & $\P$ \\
$ca\_ip$& (\minusR)  & $\N$ & $\P$ & $\P$\\
$ca\_oop$& (\minusR)  & $\N$ & $\P$ & $\P$\\
$sa\_ip$ &(\minusR)  & $\N$ & $\P$ & $\P$\\
$sa\_oop$ & (\minusR)  & $\N$ & $\P$ & $\P$\\
\bottomrule
\end{tabular}
\vspace{-5pt}
\subcaption{Results for the Spectre-RSB programs under the $\semr$ semantics against unpatched programs (column ``None''), programs patched with \texttt{lfence} (column ``Fence''), and programs patched with  the modified \texttt{retpoline} defense proposed in \cite[\S6.1]{ret2spec} (column ``Mod. Retpoline'').}
\label{t:table-v5}
\end{subtable}

\begin{subtable}[t]{.47\linewidth}
\centering
\begin{tabular}{llcccc}
\toprule
\multirow{2}{*}{Test case}  &  &  \multicolumn{3}{c}{$\semj$} \\ \cline{3-5}
&  & None & Fence & Retpoline \\
\midrule
caseJ01 &(\minusR)  & $\N$ & $\P$ & $\P$ \\ %
caseJ02& (\minusR)  & $\N$ & $\P$ & $\P$\\
\bottomrule
\end{tabular}
\subcaption{Results for the Spectre-BTB programs under the $\semj$ semantics against unpatched programs (column ``None''), programs patched with \texttt{lfence} (column ``Fence''), and programs patched with the \texttt{retpoline} defense proposed in \cite{retpoline} (column ``Retpoline'').}
\label{t:table-v2}
\end{subtable}
\hspace{1em}
\begin{subtable}[t]{.47\linewidth}
\centering
\begin{tabular}{llccc}
\toprule
\multirow{2}{*}{Test case}  &  &  \multicolumn{2}{c}{$\semsls$} \\ \cline{3-4}
&  & None & Fence \\
\midrule
caseSLS01 &(\minusR)  & $\N$ & $\P$ \\ %
caseSLS02 &(\minusR)  & $\N$ & $\P$ \\
\bottomrule
\end{tabular}
\subcaption{Results for the Spectre-SLS programs under the $\semsls$ semantics against unpatched programs (column ``None''), programs patched with \texttt{lfence} (column ``Fence'').}
\label{t:table-sls}
\end{subtable}

\vspace{-10pt}
\caption{Result of the analysis of our benchmarks for $\semj$, $\sems$, $\semr$ and $\semsls$. 
For each program,  $\N$ denotes that \tool finds a violation of SNI under the corresponding semantics, whereas $\P$ denotes that \tool proves the program secure under the semantics.
Next to each program, we report if the program is \plusG ecure or \minusR nsecure in its unpatched version.
}
\label{tab:evaluation}
\end{figure*}

\textbf{Results for Spectre-BTB}
\Cref{t:table-v2} reports the analysis of the programs in the \textbf{Spectre-BTB} benchmark.
Using the $\semj$ semantics, \tool{} detected leaks (i.e., violations of \SNI) in all unpatched programs.
Furthermore, \tool{} proved that the manually patched programs using \texttt{lfence} and the automatically patched programs using the retpoline countermeasure satisfy SNI, i.e., they are free of speculative leaks.

\subsubsection{Results for Spectre-STL}
\Cref{t:table-v4} reports the results of analysing the programs in the \textbf{Spectre-STL} benchmark.\footnote{We had to slightly modify programs 02, 05, and 06 due to limitations of \tool{}'s x86 front-end when dealing with global values (programs 05 and 06) and 32-bit addressing (program 02). We had to limit the speculation window, due to vanilla \tool{}'s limitations in symbolic execution, when analyzing program 09, which contains a loop.}
Using the $\sems$ semantics, \tool{} successfully detected leaks (i.e., violations of \SNI) in all unpatched programs, except programs 03, 09, and 12 which do not contain speculative leaks (consistently with other analysis results~\cite{ST_binsec,cats}).
Observe that Binsec/Haunted~\cite{ST_binsec} flags program 13 as secure since the program can \emph{only} speculatively leak initial values from the stack, which Binsec/Haunted treats as public by default~\cite{hauntedBugReport}.
Since we assume initial memory values to be secret (like~\citet{cats}), \tool{} correctly detected the leak in program 13.
\tool{} also successfully proved that all patched programs (where an \texttt{lfence} is added between $\storeC{}$ instructions) satisfy \SNI{} and are free of speculative leaks.

\subsubsection{Results for Spectre-RSB}
\Cref{t:table-v5} reports the analysis results on the \textbf{Spectre-RSB} programs.
Using $\semr$, \tool{} successfully detected leaks in all unpatched programs.
Moreover, \tool{} successfully proved that the patched programs where a \texttt{lfence} instruction is added after every $\callC{}$ satisfy \SNI{}, i.e., they are free of speculative leaks.
\tool{} also successfully proved secure the programs patched using the modified \texttt{retpoline} defense proposed by \citet{ret2spec}, which replaces return instructions with a construct that traps the speculation in an infinite loop.

\subsubsection{Results for Spectre-SLS}
\Cref{t:table-sls} reports the analysis results on the \textbf{Spectre-SLS} programs.
Using $\semsls$, \tool{} successfully detected leaks in all unpatched programs and proved the security (i.e. SNI) of the patched programs where an \textsc{lfence} instruction was added after every return instruction.

%% file: src/Evaluations/case-study-combinations.tex
\subsection{Case Study: Detecting Leaks w.r.t. Combined Speculative Semantics}\label{sec:case-study-comb}

Using \tool{}, we analyze a corpus of 18 microbenchmarks containing speculative leaks generated by a combination of speculation mechanisms (for all composed semantics from \Cref{sec:comb-in}).
In particular, we consider one microbenchmark for each of all possible combinations of the $\semb$, $\semj$, $\sems$, $\semr$ and $\semsls$ semantics, i.e., \nrComb{} combined semantics (as mentioned, combinations containing both $\semr$ and $\semsls$ are not possible).
With these experiments, we aim to show that our combined semantics can detect novel leaks that are otherwise undetectable when considering single speculation mechanisms in isolation.

\newcommand{\nrProgramsCombination}{18}
\subsubsection{Benchmarks}
For each of the 18 combined semantics, we manually crafted an example \muasm{} program that has a speculative leak under that specific combination.
  For each program, we also analyze a manually patched version where \texttt{lfence} instructions prevent speculative leaks.
  We refer to this benchmark, consisting of 36 \muasm{} programs, as \textbf{Spectre-Comb}.

\subsubsection{Experimental Setup}
For each \muasm{} program in \textbf{Spectre-Comb}, we run \tool{} with all individual speculative semantics as well as all their combinations.
All experiments were run on a laptop with a Dual Core Intel Core i5-7200U CPU and 8GB of RAM.

\setlength\tabcolsep{1.5pt}
\begin{table*}
    \begin{subtable}{\textwidth}
    \centering
    \begin{tabular}{*{15}{c}}
    \toprule
        Program  & $\semb$  & $\semj$ & $\sems$ & $\semr$ & $\semsls$ & $\sembj$ & $\sembs$ & $\sembr$ & $\sembsls$  & $\semjs$ & $\semjr$ & $\semjsls$ & $\semsr$ & $\semssls$ \\
        \midrule
        comb12 & $\P$ & $\P$ & $\P$ & $\P$ & $\P$ & $\N$ & $\P$ & $\P$ & $\P$ & $\P$ & $\P$ & $\P$ & $\P$ & $\P$ \\
        comb14 & $\P$ & $\P$ & $\P$ & $\P$ & $\P$ & $\P$ & $\N$ & $\P$ & $\P$ & $\P$ & $\P$ & $\P$ & $\P$ & $\P$ \\
        comb15 & $\P$ & $\P$ & $\P$ & $\P$ & $\P$ & $\P$ & $\P$ & $\N$ & $\P$ & $\P$ & $\P$ & $\P$ & $\P$ & $\P$ \\
        comb16 & $\P$ & $\P$ & $\P$ & $\P$ & $\P$ & $\P$ & $\P$ & $\P$ & $\N$ & $\P$ & $\P$ & $\P$ & $\P$ & $\P$ \\
        comb24 & $\P$ & $\P$ & $\P$ & $\P$ & $\P$ & $\P$ & $\P$ & $\P$ & $\P$ & $\N$ & $\P$ & $\P$ & $\P$ & $\P$ \\
        comb25 & $\P$ & $\P$ & $\P$ & $\P$ & $\P$ & $\P$ & $\P$ & $\P$ & $\P$ & $\P$ & $\N$ & $\P$ & $\P$ & $\P$ \\
        comb26 & $\P$ & $\P$ & $\P$ & $\P$ & $\P$ & $\P$ & $\P$ & $\P$ & $\P$ & $\P$ & $\P$ & $\N$ & $\P$ & $\P$ \\
        comb45 & $\P$ & $\P$ & $\P$ & $\P$ & $\P$ & $\P$ & $\P$ & $\P$ & $\P$ & $\P$ & $\P$ & $\P$ & $\N$ & $\P$ \\
        comb46 & $\P$ & $\P$ & $\P$ & $\P$ & $\P$ & $\P$ & $\P$ & $\P$ & $\P$ & $\P$ & $\P$ & $\P$ & $\P$ & $\N$ \\
        comb124 & $\P$ & $\P$ & $\P$ & $\P$ & $\P$ & $\P$ & $\P$ & $\P$ & $\P$ & $\P$ & $\P$ & $\P$ & $\P$ & $\P$\\
        comb125 & $\P$ & $\P$ & $\P$ & $\P$ & $\P$ & $\P$ & $\P$ & $\P$ & $\P$ & $\P$ & $\P$ & $\P$ & $\P$ & $\P$\\
        comb126 & $\P$ & $\P$ & $\P$ & $\P$ & $\P$ & $\P$ & $\P$ & $\P$ & $\P$ & $\P$ & $\P$ & $\P$ & $\P$ & $\P$\\
        comb145 & $\P$ & $\P$ & $\P$ & $\P$ & $\P$ & $\P$ & $\P$ & $\P$ & $\P$ & $\P$ & $\P$ & $\P$ & $\P$ & $\P$\\
        comb146 & $\P$ & $\P$ & $\P$ & $\P$ & $\P$ & $\P$ & $\P$ & $\P$ & $\P$ & $\P$ & $\P$ & $\P$ & $\P$ & $\P$\\
        comb245 & $\P$ & $\P$ & $\P$ & $\P$ & $\P$ & $\P$ & $\P$ & $\P$ & $\P$ & $\P$ & $\P$ & $\P$ & $\P$  & $\P$\\
        comb246 & $\P$ & $\P$ & $\P$ & $\P$ & $\P$ & $\P$ & $\P$ & $\P$ & $\P$ & $\P$& $\P$ & $\P$ & $\P$ & $\P$ \\
        comb1245 & $\P$ & $\P$ & $\P$ & $\P$ & $\P$ & $\P$ & $\P$ & $\P$ & $\P$ & $\P$ & $\P$ & $\P$ & $\P$ & $\P$\\
        comb1246 & $\P$ & $\P$ & $\P$ & $\P$ & $\P$ & $\P$ & $\P$ & $\P$ & $\P$ & $\P$ & $\P$ & $\P$ & $\P$ & $\P$\\
        \bottomrule
    \end{tabular}
    \end{subtable}
    
\vspace{2em}

    \setlength\tabcolsep{1.5pt}
    \begin{subtable}{\textwidth}
    \centering
    \begin{tabular}{*{10}{c}}
    \toprule
    Program & $\sembjs$ & $\sembjr$ & $\sembjsls$ & $\sembsr$ & $\sembssls$ & $\semjsr$ & $\semjssls$ & $\sembjsr$ & $\sembjssls$\\
    \midrule
        comb12 & $\N$ & $\N$ & $\N$ & $\P$ & $\P$ & $\P$ & $\P$ & $\N$ & $\N$ \\
        comb14 & $\N$ & $\P$ & $\P$ & $\N$ & $\N$ & $\P$ & $\P$ & $\N$ & $\N$ \\
        comb15 & $\P$ & $\N$ & $\P$ & $\N$ & $\P$ & $\P$ & $\P$ & $\N$ & $\P$ \\
        comb16 & $\P$ & $\P$ & $\N$ & $\P$ & $\N$ & $\P$ & $\P$ & $\P$ & $\N$ \\
        comb24 & $\N$ & $\P$ & $\P$ & $\P$ & $\P$ & $\N$ & $\N$ & $\N$ & $\N$ \\
        comb25 & $\P$ & $\N$ & $\P$ & $\P$ & $\P$ & $\N$ & $\P$ & $\N$ & $\P$ \\
        comb26 & $\P$ & $\P$ & $\N$ & $\P$ & $\P$ & $\P$ & $\N$ & $\P$ & $\N$ \\
        comb45 & $\P$ & $\P$ & $\P$ & $\N$ & $\P$ & $\N$ & $\P$ & $\N$ & $\P$ \\
        comb46 & $\P$ & $\P$ & $\P$ & $\P$ & $\N$ & $\P$ & $\N$ & $\P$ & $\N$ \\
        comb124 & $\N$ & $\P$ & $\P$ & $\P$ & $\P$ & $\P$ & $\P$ & $\N$ & $\N$\\
        comb125 & $\P$ & $\N$ & $\P$ & $\P$ & $\P$ & $\P$ & $\P$ & $\N$ & $\P$\\
        comb126 & $\P$ & $\P$ & $\N$ & $\P$ & $\P$ & $\P$ & $\P$ & $\P$ & $\N$ \\
        comb145 & $\P$ & $\P$ & $\P$ & $\N$ & $\P$ & $\P$ & $\P$ & $\N$ & $\P$ \\
        comb146 & $\P$ & $\P$ & $\P$ & $\P$ & $\N$ & $\P$ & $\P$ & $\P$ & $\N$ \\
        comb245 & $\P$ & $\P$ & $\P$ & $\P$ & $\P$ & $\N$ & $\P$ & $\N$ & $\P$  \\
        comb246 & $\P$ & $\P$ & $\P$ & $\P$ & $\P$ & $\P$ & $\N$ & $\P$ & $\N$ \\
        comb1245 & $\P$ & $\P$ & $\P$ & $\P$ & $\P$ & $\P$ & $\P$ & $\N$ & $\P$  \\
        comb1246 & $\P$ & $\P$ & $\P$ & $\P$ & $\P$ & $\P$ & $\P$ & $\P$ & $\N$  \\
        \bottomrule
    \end{tabular}
    \end{subtable}
    \caption{Results of the analysis.
For each program,  $\N$ denotes that \tool finds a violation of SNI whereas $\P$ denotes that \tool proves the program secure under the corresponding semantics.
The different programs were devised to be vulnerable under a specific combination. The numbers correspond to that specific version $1 = \Bv$, $2 = \Jv$, $4 = \Sv$, $5 = \Rv$ and $6 = \SLSv$. For example, comb125 was devised to be vulnerable under $\sembjr$.
}\label{t:table-comb}
\end{table*}

\subsubsection{Results for Spectre-Comb}
\Cref{t:table-comb} reports the results of our analysis on the \textbf{Spectre-Comb} programs, which involve leaks arising from a combination of multiple speculation mechanisms.

\tool{} equipped with the individual semantics $\semb$, $\semj$, $\sems$, $\semr$ and $\semsls$ is not able to detect the speculative leaks in any of the \nrProgramsCombination{} programs and, therefore, proves them secure.
This is expected since the programs contain leaks that arise from a combination of semantics.

\tool{} can successfully identify leaks in all programs when using the semantics corresponding to that example. For example, \textbf{comb25} is proven insecure by the respective semantics $\semsr$.

Furthermore, as \Cref{fig:instantiate} suggests, all 'stronger' semantics (higher in the combination lattice)  also detect the vulnerability in the programs made for a 'weaker' semantics (lower in the combination lattice) and the 'weaker' ones do not detect the vulnerability of programs made for 'stronger' semantics. For example, \textbf{comb25} is also proven to be insecure by \tool when using $\sembjr$, $\semjsr$, $\sembjsr$, while the program \textbf{comb1245} can only be proven insecure by $\sembjsr$.

Semantics that are in no particular order in the lattice like $\sembjsr$ and $\sembjssls$ are also not able to detect the vulnerability in \textbf{comb1246} and \textbf{comb1245} respectively. This is expected since they each miss one speculative mechanism necessary to detect the vulnerability. 

Because there is no combination of all our source speculative semantics $\semb$, $\semj$, $\sems$, $\semr$ and $\semsls$, there is also no unique combination that can find the leaks in all programs in \textbf{Spectre-Comb}.
However, using $\sembjsr$ and $\sembjssls$, \tool is able to successfully detect leaks in all programs.

We also analyzed programs manually patched with \texttt{lfence} statements. 
The examples are not shown here for brevity and can be found with the other code snippets at~\cite{tool-og}.
As before, \tool{} successfully proves the security of all patched programs.
Even for leaks that arise from multiple speculation mechanisms, it is often sufficient to insert a single \texttt{lfence} to secure the entire program, e.g., an  \texttt{lfence} after the $\jzC{}$ instruction in \textbf{comb15}
is enough to make the program \SNI{} with respect to $\sembsr$.

%% file: src/discussion.tex
\section{Discussion}\label{sec:limitations}

\subsection{Scope of the Models}
Lifting the results of the security analysis for our speculative semantics to real-world CPUs is only possible to the extent that these semantics capture the information flows in the target system.
Thus, \tool{}'s result may incorrectly classify programs as secure (if our semantics do not capture information flows happening in real-world CPUs) or insecure (if our semantics admit speculations that are impossible on real systems).

We capture ``leakage into the microarchitectural state'' using the relatively powerful observer of the program execution that sees the location of memory accesses and the jump targets.
This observer could be replaced by a weaker one, which accounts for more detailed models of a CPU's memory hierarchy, and \tool could be adapted accordingly, e.g. by adopting the cache models from CacheAudit~\cite{cacheAudit}.
We believe, however, that highly detailed models are not actually desirable for several reasons: 
\begin{inparaenum}[(a)]
	\item they encourage brittle designs that break under small changes to the model,
	\item they have to be adapted frequently, and
	\item they are hard to understand and reason about for compiler developers and hardware engineers.
\end{inparaenum}
The ``constant-time'' observer model adopted in this paper has  proven to offer a good tradeoff between precision and robustness~\cite{MolnarPSW05,AlmeidaBBDE16}.

\subsection{Other Speculation Mechanisms}
There are many speculation mechanisms beyond those modeled in the speculative semantics $\semb$, $\sems$, $\semj$, $\semr$ and $\semsls$ from \Cref{sec:inst-spec-semantics-single}:
\begin{asparaitem}

\item CPUs speculate over $\retC{}$ instructions in different ways.
For instance, there are many different ways of implementing return stack buffers (e.g., cyclic versus acyclic RSBs~\cite{ret2spec} or RSBs that fall back to indirect branch prediction~\cite{retbleed}).
This kind of speculation can be modeled by modifying the \Cref{tr:v5-ret-paper} rule in $\semr$.

\item Many proposals for value prediction over different kinds of instructions exist~\cite{val_pred, val_pred1, val_pred2}. 
While naive speculative semantics might have to explore \emph{all} possible values as prediction, semantics that model specific prediction mechanisms might restrict the set of predicted values (thereby leading to a more tractable analysis).
\end{asparaitem}

We expect that most of these mechanisms can be modeled as speculative semantics satisfying our well-formedness conditions.
Hence, they could work with our composition framework. %

\subsection{Limitations of Composition}
Our composition framework has two main limitations:
\begin{asparaenum}
\item The metaparameter $Z$ is expressed in terms of \muasm{} instructions, i.e., the smallest unit of computation in our framework.
Since $Z$ restricts how the composed semantics delegates execution to its sources, this limit the expressiveness of composed semantics.
For instance, $\semsr$ cannot speculate over the \emph{implict} $\storeC{}$ writing the return address to the stack that happens as part of $\callC{}$ instructions.
\item Our framework does not support combinations where a single instruction performs speculation-relevant changes in both source semantics.
For instance, consider a combination of $\semr$ with $\semsls$.
Here, both semantics start different speculative transactions on executing $\retC{}$ instructions.
However, instantiating $Z$ as $(\emptyset, \emptyset)$, which enables both speculations, violates the confluence well-formedness condition for the composed semantics, whereas setting $Z = (x,y)$ so that only one of $x$ and $y$ is $\retC{}$ would only capture one of the two speculation mechanisms.

\end{asparaenum}
We leave addressing both limitations as future work.

\subsection{Different flavours of Non-Interference}\label{ssec:ni-variants}

Here, we discuss the relation of SNI with other well-known notions of security against side-channel leaks.

\subsubsection{General Non-Interference}
We start by introducing General Non-Interference (GNI), a notion of  security capturing that secrets do not leak through side channels.
This notion is adapted from the hardware-software contracts framework of \citet{ST_spectector2} into our setting.

\begin{definition}[General Non-Interference (GNI)]\label{def:gni}
    Program $p$ satisfies GNI (denoted $\gnix{}$) for a semantics $x$ iff for all $\sigma$, $\sigma'$, 
    {if} $\sigma \backsim_{\pol} \sigma'$
    {then} $\behx{p, \sigma} = \behx{p, \sigma'}$.
\end{definition}

In a nutshell, a program $p$ satisfies GNI for a given semantics $x$ if any pair of low-equivalent initial configurations $\sigma$ and $\sigma'$ results in the same observations, i.e., $\behx{p, \sigma} = \behx{p, \sigma'}$.

Observe that GNI is an \emph{absolute security property}~\cite{sok:spectre_defense}, that is, it ensures that \emph{all} observations are the same for any two executions starting from low-equivalent configurations.
In contrast, SNI is a \emph{relative security property}~\cite{sok:spectre_defense}. 
That is, it ensures that speculatively executed instructions (and, thus, speculative observations) do not leak more information than what is already leaked by the program's non-speculative execution. 
Naturally, satisfying GNI implies satisfying SNI for the same semantics $x$:

\begin{restatable}[GNI implies SNI]{corollary}{nisni}\label{thm:ni-sni}
    If $\gnix{}$ then $\snix{}$.
\end{restatable}

\subsubsection{Relation with constant-time variants}

A common defense against several side-channel timing attacks is the constant-time programming discipline~\cite{MolnarPSW05, AlmeidaBBDE16}. 
This programming discipline requires that a program's control flow and memory access patterns are strictly independent of sensitive data. 

Following~\cite{AlmeidaBBDE16}, constant-time programming can be modeled as a non-interference property by requiring that low-equivalent executions  result in the same architectural control-flow and in the same sequence of memory accesses.
That is, we can instantiate constant-time security by requiring the equivalence of the non-speculative trace, as indicated next:\looseness=-1

\begin{definition}[Non-Speculative Constant-Time (CT)]\label{def:ct}
Program $p$ satisfies CT (denoted $\seqct{}$) iff
    for all $\sigma$, $\sigma'$, 
    {if} $\sigma \backsim_{\pol} \sigma'$
    {then} $\behNs{p, \sigma} = \behNs{p, \sigma'}$.
\end{definition}

Observe that CT is an instantiation of GNI w.r.t. the non-speculative semantics, i.e., $\seqct{} \Leftrightarrow \gni{NS}$.
Note also that we named \Cref{def:ct} as ``Sequential Constant-Time'' rather than the more standard ``Constant-Time'' to stress that CT traditionally applies only to the architectural, non-speculative semantics.

To account for the effects of speculatively executed instructions, several works~ \cite{ST_constantTime_Spec, ST_jasmin, ST_jasmin2, ST_binsec, ST_blade} proposed security conditions that also restrict leaks through speculatively executed instructions.
That is, all these works enforce variants of GNI, differing in the underlying speculative semantics or in what observations are part of the trace.
For instance, in \citet{ST_constantTime_Spec} the low-projection of the final configuration is also part of the trace (in addition to the usual observations associated with the constant-time attacker) to ensure the absence of explicit leaks.

Crucially, we can precisely connect CT, SNI, and GNI. 
That is, general non-interference w.r.t. a speculative semantics $x$ is exactly the conjunction of constant-time and speculative non-interference w.r.t. $x$~\cite[Proposition 4]{ST_spectector2}:

\begin{restatable}[GNI Decomposition]{proposition}{snigni}\label{prop:sni-gni}
    Given a policy $\pol$ and a program $p$:
    $\seqct{} \land \snix{} \Leftrightarrow \gnix{}$
\end{restatable}

\Cref{prop:sni-gni} allows us to decompose the check for GNI into two independent checks.
An analysis only needs to prove that the program is constant-time (SeqCT) and that it satisfies speculative non-interference (SNI).

\subsubsection{Extending \tool{} with support for GNI}
We conclude by discussing how \tool{} can be extended to check GNI.
As stated above, SeqCT can be obtained by instantiating GNI with the non-speculative semantics, so this extension can be used also to check the classic constant-time property.

Recall that \tool algorithm from \Cref{sec:impl-symbolic} searches for pairs of execution traces that have equivalent non-speculative behavior but differing speculative observations.

To support GNI, we need to modify \memcheck{} and \pccheck{} as indicated in \Cref{fig:tool-ext}.
Specifically, we enforce that the entire symbolic trace $\tauStack$ can only produce identical concrete observations for all low-equivalent inputs.

For this, we modify \memcheck{} to check the satisfiability of the following formula:
\begin{center}
    $\pathCond{{\tauStack}}_{1 \wedge 2} \land \policyEqv{\policy} \land
\neg \cstrs{\tauStack}$.    
\end{center}
In particular, $\pathCond{\tauStack}_{1 \wedge 2}$ ensures that the same symbolic path is followed, $\policyEqv{\policy}$ ensures that the initial states are low-equivalent, and $\neg \cstrs{\tauStack}$ asserts that at least one observation (either architectural or speculative) associated with memory accesses differ among the two runs.
That is, this formula is satisfiable only if there are two concrete traces associated with the symbolic trace $\tauStack$ where memory operation accesses a different address.

The \pccheck{} procedure is modified analogously to ensure that all control-flow observations (architectural or speculative) are identical between traces.

Thus, with minor modifications to the \tool algorithm, we are able to check for GNI-style properties for all the semantics discussed in the paper.
The full algorithm for checking GNI can be found in \Cref{app:tool-variants}.

\begin{figure}
    \centering
    \begin{minipage}[t]{0.46\textwidth}
    \begin{algorithm}[H]
        \caption{$\memcheck$ for GNI}
        \label{algorithm:tool-seqct-short}
        \begin{algorithmic}[1]
     \Procedure{$\memcheck$}{$\tauStack, \policy$}
        		\State{$\psi \gets \pathCond{\tauStack}_{1 \wedge 2} \wedge \policyEqv{\policy} \wedge $}
        		\Statex{$\qquad \qquad \qquad \neg \cstrs{\tauStack}$}
    			\State{\Return{$\textsc{Satisfiable}(\psi)$}}
    
        \EndProcedure
    
        \end{algorithmic}
    \end{algorithm}
    \end{minipage}
    \hspace{10pt}
    \begin{minipage}[t]{0.46\textwidth}
    \begin{algorithm}[H]
        \caption{$\pccheck$ for GNI}
        \label{algorithm:tool-gni-short}
        \begin{algorithmic}[1]
    
        \Procedure{$\pccheck$}{$\tauStack, \policy$}
        		\For{each prefix $\nu \concat \symPcObs{\mathit{se}}$ of $\tauStack$}
    				\State{$\psi \gets \pathCond{\nspecProject{\tauStack}\concat \nu}_{1 \wedge 2} \wedge \policyEqv{\policy}$}
    				\Statex{$\qquad \qquad \qquad \wedge \neg \sameSymPc{se}$}
    				\If{$\textsc{Satisfiable}(\psi)$}
    					\State{\Return{$\top$}}
    				\EndIf
    			\EndFor
        		\State{\Return{$\bot$}}
        \EndProcedure
        \end{algorithmic}
    \end{algorithm}
    \end{minipage}
    \caption{The modified versions of \memcheck{} and \pccheck{} for GNI}
    \label{fig:tool-ext}
\end{figure}

%% file: src/related_work.tex
\section{Related Work}\label{sec:rw}

\tightpar{Speculative Execution Attacks}
After Spectre~\cite{spectre} has been disclosed to the public in 2018, researchers have identified many other speculative execution attacks~\cite{spectreRsb,ret2spec,S_smotherSpectre,S_trans_troj,barberis2022branch, wikner_phantom_2023}.
These attacks differ in the exploited speculation sources~\cite{ret2spec, spectreRsb, S_specv4}, the covert channels~\cite{S_specPrime, S_netSpectre, S_lazyFP, port_cont, platypous} used, or the target platforms~\cite{chen2018sgxpectre}. 
We refer the reader to~\citet{transientfail} for a survey of existing attacks.

\tightpar{Security Properties for Speculative Leaks}
Researchers have proposed many program-level properties for security against speculative leaks, which  can be classified in three main groups~\cite{sok:spectre_defense}:
\begin{asparaenum}
\item Non-interference definitions ensure the security of speculative \emph{and} non-speculative instructions. 
For instance, speculative constant-time~\cite{ST_constantTime_Spec, ST_jasmin, ST_jasmin2, ST_binsec, ST_blade} extends the constant-time security condition to account also for transient instructions.\looseness=-1 
\item Relative non-interference definitions~\cite{ST_tpod,ST_inspectre,ST_specusym, spec_dec} ensure that transient instructions do not leak more information than what is leaked by non-transient instructions.  
The notion of Speculative Non-Interference (SNI) introduced in this paper falls into this category, as it restricts the information leaked by speculatively executed instructions relative to the program's non-speculative behavior.

\item Definitions that formalise security as a safety property~\cite{cats,S_sec_comp}, which may over-approximate definitions from the groups above.\looseness=-1
\end{asparaenum}

Finally, we highlight the relationship between Speculative Constant-Time (SCT) and our properties SNI and GNI.
As shown by \citet{ST_constantTime_Spec} if the program is sequentially constant-time, SNI, and the resulting architectural states are low-equivalent, then the program is also SCT.
In relation to GNI, SCT enforces the exact same trace equivalence but imposes the additional constraint that final architectural states must be low-equivalent.
We can verify SCT on top of our \tool-GNI check by explicitly exposing the final architectural state as a terminal observation within the execution trace.

\tightpar{Operational Semantics for Speculative Leaks}
In the last few years, there has been a growing interest in developing formal models and principled program analyses for detecting leaks caused by speculatively executed instructions.
We refer the reader to~\cite{sok:spectre_defense} for a comprehensive survey.
In the following, we discuss the approaches that are particularly relevant to our paper.

Some approaches explicitly model microarchitectural components like multiple pipeline stages, caches, and branch predictors. 
For instance, KLEESpectre~\cite{kleeSpectre} and SpecuSym~\cite{ST_specusym} consider a semantics that explicitly models the cache, enabling reasoning about the cache content.
\citet{R_sem_model2} go a step further and model a multi-stage pipeline with explicit cache and branch predictor.
Their semantics can only model speculation over branch instructions since it lacks store-forwarding or RSB.

\citet{ST_constantTime_Spec}'s semantics model speculation over branch instructions, store-bypasses, and return instructions.
Differently from our semantics, their 3-stage pipeline semantics explicitly models several microarchitectural components like a reorder buffer and an RSB.
Their tool detects violations of speculative constant-time induced by speculation over branch instructions and store-bypasses.

\citet{ST_jasmin} extend the Jasmin~\cite{ST_jasmin2} cryptographic verification framework to reason about speculative constant-time and supports speculation over store-bypasses and branch instructions.
\citet{jasmin_typed} equip Jasmin with an information flow control type system that enforces SCT. However, only against \specb attacks.

Binsec/Haunted~\cite{ST_binsec}  detect violations of speculative constant-time due to speculation over store-bypasses and branch instructions.
For this, they explicitly model the store buffer, which $\sems$ abstracts away.
Blade \cite{ST_blade} uses a JIT-step semantics that translates high-level commands (from a WHILE-language) to low-level machine instructions while tracking control-flow predictions. This allows for source-level reasoning and their semantics captures speculation over branch instructions.

While several of these models support multiple speculation mechanisms, these mechanisms are \emph{hard-coded} and no existing approach provides a composition framework or extensible ways of extending the main theoretical results to new mechanisms ``for free''.
Moreover, while we could have used other semantics as a basis for our framework, this would have resulted in more difficult proofs (since semantics like the one in~\cite{ST_constantTime_Spec} are significantly more complex than ours).\looseness=-1

\tightpar{Axiomatic Semantics for Speculative Leaks}
A few approaches formalise the effects of speculatively executed instructions using axiomatic semantics inspired by work on weak memory models.
For instance, \citet{ST_abs_sem} and \citet{R_sem_model} capture the effects of branch speculation but both lack program analyses. 

\citet{cats} illustrate how one can model leaks resulting from speculation over branch instructions and store-bypasses using the CAT modeling language for memory consistency, and they present a bounded model checking analysis for detecting speculative leaks.
Interestingly, they talk about composing several of their semantics \cite[\S IV.F]{cats}, which should allow them to detect vulnerabilities like \Cref{lst:v1-v4-combined} (which we detect under $\sembs$).
However, they do not formally characterize compositions and, therefore, they cannot derive interesting results ``for free'' about the composed semantics (like we do in \Cref{thm:comp-sss-paper}).
Moreover, even though they state that composability is an advantage of axiomatic models, our framework (and tool implementation) shows that composability can be done with operational semantics as well.

\tightpar{Secure Compilation for Speculative Leaks}
In addition to program analyses like the ones described above, researchers have proposed compiler passes to prevent and mitigate speculative leaks.
For instance, \citet{S_sec_comp} and \citet{fabian2024lift} study the security of compiler-level countermeasures implemented in major compilers using Secure Compilation theory, whereas \citet{ST_blade} propose a compiler pass for automatically patching speculative leaks using a type system to track speculative leaks.

\citet{serberus} propose a set of compiler passes that, together with hardware support (e.g. Intel CET-IBT, a shadow stack for return addresses) offers protection against \specb, \specj, \specs, \specr, and predictive store forwarding \cite{psf}.
We note that \citet[Section~3.3]{serberus} showed a counterexample that our semantics $\sems$ does not catch. 
When devising our speculative semantics, we encountered a trade-off: keeping the related \tool implementation tractable (albeit imprecise) or making the semantics precise (but not implementable due to state explosion of the tool). 
We decided to model the speculation in that way to keep the analysis tractable. 
Furthermore, our case study in \Cref{sec:case-study-semantics} was used by other detection tools as well \cite{ST_binsec, cats} and our semantics $\sems$ detects all vulnerabilities there. 
Thus, we think this is a reasonable trade-off between precision and performance. 

\tightpar{Detecting Leaks through Testing}
The Revizor testing tool \cite{revizor, revizor2, revizor_23, Oleksenko2026EnterExitPageFaultLeak} and SpecFuzz \cite{ST_specfuzz} use fuzz testing to find vulnerabilities caused by speculation in CPUs.
Similarly, SpecDoctor by \citet{specDoc} uses differential fuzzing to fuzz for transient vulnerabilities on the RTL level.
\citet{marinaro} extend Scam-V \cite{scamv} to optimize software mitigations for \specb using a relational testing approach.

AMuLeT \cite{amulet} adapts Revizor to the design phase. AMuLeT instruments microarchitectural simulators (e.g., gem5) to check secure speculation countermeasures against leakage contracts.
LMTest \cite{lmspec} provides a framework for testing cryptographic implementations against parameterized leakage models (defined in a DSL called LmSpec). It generates test cases to detect secret-dependent leaks in the presence of proposed microarchitectural optimizations.

Unlike \tool, these approaches cannot detect the absence of leaks.

%% file: src/future_work.tex
\section{Conclusion}\label{sec:conc}
This paper presented an approach to automatically detect speculative leaks in programs. 
We introduced speculative non-interference, the first semantic notion of security against speculative execution attacks, and we defined multiple speculative semantics to capture 5 classes of Spectre attacks.
Next, we defined a general framework to reason about the composition of different speculative semantics and instantiated the framework with our speculative semantics.  Our framework yields safety of the composed semantics (almost) for free, given the safety of its parts.
We utilize our theoretical development and implement a program analysis tool called \tool that automatically detects speculative leaks or proves their absence, for our speculative semantics and their combinations.

\begin{acks}
\newcounter{thesponsor}
\setcounter{thesponsor}{0}

This work is supported by 
\stepcounter{thesponsor}the \grantsponsor{\arabic{thesponsor}}{Spanish Ministry of Science, Innovation, and University}{https://www.ciencia.gob.es/} under the Ram\'on y Cajal grant \grantnum{\arabic{thesponsor}}{RYC2021-032614-I};
\stepcounter{thesponsor}the \grantsponsor{\arabic{thesponsor}}{Spanish Ministry of Science, Innovation, and University}{https://www.ciencia.gob.es/} under the project \grantnum{\arabic{thesponsor}}{PID2022-142290OB-I00 ESPADA};
\stepcounter{thesponsor}the \grantsponsor{\arabic{thesponsor}}{Spanish Ministry of Science, Innovation, and University}{https://www.ciencia.gob.es/} under the Europa Excelencia project \grantnum{\arabic{thesponsor}}{EUR2025-164828 SINTRAZAS};
\stepcounter{thesponsor}the \grantsponsor{\arabic{thesponsor}}{Spanish Ministry of Science, Innovation, and University}{https://www.ciencia.gob.es/} under the project \grantnum{\arabic{thesponsor}}{CEX2024-001471-M};
\stepcounter{thesponsor}the \grantsponsor{\arabic{thesponsor}}{European Union}{https://erc.europa.eu/} under the Horizon Europe projects \grantnum{\arabic{thesponsor}}{101230068 PRIMULA} and \grantnum{\arabic{thesponsor}}{101020415 SafeSecS}.
\end{acks}

%% file: src/Appendix/code-case-studies.tex
\section{Code from Case Studies}\label{sec:code}

\subsection{Example \#8}\label{secs:example8}

In Example \#8, the bounds check of \Cref{lst:v1-vanilla}
is implemented using a conditional operator:
\begin{lstlisting}[style=Cstyle]
temp &= B[A[y<size?(y+1):0]*512];
\end{lstlisting}
When compiling the example without countermeasures or optimizations,
the conditional operator is translated to a branch instruction (cf.~line 4), which is a source of speculation. Hence, the resulting program
contains a speculative leak, which \tool{} correctly detects.

\begin{lstlisting}[style=ASMstyle]
	 mov    size, %
	 mov    y, %
	 cmp    %
	 jae    .L1
	 add    $1, %
	 jmp    .L2
.L1:
	 xor    %
	 jmp    .L2
.L2:
	 mov    A(%
	 shl    $9, %
	 mov    B(%
	 mov    temp, %
	 and    %
	 mov    %
\end{lstlisting}

In the \unp{} \opt{} mode, the
conditional operator is translated as a conditional move
(cf. line 6), for which \tool{} can prove security.

\begin{lstlisting}[style=ASMstyle]
 mov    size, %
 mov    y, %
 xor    %
 cmp    %
 lea    1(%
 cmova  %
 mov    A(%
 shl    $9, %
 mov    B(%
 and    %
\end{lstlisting}

\subsection{Example \#15 in \slh{} mode}\label{ssec:example15}
Here, the adversary provides the input via the pointer \inlineCcode{*y}:
\begin{lstlisting}[style=Cstyle]
 if (*y < size)
    temp &= B[A[*y] * 512];
\end{lstlisting}
In the \unopt{} \slh{} mode, 
\clang{} hardens the address used for performing the memory access \inlineCcode{A[*y]} in lines 8--12, but not the resulting value, which is stored in the register \inlineASMcode{\%cx}.
However, the value stored in  \inlineASMcode{\%cx} is used to perform a second memory access at line 14.
An adversary can exploit the second memory access to speculatively leak the content of \inlineCcode{A[0xFF...FF]}.
In our experiments, \tool{} correctly detected such leak.
\begin{lstlisting}[style=ASMstyle]
 mov    $0, %
 mov    y, %
 mov    (%
 mov    size, %
 cmp    %
 jae     END
 cmovae $-1, %
 mov    y, %
 mov    (%
 mov    %
 or     %
 mov    A(%
 shl    $9, %
 mov    B(%
 mov    temp, %
 and    %
 mov    %
\end{lstlisting}
In contrast, when Example~\#15 is compiled with the \opt{} flag, \clang{} correctly hardens  \inlineCcode{A[*y]}'s result (cf.~line 10).
This prevents information from flowing into the microarchitectural state during speculative execution.
Indeed, \tool{} proves that the program satisfies speculative non-interference.
\begin{lstlisting}[style=ASMstyle]
 mov    $0, %
 mov    y, %
 mov    (%
 mov    size, %
 cmp    %
 jae    END
 cmovae $-1, %
 mov    A(%
 shl    $9, %
 or     %
 mov    B(%
 or     %
 and    %
\end{lstlisting}

%% file: src/Appendix/spectector_variants.tex
\section{Variants of Spectector}\label{app:tool-variants}

In \Cref{ssec:ni-variants} we discussed the variation of the \tool algorithm to check for GNI. Here, we will present the full algorithm as well as the algorithm for checking sequential constant-time:

\begin{algorithm}
    \caption{\tool{} SeqCT}
    \label{algorithm:tool-seqct}
    \begin{algorithmic}[1]
	\Require A program $p$, a security policy $\policy$, a speculative window  $w \in \Nat$.
	\Ensure \textsc{Secure} if $p$ satisfies SeqCT with respect to the policy $\policy$; \textsc{Insecure} otherwise
		\Statex{}
	\Procedure{\tool}{$p,\policy,w$}
		\For{each symbolic run $\tauStack \in \behxa{p}$}
			\If{$\memcheck(\tauStack,P)  \vee \pccheck(\tauStack,P)$}
				\State{\Return{\textsc{Insecure}}}
			\EndIf
		\EndFor
		\State{\Return{\textsc{Secure}}}
        \EndProcedure
    \Statex{}
    \Procedure{$\memcheck$}{$\tauStack, \policy$}
    		\State{$\psi \gets \pathCond{\tauStack}_{1 \wedge 2} \wedge \policyEqv{\policy} \wedge$}
    		\Statex{$\qquad \qquad \qquad \neg \cstrs{\nspecProject{\tauStack}}$}
			\State{\Return{$\textsc{Satisfiable}(\psi)$}}

    \EndProcedure
    \Statex{}
    \Procedure{$\pccheck$}{$\tauStack, \policy$}
    		\For{each prefix $\nu \concat \symPcObs{\mathit{se}}$ of $\nspecProject{\tauStack}$}
				\State{$\psi \gets \pathCond{\nspecProject{\tauStack}\concat \nu}_{1 \wedge 2} \wedge \policyEqv{\policy} \wedge$}
				\Statex{$\qquad \qquad \qquad \neg \sameSymPc{se}$}
				\If{$\textsc{Satisfiable}(\psi)$}
					\State{\Return{$\top$}}
				\EndIf
			\EndFor
    		\State{\Return{$\bot$}}
    \EndProcedure
    \end{algorithmic}
\end{algorithm}

\begin{algorithm}
    \caption{\tool{} GNI}
    \label{algorithm:tool-gni}
    \begin{algorithmic}[1]
	\Require A program $p$, a security policy $\policy$, a speculative window  $w \in \Nat$.
	\Ensure \textsc{Secure} if $p$ satisfies general non-interference with respect to the policy $\policy$ and speculative semantics $\semx$; \textsc{Insecure} otherwise
		\Statex{}
	\Procedure{\tool}{$p,\policy,w$}
		\For{each symbolic run $\tauStack \in \behxa{p}$}
			\If{$\memcheck(\tauStack,P)  \vee \pccheck(\tauStack,P)$}
				\State{\Return{\textsc{Insecure}}}
			\EndIf
		\EndFor
		\State{\Return{\textsc{Secure}}}
        \EndProcedure
    \Statex{}
    \Procedure{$\memcheck$}{$\tauStack, \policy$}
    		\State{$\psi \gets \pathCond{\tauStack}_{1 \wedge 2} \wedge \policyEqv{\policy} \wedge $}
    		\Statex{$\qquad \qquad \qquad \neg \cstrs{\tauStack}$}
			\State{\Return{$\textsc{Satisfiable}(\psi)$}}

    \EndProcedure
    \Statex{}
    \Procedure{$\pccheck$}{$\tauStack, \policy$}
    		\For{each prefix $\nu \concat \symPcObs{\mathit{se}}$ of $\tauStack$}
				\State{$\psi \gets \pathCond{\nspecProject{\tauStack}\concat \nu}_{1 \wedge 2} \wedge \policyEqv{\policy}$}
				\Statex{$\qquad \qquad \qquad \wedge \neg \sameSymPc{se}$}
				\If{$\textsc{Satisfiable}(\psi)$}
					\State{\Return{$\top$}}
				\EndIf
			\EndFor
    		\State{\Return{$\bot$}}
    \EndProcedure
    \end{algorithmic}
\end{algorithm}

%% file: src/Appendix/trace-projections.tex
\section{Trace Projections}\label{app:trace-proj}
Here, we formalize the speculative projections $\specProjectx{}$ for each of our semantics and the non-speculative projection $\nspecProject{}$.

\para{Non-speculative Trace Projection}

Given a trace $\tau$, its non-speculative projection contains only the observations that are produced by committed transactions; in other words, rolled-back transactions are removed in the projection.
Formally, $\nspecProject{\tau}$ is defined as follows:

\begin{definition}[Non-speculative projection]
We define the non-speculative projection mutually recursive as 
\begin{align*}
    \nspecProject{\empTr} =& \empTr \\
    \nspecProject{\tauStack \cdot \commitObsx{n}} =& \nspecProject{\tauStack} \\
     \nspecProject{\tauStack \cdot \startObsx{n}} =& \nspecProject{\tauStack} \\
    \nspecProject{\tauStack \cdot \rollbackObsx{id} } =& \nspecProjectHelper{\tauStack, \id} \\
    \nspecProject{\tauStack \cdot \tau} =& \nspecProject{\tauStack} \cdot \tau ~\text{otherwise} \\
\end{align*}

The $\nspecProjectHelper$ is defined as
\begin{align*}
    \nspecProjectHelper{\empTr, \id} &= \empTr  \\
    \nspecProjectHelper{\tauStack \cdot \startObsx{\id}, \id} &= \nspecProject{\tauStack}\\
    \nspecProjectHelper{\tauStack \cdot \tau, \id} &= \nspecProjectHelper{\tauStack, \id} ~\text{otherwise}
\end{align*}
\end{definition}

\para{Speculative Trace Projections}

Given a speculative trace $\tau$, its speculative projection contains only the observations produced by rolled-back transactions.

\begin{definition}[Speculative Projection]

\begin{align*}
    \specProject{\empTr} =&~ \empTr \\
    \specProject{\tauStack \cdot \rollbackObsx{i} } =&~ \specProjectHelper{\tauStack, i} \\
    \specProject{\tauStack \cdot \tau} =&~ \specProject{\tauStack} ~~\text{otherwise}
\end{align*}

\begin{align*}
    \specProjectHelper{\empTr, \id} &= \empTr  \\
    \specProjectHelper{\tauStack \cdot \startObsx{\id}, \id} &= \specProject{\tauStack}\\ %
    \specProjectHelper{\tauStack \cdot \startObsx{\id'}, \id} &= \specProjectHelper{\tauStack, \id} ~~ \text{if $id \neq id'$}\\
    \specProjectHelper{\tauStack \cdot \rollbackObsx{\id'}, \id} &= \specProjectHelper{\tauStack, \id} ~~ \text{if $id \neq id'$}\\
    \specProjectHelper{\tauStack \cdot \commitObsx{\id'}, \id} &= \specProjectHelper{\tauStack, \id}\\
    \specProjectHelper{\tauStack \cdot \tau, \id} &= \specProjectHelper{\tauStack, \id} \cdot \tau ~~\text{otherwise}
\end{align*}
\end{definition}